\documentclass[11pt]{article}
\pdfoutput=1

\usepackage{amsmath,amssymb,amsthm,amscd,graphicx}
\usepackage{psfrag}
\usepackage{xcolor}
\usepackage[english]{babel}
\usepackage{float}\input epsf.sty
\usepackage{slashed}
\usepackage{cite}
\usepackage[small]{caption}
\usepackage{graphicx}
\usepackage[export]{adjustbox}

\makeatletter
\def\section{\@startsection {section}{1}{\z@}{-3.5ex plus -1ex minus
 -.2ex}{2.3ex plus .2ex}{\large\bf}}
\def\subsection{\@startsection{subsection}{2}{\z@}{-3.25ex plus -1ex
minus -.2ex}{1.5ex plus .2ex}{\normalsize\bf}}
\makeatother
\makeatletter

\@addtoreset{equation}{section}

\makeatother

\usepackage{color}

\theoremstyle{definition}

\newcommand{\CF}{{\cal F}}

\newcommand{\CL}{{\cal L}}

\newcommand{\CN}{{\cal N}}
\newcommand{\CO}{{\cal O}}

\def\IZ{{\mathbb Z}}
\def\IR{{\mathbb R}}

\newcommand{\tr}{{\rm Tr}}
\newcommand{\re}{{\rm e}}
\newcommand{\ri}{{\rm i}}
\newcommand{\rd}{{\rm d}}
\newcommand{\rk}{{\rm k}}
\newcommand{\rf}{{\rm f}}

\newcommand{\mQ}{\mathsf{Q}}

\newcommand{\mH}{\mathsf{H}}

\newcommand{\be}{\begin{equation}}
\newcommand{\ee}{\end{equation}}
\newcommand{\ba}{\begin{aligned}}
\newcommand{\ea}{\end{aligned}}
\newcommand{\ben}{\begin{eqnarray}\displaystyle}
\newcommand{\een}{\end{eqnarray}}

\newdimen\tableauside\tableauside=1.0ex
\newdimen\tableaurule\tableaurule=0.4pt
\newdimen\tableaustep
\def\phantomhrule#1{\hbox{\vbox to0pt{\hrule height\tableaurule width#1\vss}}}
\def\phantomvrule#1{\vbox{\hbox to0pt{\vrule width\tableaurule height#1\hss}}}
\def\sqr{\vbox{%
  \phantomhrule\tableaustep
  \hbox{\phantomvrule\tableaustep\kern\tableaustep\phantomvrule\tableaustep}%
  \hbox{\vbox{\phantomhrule\tableauside}\kern-\tableaurule}}}
\def\squares#1{\hbox{\count0=#1\noindent\loop\sqr
  \advance\count0 by-1 \ifnum\count0>0\repeat}}
\def\tableau#1{\vcenter{\offinterlineskip
  \tableaustep=\tableauside\advance\tableaustep by-\tableaurule
  \kern\normallineskip\hbox
    {\kern\normallineskip\vbox
      {\gettableau#1 0 }%
     \kern\normallineskip\kern\tableaurule}%
  \kern\normallineskip\kern\tableaurule}}
\def\gettableau#1{\ifnum#1=0\let\next=\null\else
\squares{#1}\let\next=\gettableau\fi\next}

\usepackage[colorlinks,linkcolor=black,citecolor=blue,urlcolor=blue,linktocpage,pagebackref]{hyperref}

\allowdisplaybreaks

\begin{document}

\thispagestyle{empty}

\begin{center}

	\vspace*{-.6cm}

	\begin{center}

		\vspace*{1.1cm}

		{\centering \Large\textbf{Resurgence and $1/N$ Expansion in Integrable Field Theories}}

	\end{center}

	\vspace{0.8cm}
	{\bf Lorenzo Di Pietro$^{a,b}$, Marcos Mari\~no$^{c}$, Giacomo Sberveglieri$^{d,b}$ and Marco Serone$^{d,b}$}

	\vspace{1.cm}
	
	${}^a\!\!$
	{\em  Dipartimento di Fisica, Universit\`a di Trieste, \\ Strada Costiera 11, I-34151 Trieste, Italy}
		
	\vspace{.3cm}

	${}^b\!\!$
	{\em INFN, Sezione di Trieste, Via Valerio 2, I-34127 Trieste, Italy}
	
		\vspace{.3cm}

	${}^c\!\!$
{\em	D\'epartement de Physique Th\'eorique et Section de Math\'ematiques}\\
{\em Universit\'e de Gen\`eve, Gen\`eve, CH-1211 Switzerland}

	\vspace{.3cm}

	${}^d\!\!$
	{\em SISSA, Via Bonomea 265, I-34136 Trieste, Italy}

	\vspace{.3cm}

\end{center}

\vspace{1cm}

\centerline{\bf Abstract}
\vspace{2 mm}
\begin{quote}

In theories with renormalons the perturbative series is factorially divergent even after restricting to a given order in $1/N$, making the $1/N$ expansion a natural testing ground for the theory of resurgence.
We study in detail the interplay between resurgent properties and the $1/N$ expansion in various integrable field theories with renormalons. We focus on the free energy in the presence of a chemical potential coupled to a conserved charge, which can be computed exactly with the thermodynamic Bethe ansatz (TBA). In some examples, like the first $1/N$ correction to the free energy in the non-linear sigma model, the terms in the $1/N$ expansion can be fully decoded in terms of a resurgent trans-series in the coupling constant. 
In the principal chiral field we find a new, explicit solution for the large $N$ free energy which can be written as the median resummation of a trans-series with infinitely many, analytically computable IR renormalon corrections. However, in other examples, like the Gross-Neveu model, each term in the $1/N$ expansion includes non-perturbative corrections which can not be predicted by a resurgent analysis of the corresponding perturbative series. We also study the properties of the series in $1/N$. In the Gross-Neveu model, where this is convergent, we analytically continue the series beyond its radius of convergence and show how the continuation matches with known dualities with sine-Gordon theories. 

\end{quote}

\newpage

\tableofcontents

\section{Introduction}
 
Since its discovery \cite{Stanley:1968gx,tHooft:1973alw}, the $1/N$ expansion has played an important r\^ole as a tool to study non-perturbative aspects of quantum field theory. 
Many phenomena which are invisible in ordinary perturbation theory, like spontaneous chiral 
symmetry breaking or the dependence on the theta angle, can be discovered already in the large $N$ limit of four-dimensional 
gauge theories \cite{coleman-witten, witten-current}. In two-dimensional models one can even obtain explicit quantitative results at large $N$ for the 
chiral condensate \cite{gross-neveu} or the topological susceptibility \cite{dadda}. 

The reason that is often given for these successes is that the $1/N$ expansion ``resums" the perturbative series, 
and therefore it goes beyond what is available in perturbation theory. 
However, making this statement precise requires being more explicit about what we mean by resummation. 
It is known since the earlier work \cite{knn,Brezin:1977sv} that the factorial growth of diagrams in perturbation theory is tamed to just an exponential growth at the planar level in large $N$
matrix model QFTs. A similar phenomenon applies to higher order in $1/N$ and QFTs based on vector models.
The reduced number of diagrams at fixed order in $1/N$ leads in many theories to expressions that are analytic functions 
at the origin of the fixed large $N$ coupling constant, order by 
order in $1/N$.\footnote{By ``expressions" we refer here to physical, and hence renomalization scheme independent, quantities. At each order in $1/N$, the analyticity properties in the coupling
of unphysical quantities, such as beta-functions, depend on the scheme. For instance, in the limit of large number of flavours $n_f$, the first orders in $1/n_f$ of 4d QED $\beta$-functions 
are analytic in the $\overline{\rm MS}$ scheme, while they are non-analytic in other schemes \cite{Broadhurst:1992si}.} 
Theories of this kind include zero-dimensional (0d) matrix models, $\CN=4$ super Yang--Mills theory in 4d, 3d $O(N)$ models, or Chern--Simons--matter theories.

In many theories, however, the coefficients in the perturbative series still grow factorially after restricting 
oneself to a fixed order in $1/N$. This happens when the growth is not dominated by the proliferation of diagrams, 
but by integration over momenta. The $1/N$ expansion can tame the first growth, but not the second one. This is the phenomenon 
of renormalons (see e.g. \cite{beneke} for a review). Therefore, the way the $1/N$ expansion resums the perturbative series in a theory with renormalons 
must be very different from what happens in the theories without renormalons that have attracted more attention. 

The most general framework to resum perturbative series is the theory of resurgence, which has been extensively studied in 
recent years (see e.g. \cite{mmlargen,abs} for a review and references). 
In this theory, conventional perturbative series have to be extended to 
more general objects called trans-series, which include exponentially small 
corrections. This trans-series can be obtained from the perturbative sector by a detailed study of 
the singularities in the Borel plane. There is growing evidence that many exact quantities in 
quantum theory can be obtained as Borel resummations of these trans-series. These include energy 
levels in quantum mechanics \cite{voros-quartic,ddpham,zjj1, zjj2,power,Serone:2016qog}, $1/N$ expansions in matrix 
models \cite{mmnp,csv}, and perturbative expansions in some quantum field theories \cite{gmp,mr-long,mr-hub,borinsky-dunne,abbh1,abbh2,ggm2}. 

We can now ask the following question. In a theory which admits a $1/N$ expansion and has renormalons, 
each order in the $1/N$ series is a non-perturbative function which 
resums perturbation theory. Can we decode each of these functions in terms of the 
conventional perturbative series, plus its associated trans-series? Can we in principle recover
each of these non-perturbative functions from the perturbative series? In other words, what is the interplay between the resurgent structure of perturbation theory and the $1/N$ expansion? 

An additional reason to ask this question is the following. The resurgent structure of fully-fledged quantum field theories is quite intricate. We 
know however that quantum field theories tend to become simpler and more tractable in the $1/N$ expansion. One can then hope that by looking at the 
large $N$ limit one will find somewhat simpler resurgent structures which can be studied analytically. 

Another, more difficult question concerns the nature of the $1/N$ expansion itself. It is well-known that, 
even in zero-dimensional models, this expansion grows 
factorially or doubly-factorially, and the resummation and resurgent properties of this expansion have been studied in detail in toy theories, like 
matrix integrals (see \cite{mmlargen} for a review and references). There has been much less progress in quantum field theory, 
due among other things to the difficulty of going to large order in the $1/N$ expansion.

In order to address these questions as concretely as possible, it is convenient to look at 
models which have renormalons and at the same time can be studied 
in detail, both in perturbation theory and in the $1/N$ expansion. The ideal candidates for such a 
study are asymptotically free theories in two dimensions which are 
integrable, i.e. their S-matrices are known exactly. In this paper we will focus on three 
classical examples: the $O(N)$ non-linear sigma model (NLSM), the $SU(N)$ principal 
chiral field (PCF), and the $O(N)$ Gross--Neveu (GN) model.\footnote{As a matter of fact, the existence of renormalon singularities has been analytically 
established only in integrable models at large $N$. They were found in the GN model in \cite{gross-neveu} and studied in some detail in the NLSM, see e.g. \cite{David:1982qv,David:1983gz,Novikov:1984ac}.}
 It was noted long ago by 
Polyakov and Wiegmann that a Thermodynamic Bethe Ansatz (TBA) can be used to compute exactly the free energy of these theories in the presence of an 
external field coupled to 
a conserved current \cite{pw}. In addition, when the external field is large, one can use asymptotic 
freedom to calculate this 
observable in perturbation theory, and this was exploited in \cite{hmn,hn,pcf,fnw1,fnw2,h-pcf,eh-ssm,eh-scpn} to obtain 
the relation between the 
mass gap and the dynamically generated scale (see \cite{eh-review} for a review). In addition, a powerful 
method developed in \cite{volin, volin-thesis} makes it possible 
to extract the perturbative series for the free energy at very high orders. This has led to many 
quantitative studies of renormalon physics and resurgence in relativistic 
\cite{mr-ren, abbh1,abbh2} and non-relativistic \cite{mr-ll,mr-long, mr-hub, mr-roads} integrable 
quantum field theories. 

These quantum integrable models have been also studied in the $1/N$ expansion 
\cite{fnw2,fkw1,fkw2,zarembo, ksz,mmr}. Our aim in this paper is to elaborate on and 
extend this line of research in order to answer in detail the questions 
raised above. 
Before diving into all the details of the 2d QFT models, we start in section \ref{sec:OI} by considering the $0d$ reduction 
of certain large $N$ vector models.
We will show explicitly that each order in $1/N$ is analytic in the 't Hooft coupling, the large $N$ expansion is factorially divergent, while 
the reduction to 0d of the free energy ${\cal F}(h)$ defined in \eqref{eq:freeenergy} turns out to be analytic at $N=\infty$.
In section \ref{sec-fe-int} we come back to field theory and introduce the key observable we will consider in this paper, the free energy  ${\cal F}(h)$
as a function of a chemical potential $h$. We review how this can be computed by using the TBA in 2d integrable QFTs.
The main results of the paper are reported in sections \ref{sec:NLSM}, \ref{sec:PCF}, and \ref{sec:GN}.

In section \ref{sec:NLSM} we consider the NLSM. We compute ${\cal F}(h)$ at the leading and next-to-leading order in the $1/N$ expansion, which 
resums an infinite number of renormalon diagrams appearing in ordinary perturbation theory, 
described in detail in \cite{mmr}. We extract the exact answer both from a direct QFT calculation and from the TBA equations. At this order in $1/N$ 
there is a single IR renormalon singularity and the exact answer is obtained by the so-called median Borel 
resummation of the perturbative series.  We have also studied the properties of the $1/N$ expansion of ${\cal F}(h)$ by exploiting the TBA
to generate several terms. Our explicit results do not show factorial growth and are inconclusive. Either the asymptotic regime
has not been reached yet or the $1/N$ series is actually convergent. A similar analysis has also been made in the PCF model, with the same inconclusive result.

In section \ref{sec:PCF} we consider the PCF model. We find a new, explicit solution for ${\cal F}(h)$ at leading order in $1/N$ from TBA, and 
for the choice of charges used in \cite{pcf}. The exact answer can be understood as the median resummation of a non-trivial trans-series which can be obtained analytically and has an infinite number of IR renormalon singularities (in contrast to the solution of \cite{fkw1,fkw2}, which has a single IR renormalon singularity, and similar 
to the numerical results obtained in \cite{abbh1,abbh2} for the $O(4)$ NLSM). Therefore, in this case the large $N$ limit provides an explicit, analytic, yet non-trivial example of resurgence and median resummation in a model with infinitely many IR renormalon corrections.  

In section \ref{sec:GN} we consider the GN model. In this case, a new phenomenon appears: at each order in the $1/N$ expansion, 
the exact answer includes an infinite number of non-perturbative corrections. While ambiguities in imaginary terms nicely cancel between one series and the next in the trans-series, as expected from resurgence,
real non-perturbative corrections can {\it not} be obtained from the resurgent properties of the perturbative series. Therefore, in this case 
there is a tension between resurgence and the $1/N$ expansion. The $1/N$ series of ${\cal F}(h)$ in the GN model turns out to be convergent, with a finite radius of convergence.
Using the TBA, we generate many terms in the $1/N$ series. The latter can be analytically continued beyond its radius of convergence. Interestingly, the analytic continuation of this series gives reliable results for small $N$, such as $N=4$ and $N=2$, which are in agreement with the well-known dualities between these models and sine-Gordon theories. 

In section \ref{sec:conclusions} we conclude with a detailed discussion of our findings in the more general context of the theory of resurgence, and we present various directions for future work. We report in appendix \ref{app:OI} a few technical details needed to reproduce some results of section \ref{sec:OI}. In appendix \ref{app:analytic} we present the explicit form of the kernels entering the TBA equations \eqref{chi-ie} and discuss their analyticity properties in $1/N$.

  \section{Ordinary integrals at large $N$}

\label{sec:OI}

Before analyzing the 2d QFT models it is useful to consider ordinary integrals, where we can get complete analytic results
and show the generic divergent nature of the $1/N$ perturbative series. 

A notable example is given by the 0d reduction of the large $N$ quartic vector models, given by 
\be
I(m,g) = \frac{1}{(2\pi)^{N/2}} \int_{-\infty}^{+\infty} \! \!\! \rd^N \! \boldsymbol{x}  \; \re^{-f(\boldsymbol{x}, m, g)}\,.
\label{eq:OI1}
\ee
Here $\boldsymbol{x}=(x_1, \ldots, x_N)$ is a set of $N$ real variables, 
\be 
f(\boldsymbol{x},m,g) = \frac m2\boldsymbol{x}\cdot \boldsymbol{x}  + \frac g N (\boldsymbol{x}\cdot \boldsymbol{x} )^2\,,
\label{eq:OI2}
\ee
with $m\in \mathbb{R}$ and $g>0$.  We can trivially rescale $m$, so we get three different cases:
$m=1$, $m=-1$, and $m=0$. The integral in \eqref{eq:OI1} can be computed analytically, but we won't need its exact expression. 
The large order behavior of the $1/N$ expansion can be obtained by 
using steepest descent methods, see appendix \ref{app:OI} for details. We have
\be
I(m,g) \sim 2^{-N/2} \re^{-N K(z_c)} |K^{\prime\prime}(z_c)|^{-1/2} \bigg(1+ \sum_{p=1}^\infty \frac{c_{p}}{N^{p} }\bigg)\,,
\label{eq:HS6}
\ee
where the symbol $\sim$ in \eqref{eq:HS6} reminds us that the right-hand side is a divergent asymptotic series, and $K(z)$ is the function defined in \eqref{eq:HS3}.
For $p\gg1$ the coefficients $c_p$ in \eqref{eq:HS6} read 
\be
c_p \approx  \frac{\hat  I_c}{\pi}  \Gamma(p) \rho^{-p} \sin (p \theta)\,,
\label{eq:OI15}
 \ee
 where $\hat I_c$, $\rho$ and $\theta$ are explicit functions of $g$ and $m$ reported in \eqref{eq:HS8} and \eqref{eq:OI14}. As discussed in the appendix \ref{app:OI}, the $1/N$ expansion of  \eqref{eq:OI1} is divergent asymptotic and 
Borel resummable for any real value of $m$ and $g>0$.  This result should be contrasted with what we would get by expanding in $g$ at {\it fixed N}. In this case, taking $N$ to be odd, we can use radial coordinates with radial variable $r$, so that in a $g$ expansion the relevant saddle points of the integral \eqref{eq:OI1} are those of the function 
\be
f(r) = \frac{m}{2} r^2 + \frac{1}{4}r^4\,.
\ee
The qualitative and quantitative behaviors of such series are well known. In particular, for $m=1$ we get one real critical point at $r=0$ and a Borel resummable expression, for $m=-1$ three real critical points and Borel summability is lost, while for $m=0$ the three critical points are degenerate and no expansion is possible.

Given the relations \eqref{eq:HS8} and \eqref{eq:OI14}, we can easily get the analyticity properties of the coefficient term $c_p$ as a function of the coupling $g$. 
In particular, we see that $c_p=c_p(g)$ are {\it analytic} at $g=0$ for any $p$ and go like
\be
\lim_{g\rightarrow 0} c_p(g)  \sim  g^{p+1} +  {\cal O}(g^{p+2}) \,.
\ee

Other useful examples are given by the 0d reductions of two of the three models considered in this paper, namely the NLSM and the GN models.

The 0d reduction of the non-linear sigma model is essentially the $S^{N-1}$ sphere. We can define
\be
\re^{-F_{\rm NLSM}(0)} \equiv \int_{-\infty}^{+\infty} \!\!\rd^{N}\!  \boldsymbol{x} \, \delta(\boldsymbol{x}\cdot \boldsymbol{x}-N) = \frac{1}{2} \Omega_{N}  N^{\frac{N-2}{2}}\,,
\label{eq:NLSMToy1}
\ee
with $\Omega_N =2\pi^{N/2}/\Gamma(N/2)$ the volume of the $S^{N-1}$ sphere.  The large $N$ expansion reduces essentially to the 
Stirling approximation of the Gamma function, which is well-known 
to be divergent asymptotic. In presence of a chemical potential $h$, the vacuum energy becomes  
 \be
 \ba
\re^{-F_{\rm NLSM}(h)}& \equiv   \int_{-\infty}^{+\infty} \!\!\rd^{N}\!  \boldsymbol{x} \, \delta(\boldsymbol{x}\cdot \boldsymbol{x} -N) \re^{\frac{h^2}{2} (x_1^2+x_2^2)} \\
& =  \left(-\frac{h^2}{2}\right)^{\frac{2-N}{2}} \frac{\pi^{\frac{N}{2}} }{\Gamma(\frac{N-2}{2})}  \re^{-\frac{h^2 N}{ 2}} \gamma\left(\frac{N-2}{2},-\frac{h^2 N}{2}\right)  \,,
\label{eq:NLSMToy7}
\ea
\ee
where $\gamma(a,z) = \Gamma(a) - \Gamma(a,z)$ is the incomplete Gamma function.
The behavior of the $1/N$ expansion is now determined by the expansion of $\gamma(a,z)$ for large $a$ and $z$, at fixed ratio $z/a$. This can be found e.g. in \cite{DLMF}, see eq.(8.11.6). After simple algebraic manipulations, we have
\be
\re^{-(F_{\rm NLSM}(h) -F_{\rm NLSM}(0))} =  \sum_{k=0}^\infty \frac{Q_k(h^2) }{(1+h^2)^{2k+1}}\frac{1}{N^k} \,,
\label{eq:NLSMToy6}
\ee
where $Q_k(h^2)$ are polynomials of degree $k$ in $h^2$ for $k>0$ and $Q_0=1$. It can be shown that the above series is absolutely convergent for any real $h$ 
for $N>3$. Interestingly enough, while $F_{{\rm NLSM}}(h)$ and $F_{{\rm NLSM}}(0)$ are separately non-analytic at $N=\infty$, their difference $F_{{\rm NLSM}}(h) - F_{{\rm NLSM}}(0)$ is a well-defined and analytic function. 

The 0d reduction of the Gross-Neveu model is given by the following Grassmann integral:
\be
\re^{-F_{{\rm GN}}(0)} \equiv \sqrt{\frac{N}{2\pi}} \int \rd^N\!\boldsymbol{\chi} \rd^N\!\bar{\boldsymbol{\chi}} \, \re^{\frac{1}{2N} (\bar{\boldsymbol{\chi}}\cdot \boldsymbol{\chi})^2}\,,
\label{eq:GNToy1}
\ee
where $\boldsymbol{\chi}=(\chi_1,\chi_2,\ldots, \chi_{2N})$ is a set of $2N$ complex Grassmannian variables.
Introducing an Hubbard-Stratonovich like parameter as in \eqref{eq:HS1} we get
\be
\re^{-F_{{\rm GN}}(0)} = 2^{N-\frac{1}{2}} \frac{\Gamma\big(N+\frac12\big)}{N^{N+\frac 12}}\,.
\label{eq:GNToy2}
\ee
The large $N$ expansion of this result is again manifestly divergent asymptotic. In presence of a chemical potential $h$, the vacuum energy becomes  
\be
\re^{-F_{{\rm GN}}(h)} \equiv \sqrt{\frac{N}{2\pi}} \int \!\! \rd^N\!\boldsymbol{\chi} \rd^N\!\bar{\boldsymbol{\chi}} \, \re^{\frac{1}{2N} (\bar{\boldsymbol{\chi}}\cdot \boldsymbol{\chi})^2+ h \sum_{i=1,2}\bar \chi_i \chi_i} 
= \Big(\frac{2}{N}\Big)^N \frac{(2+h)^2N-1}{\sqrt{2N}} \Gamma\left(N-\frac12\right) \,,
\label{eq:GNToy3}
\ee
where the last line is readily computed again introducing an Hubbard-Stratonovich like parameter. We finally have
\be
\re^{-(F_{{\rm GN}}(h) - F_{{\rm GN}}(0))} =  1+ \frac{N}{2N-1} h^2\,.
\label{eq:GNToy5}
\ee 
Like in the NLSM case, $F_{{\rm GN}}(h)$ and $F_{{\rm GN}}(0)$ are separately non-analytic at $N=\infty$, but their difference $F_{{\rm GN}}(h) - F_{{\rm GN}}(0)$ is a well-defined, simple and analytic function. 

Summarizing, we have found that the $1/N$ expansion in 0d reductions of large N vector models is asymptotic, and 
each coefficient in the $1/N$ expansion is analytic in the t' Hooft coupling. In agreement with what was anticipated
in the introduction, the factorial growth of diagrams in perturbation theory is reduced to exponential growth, 
order by order in $1/N$. In contrast, the coefficients in the $1/N$ expansion we will compute 
in the 2d models will generally be non-analytic in the t' Hooft coupling because (and only because) 
of the presence of renormalon singularities. 
We have also shown that the $1/N$ expansion of the relative free energy $F(h)-F(0)$ is better 
behaved than $F(0)$ and is convergent in the 0d reduction of both the NLSM and the GN models. 
This suggests that the relative free energy can have better convergent properties in $1/N$ also in 
the 2d models. It will be explicitly verified that the $1/N$ expansion of this quantity 
is indeed convergent in the 2d GN model, while we will not be able to draw firm conclusions on its nature in the NLSM and PCF models.


 \section{Free energy and integrability}
 \label{sec-fe-int}
 
We summarize in this section the general formulation which applies to the three integrable and asymptotically free models  considered in the paper. 
More details for each model will be spelled out in subsequent sections. 

The key observable we will study in this paper is the free energy $F(h)$ as a function of an external field $h$ coupled to a conserved charge. 
Let $\mH$ be the Hamiltonian of the theory and $\mQ$ the charge associated to a global conserved 
current. The external field $h$ can be regarded as a chemical potential, and we can consider the ensemble defined by the operator
\be
\label{HQ}
\mH- h \mQ.
\ee
The corresponding free energy per unit volume is defined by 
\be
\label{free-en}
F(h) =-\lim_{V, \beta \rightarrow \infty} {1\over V\beta } \log \, \tr \, \re^{-\beta (\mH-h \mQ)}, 
\ee
where $V$ is the volume of space and $\beta$ is the total length of Euclidean time. More precisely, the observable of interest will be the relative free energy
\be
\CF(h) \equiv  F(h)-F(0)\,.
\label{eq:freeenergy}
\ee
From now on, for simplicity, we will refer to $\CF(h)$ just as the free energy. 
  
It was pointed out in \cite{pw} that in integrable quantum field theories one can calculate  $\CF(h)$ by using the exact $S$-matrix 
and the TBA ansatz, in terms of a linear integral equation. The basic physical intuition behind is the following. Let $m$ be the mass gap of the integrable theory.
If the lightest particles in the theory are charged under $Q$, for $h>m$ the  ground state of the theory will no longer be the vacuum, but a state with 
non-vanishing number density $\rho$. The latter can be determined in terms  of Bethe roots $\chi(\theta)$ by the TBA equation 
\be
\label{chi-ie}
\chi(\theta)-\int_{-B}^B  \rd \theta' \, K(\theta-\theta') \chi (\theta')=m \cosh \theta\,, 
\ee
where $\theta$ is the particle rapidity and $\chi(\theta)$ is supported on the interval (to be determined) $[-B,B]$.

The integral kernel appearing in the Bethe ansatz equation is given by 
\be
\label{int-kernel}
K(\theta)={1\over 2 \pi \ri} {\rd \over \rd\theta} \log S(\theta),
\ee
where $S(\theta)$ is the $S$-matrix element of the particles populating the ground state. In all the cases considered in this paper
only one species of particles with definite charges are present, so $S$ is a scalar quantity.
The energy per unit length $e$ and the density $\rho$ are given by
\be
e={m \over 2 \pi} \int_{-B}^B \rd \theta \, \chi(\theta)  \cosh \theta, \qquad \rho={1\over 2 \pi} \int_{-B}^B \rd \theta\, \chi(\theta). 
\ee
The value of $B$ is fixed by the density $\rho$ and one can eventually obtain an equation of state relating $e$ to $\rho$. 
The free energy $\CF(h)$ is finally obtained by a Legendre transform of $e(\rho)$:
\be
\begin{split}
h  &\equiv \partial_\rho e(\rho) \,, \quad \quad \;\;\; {\cal F}(h)  \equiv e(\rho) - \rho h\,, \\
\rho & = - \partial_h {\cal F}(h) \,, \quad \quad  e(\rho)  = {\cal F}(h)  + \rho h\,.
\label{eq:LNPTC3}
\end{split}
\ee
In an alternative  formulation of the TBA equations, the basic quantity is a function $\epsilon (\theta)$, with support on an interval $[-B,B]$, which 
describes physically the excitation of holes. This function satisfies the integral equation 
\be
\label{intetwo}
\epsilon(\theta) -\int_{-B}^{B} \rd\theta'\, K(\theta-\theta') \epsilon(\theta')=h-m \cosh \theta\,,
\ee
where now the value of $B$ is determined by the condition
\be
\epsilon(\pm B)=0, 
\label{eq:Bcondition}
\ee
and depends on the external field $h$. The free energy is then given by
\be
\label{fh-bethe}
\CF(h)=  -{m \over 2 \pi}\int_{-B}^B \rd \theta  \, \cosh \theta \epsilon(\theta)\,.
\ee
We refer the reader to appendix \ref{app:analytic} for the explicit form of the kernel $K(\theta)$ in the three models,
for a detailed discussion of the existence and uniqueness of the solutions of \eqref{chi-ie} and \eqref{intetwo}, as well as for the  analyticity properties in $1/N$ of $K(\theta)$ in each case.

Given the free energy $\CF(h)$, we denote by $\CF_k(h)$ its coefficients in a  $1/N$ expansion (see \eqref{eq:FNLSM}-\eqref{eq:FGN} below for the precise definition for each model).
The coefficients $\CF_k(h)$ are non-perturbative functions of the external field $h$ and the mass gap $m$ in each model.  In order to recast the results 
in terms of asymptotic expansions of ordinary perturbation theory, we have to define a running coupling constant of some kind.
Let us denote by $g$ the coupling constant (before the large $N$ limit) appearing  in the Lagrangian description of our models,  
with beta function given by  
\be
\label{betaf}
\beta(g)= \mu {\rd g \over \rd \mu} =-\beta_0 g^3 - \beta_1 g^5 + {\cal O}(g^7)\,,
\ee
with $\beta_0>0$. Since in all the three models considered $\beta_0 \propto N$
we can conveniently  define a 't Hooft-like coupling as
\be
\alpha \equiv 2 \beta_0 g^2\,,
\label{eq:LNPTC4}
\ee
so that the $\beta$-function up to two loops reads
\be
\beta(\alpha) = - \alpha^2 - \xi \alpha^3 + {\cal O}(\alpha^4)\,, 
\label{eq:LNPTC5}
\ee
where
\be
\label{xi-constant}
\xi = \frac{\beta_1}{2 \beta_0^2}\,.
\ee
As it is well-known, the first two terms are renormalization scheme-independent, while all the others are not. 
A useful definition of running coupling is\footnote{Note that the definition \eqref{eq:LNPTC6} slightly differs from the one originally defined in \cite{bbbkp} and
used in subsequent works where, inside the log, $m$ is replaced by the dynamically generated mass scale $\Lambda$. The two definitions are equivalent, but for 
our purposes  \eqref{eq:LNPTC6} is more convenient.}
\be
\frac{1}{\alpha(\mu)} + \xi  \log \alpha(\mu) \equiv \log\Big( \frac{\mu}{m} \Big)\,.
\label{eq:LNPTC6}
\ee
Applying $\mu \partial_\mu$ to \eqref{eq:LNPTC6} gives
\be
\beta_\alpha^{{\rm TBA}}  = \mu \frac{d\alpha}{d\mu} = - \frac{\alpha^2} {1- \xi  \alpha } = - \alpha^2 - \xi \alpha^3 + \ldots \,.
\label{eq:LNPTC7}
\ee
So we see that $\alpha$ is a plausible coupling of the integrable model, in a renormalization scheme where the $\beta$-function has exactly the form \eqref{eq:LNPTC7}. 
We will refer to this renormalization scheme as the TBA scheme in the following. 
In the three models we expand the free energy as
\ben
\CF_{{\rm NLSM}} (h)& = &  \sum_{k \ge 0} \CF_k(h) {\Delta}^{k-1}    \sim -\frac{h^2}{4\pi} \sum_{k \ge 0}  \Phi_k(\alpha,C_\pm) \Delta^{k-1} ,  \label{eq:FNLSM} \\
 \CF_{{\rm PCF}} (h) &= &\sum_{k \ge 0} \CF_k(h) {\overline \Delta}^{k-1}   \sim -\frac{h^2}{8\pi} \sum_{k \ge 0}  \Phi_k(\alpha,C_\pm) \overline \Delta^{k-1}  \,,  \label{eq:FPCF}  \\
 \CF_{{\rm GN}} (h)& = & \sum_{k \ge 0} \CF_k(h) {\Delta}^{k} \sim - \frac{h^2}{2\pi} \sum_{k \ge 0}  \Phi_k(\alpha,C_\pm) \Delta^{k}  \,. \label{eq:FGN} 
\een
Several clarifications are in order. In \eqref{eq:FNLSM}-\eqref{eq:FGN}
\be
 \Delta \equiv  \frac{1}{N-2}\,, \quad \quad \overline\Delta \equiv  \frac{1}{N}\,,
 \label{eq:DefDDbar}
\ee
the numerical factors  have been chosen for convenience and $\alpha$ is the coupling \eqref{eq:LNPTC6} evaluated at a convenient scale $\mu \sim h$ that 
will be spelled out in detail for each model in the next sections.
In order to clearly distinguish exact quantities from their asymptotic expansions, we have denoted by ${\cal F}_k(h)$ and $\Phi_{k}(\alpha,C_\pm)$ the exact and the asymptotic expansion in $\alpha$ of the $1/N$ coefficients of ${\cal F}(h)$. The factors $\Phi_{k}$ are trans-series of the form
\be
\Phi_k(\alpha,C_\pm) = \varphi_k^{(0)}(\alpha)+ \sum_{\ell=1}^\infty \re^{-\frac{2\ell}{\alpha}} \varphi_k^{(\ell)}(\alpha, C_\pm)\,,
\label{eq:TransPhi}
\ee
where $\varphi_k^{(\ell)}$ are in general asymptotic divergent series. $\varphi_k^{(0)}$ coincides with the perturbative series, while 
$\varphi_k^{(\ell)}$ with $\ell>0$ are the non-perturbative contributions associated with the trans-series. The latter can present a two-fold ambiguity, encoded in the trans-series parameters $C_\pm$. This ambiguity is balanced by the one due to the non-Borel summability (because of the IR renormalons) of the asymptotic series $\varphi_k^{(\ell)}$ in a way that will be detailed for each model in the next sections.
The symbol $\sim$ stands for ``asymptotically equivalent to".\footnote{For simplicity, and with an abuse of notation, we have used 
the equality sign in the first relations of \eqref{eq:FNLSM}-\eqref{eq:FGN}, though the convergence of the $1/N$ expansion of ${\cal F}$ is established only for the GN model.} 
Note finally that in the NLSM and in the GN models the $\Phi_k$'s do not precisely correspond to the expansions of the ${\cal F}_k$'s because
the relation between $\alpha(\mu\sim h)$ and $h$ given by \eqref{eq:LNPTC6} depends on $\Delta$ by means of the factor $\xi$.


\section{The non-linear sigma model}
\label{sec:NLSM}

In this section we present our results for the NLSM. After reviewing general aspects of the theory, we will present two approaches to obtain the exact $\mathcal{F}(h)$ up to next-to-leading order, one based on the diagrammatic calculation in the $1/N$ expansion, and another one based on the expansion of the integral equation from the Bethe ansatz. We will then compare this non-perturbative result with the resummation of the perturbative calculation. 

\subsection{General aspects} 
The NLSM is described by the Lagrangian density 
\be
\CL={1\over 2} \partial_\mu\boldsymbol{\phi}\cdot \partial^\mu\boldsymbol{\phi}~, 
\ee
where $\boldsymbol{\phi}=\left(\phi^1,\dots,\phi^N\right)$ is an $N$-uple of real scalar fields 
satisfying the constraint
\be 
\label{ctr} 
\boldsymbol{\phi}\cdot\boldsymbol{\phi}=\frac{N}{g^2}~. 
\ee
In our conventions (\ref{betaf}) we have
\be
\label{eq:betaxi_nlsm} 
\beta_0= {1 \over 4 \pi \Delta}~, \qquad \xi=\Delta~. 
\ee
The NLSM has a global $O(N)$ symmetry. The conserved currents are given by
\be
J_{\mu}^{IJ}=\phi^{I}\partial_{\mu}\phi^{J}-\phi^{J}\partial_{\mu}\phi^{I},  
\ee
where $I,J=1,\dots,N$, and we denote by $Q^{IJ}$ the corresponding charges. As in \cite{hmn, hn}, we can add a chemical potential $h$ associated to the charge 
$Q^{12}$.  In the NLSM  it is convenient to define the 't Hooft coupling
\be
\alpha \equiv \alpha (\mu=h)\,,
\ee
where $\alpha(\mu)$ is the TBA coupling defined in \eqref{eq:LNPTC6}.

The free energy $\CF(h)$ was computed in perturbation theory in the coupling constant up to two-loops in \cite{hmn,hn,bbbkp}. 
In terms of the asymptotic expansions defined in \eqref{eq:FNLSM} and \eqref{eq:TransPhi}, we have at leading order in $1/N$ ($k=0$)
\be
\varphi_0^{(0)}(\alpha)= - {1\over \alpha}+{1\over 2}\,.
\label{eq:phi0NLSM}
\ee
At next-to-leading order ($k=1$) the series $\varphi_1^{(0)}(\alpha)$ has been calculated explicitly and at all orders in \cite{mmr}. It is obtained by selecting Feynman diagrams with the appropriate power of $N$, which turn out to be ring diagrams. The resulting series has the factorial 
growth typical of renormalon behavior, due to integration over momenta, and one finds\footnote{Notice that the constant factor is due to the different definition for $\alpha$, i.e. \eqref{eq:LNPTC6}, with respect to the one 
used in \cite{mmr}.}
\be
\label{sigma1}
\varphi_1^{(0)} (\alpha)=  3 \log(2) + \gamma_E-1 +
\frac{\alpha}{2}+ \left(\frac{1}{4}-\frac{21 \zeta (3)}{32}\right) \alpha^2 + \left(\frac{1}{4} +\frac{35 \zeta
   (3)}{32}\right) \alpha^3+ \CO\left(\alpha^4\right). 
   \ee
 We can then ask the question of how the $1/N$ expansion resums this series. We will now present two different ways of computing  $\CF_0(h)$ and
 $\CF_1(h)$ as exact functions of $h$.

 \subsection{ The $1/N$ expansion from QFT}  

The standard way to perform the $1/N$ expansion is to introduce an auxiliary field $\sigma$ which implements the 
constraint (\ref{ctr}) (see e.g. \cite{bzj, mmbook}). 
Once this is done, we obtain the action
\begin{align}\label{eq:NLSMaction}
\begin{split}
S = & \int \rd^2 x \left[\frac12 \partial_\mu \boldsymbol{\phi}_B\cdot\partial^\mu\boldsymbol{\phi}_B +\frac{\sigma_B}{\sqrt{N}}\left(\boldsymbol{\phi}_B\cdot\boldsymbol{\phi}_B -  \frac{N}{g_B^2}\right)\right] \\ 
= & \int \rd^2 x \left[\frac{Z_\phi}{2} \partial_\mu \boldsymbol{\phi}\cdot\partial^\mu\boldsymbol{\phi} + \sqrt{Z_\sigma} Z_\phi \frac{\sigma}{\sqrt{N}}\boldsymbol{\phi}\cdot\boldsymbol{\phi} - Z_g\frac{ \sqrt{N}}{g^2} \sigma +\frac{\mu^2}{2} \boldsymbol{\phi}\cdot\boldsymbol{\phi}+ C \right]~.
\end{split}
\end{align}
The subscript $B$ denotes bare quantities and the second line is a rewriting in terms of renormalized fields and counterterms. The additional couplings for $h\neq 0$ are
\begin{align}
\begin{split}
S_h & =  \int \rd^2 x \left[\ri h (\phi_B^1 \partial_\tau \phi_B^2 -\phi_B^2 \partial_\tau\phi_B^1) -\frac{h^2}{2}\left((\phi_B^1)^2 + (\phi_B^2)^2\right)\right] \\
& =\int \rd^2 x \left[\ri  h Z_\phi (\phi^1 \partial_\tau \phi^2 -\phi^2 \partial_\tau\phi^1) - Z_\phi \frac{h^2}{2}\left((\phi^1)^2 + (\phi^2)^2\right) +C_h \right] ~.
\end{split}
\end{align}
We do not need a vertex renormalization for $h$ because it couples to a conserved current. Moreover turning on $h$ does not introduce any new UV divergence, so all the counterterms in \eqref{eq:NLSMaction} can be taken independent of $h$. Fixing also the finite part of the counterterms to be $h$-independent means that we choose the same scheme for $h=0$ and $h\neq 0$. The only exception is the counterterm $C$, for which we find that a finite $h$-dependent shift $C_h$ is needed in order to satisfy a certain ``renormalization condition'' on the observable that we explain below. 

In order to compute the vacuum energy up to NLO, we need to plug in the action the VEVs including their $1/N$ corrections
\begin{align}
\begin{split}\label{eq:VEVs}
\sigma & = \sqrt{N} (\Sigma +\frac{1}{N}\delta \Sigma) +\hat{\sigma}~,~~\boldsymbol{\phi}  = \sqrt{N} (\boldsymbol{\Phi}+\frac{1}{N}\delta \boldsymbol{\Phi})+\widehat{\boldsymbol{\phi}}~,
\end{split}
\end{align}
where capital letters denote the VEVs and hatted fields denote the fluctuations. Similarly we plug the expansion of the renormalization constants, out of which only $Z_g$ and $C$ are non-trivial already at leading order
\begin{align}
\begin{split}\label{eq:cts}
Z_g & = Z_g^{(0)} + \frac{1}{N}\delta Z_g~,~~C = N C^{(0)}+ \delta C~,~~C_h = \delta C_h~,\\
Z_\phi & = 1 + \frac{1}{N}\delta Z_\phi~,~~Z_\sigma  = 1 + \frac{1}{N}\delta Z_\sigma~,~~\mu^2  =  \frac{1}{N}\delta \mu^2~.
\end{split}
\end{align}
It will be convenient to collect some combinations of counterterms
\begin{align}
\begin{split}\label{eq:ctcollect}
\delta m^2& \equiv \delta\mu^2 + 2 \Sigma \,\delta Z_3  ~,\\
\delta Z_3 & \equiv \delta Z_\phi + \frac{1}{2}\delta Z_\sigma~.
\end{split}
\end{align}
$\delta m^2$ is the total counterterm for the mass-squared coupling, and $\delta Z_3$ the counterterm for the cubic coupling. As we will show, $h\neq 0$ induces a non-zero VEV for $\boldsymbol{\Phi}$ in the $12$ plane, which in turn induces a quadratic mixing between $\hat{\sigma}$, $\hat{\phi}^1$ and $\hat{\phi}^2$. In the following we will draw diagrams with the following conventions
\begin{itemize}
\item[]{$\widehat{\boldsymbol{\phi}}$ propagator:~\includegraphics[width=.1\columnwidth , valign=c]{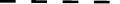}}~~;
\item[]{$\hat{\sigma}$ propagator for $h=0$:~\includegraphics[width=.1\columnwidth , valign=c]{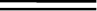}~~;~~$\hat{\sigma}$-$\hat{\phi}^1$-$\hat{\phi}^2$ mixed propagator for $h\neq 0$:~\includegraphics[width=.1\columnwidth , valign=c]{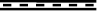}}~~;
\item[]{$\hat{\sigma}$ tadpole vertex:~\includegraphics[width=.01\columnwidth , valign=c]{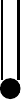}~~;~~VEV insertion:~\includegraphics[width=.02\columnwidth , valign=c]{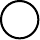}~~;~~Counterterm:~~\includegraphics[width=.02\columnwidth , valign=c]{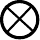}}~~.
\end{itemize}

\subsubsection{Leading order}

The LO (leading order) vacuum energy density for $h=0$ is given by the following diagrams
\begin{equation}
N F_0 = ~~\includegraphics[width=.1\columnwidth , valign=c]{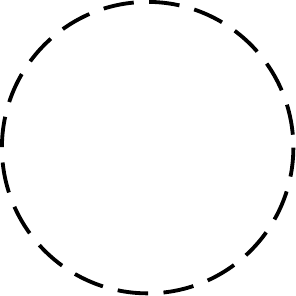}~~+~~\includegraphics[width=.04\columnwidth , valign=c]{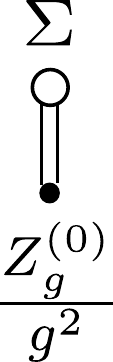} ~~+~~\raisebox{-2.5mm}{\includegraphics[width=.04\columnwidth , valign=c]{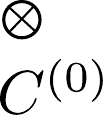}}~,
\end{equation}
subject to the vanishing tadpole condition (this is equivalent to minimizing the effective potential)
\begin{equation}\label{eq:Ntadfd}
\raisebox{1.5 mm}{\includegraphics[width=.1\columnwidth , valign=c]{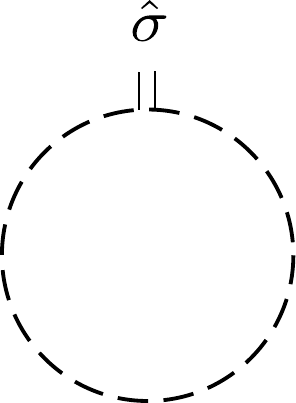}} ~~+~~\raisebox{-1 mm}{\includegraphics[width=.04\columnwidth , valign=c]{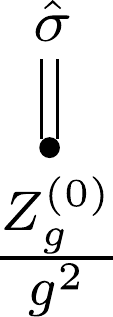}} ~= 0~. 
\end{equation}
We did not write explicitly the tadpole condition for $\hat{\phi}^I$ but it is easily checked that it is solved by $\boldsymbol{\Phi}= 0$. Evaluating the diagrams in the tadpole condition we find
\begin{equation}\label{eq:Ntad}
  -\int \frac{\rd^2 p}{(2\pi)^2} \frac{1}{p^2 + 2 \Sigma} + \frac{Z^{(0)}_g}{g^2} = 0~.
\end{equation}
The solution to this equation gives us the physical mass-squared of the scalars at leading order, that we denote as $m^2_0$, i.e. $\Sigma = \frac{m^2_0}{2}$. Plugging in the diagrams for the vacuum energy we obtain
\begin{align}
\begin{split}
F_0 & =\left[\frac{1}{2}\int \frac{\rd^2 p}{(2\pi)^2} \log[p^2 + 2 \Sigma] -  \frac{Z^{(0)}_g}{g^2} \Sigma+ C^{(0)}\right]_{\Sigma  =\frac{m^2_0}{2}} \\
& = \frac{1}{2}\int \frac{\rd^2 p}{(2\pi)^2} \log[p^2 + m^2_0] - \frac{m^2_0}{2} \int \frac{\rd^2 p}{(2\pi)^2} \frac{1}{p^2 + m^2_0} + C^{(0)}~.
\end{split}
\end{align}

The LO vacuum energy density for $h\neq 0$ is given by
\begin{equation}
N F_0(h) = ~~\includegraphics[width=.1\columnwidth , valign=c]{Figs/VacNh0Bubble.pdf}~~+~~\includegraphics[width=.14\columnwidth , valign=c]{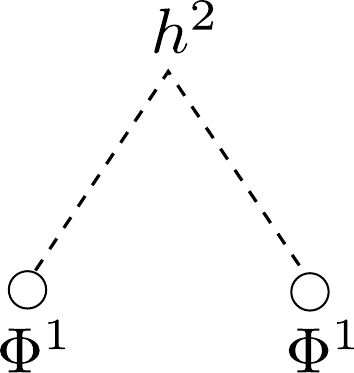}~~+~~\includegraphics[width=.14\columnwidth , valign=c]{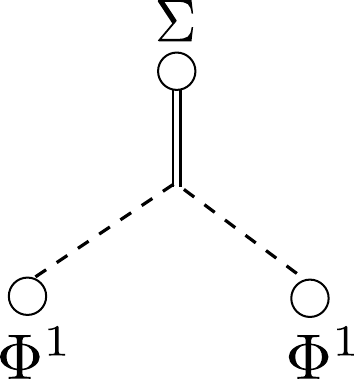}~~+~~\includegraphics[width=.04\columnwidth , valign=c]{Figs/VacNh0Tad.pdf} ~~+~~\raisebox{-2.5mm}{\includegraphics[width=.04\columnwidth , valign=c]{Figs/VacNh0ct.pdf}}~,
\end{equation}
subject to the tadpole conditions (now we cannot avoid considering also the tadpole for $\hat{\phi}$, without loss of generality we assume the VEV to be in the direction $1$)
\begin{align}
& \raisebox{1.5 mm}{\includegraphics[width=.1\columnwidth , valign=c]{Figs/TadNh0Bubble.pdf}} ~~+~~\raisebox{1.5 mm}{\includegraphics[width=.14\columnwidth , valign=c]{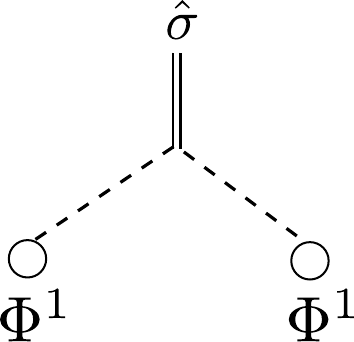}}~~+~~\raisebox{-1 mm}{\includegraphics[width=.04\columnwidth , valign=c]{Figs/TadNh0Tad.pdf}} ~= 0~,\\ 
& \includegraphics[width=.14\columnwidth , valign=c]{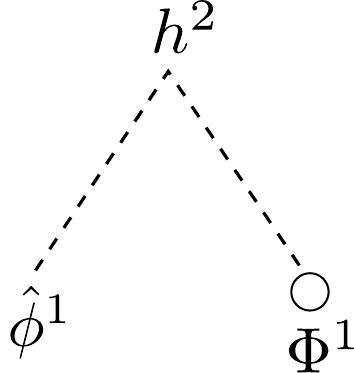}~~+~~\includegraphics[width=.14\columnwidth , valign=c]{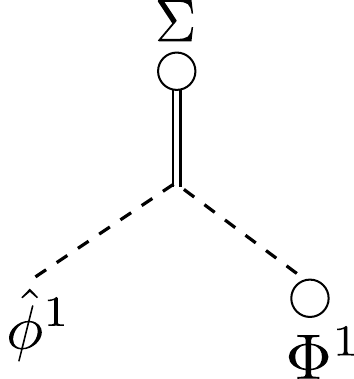}~= 0~.
\end{align}
Assuming $\Phi^1\neq 0$ the tadpole condition for $\hat{\phi}$ has a unique solution for $\Sigma$, namely $\Sigma =\frac{h^2}{2}$. Plugging this in the tadpole condition for $\hat{\sigma}$ we obtain
\begin{equation}
\left.-\int \frac{\rd^2 p}{(2\pi)^2} \frac{1}{p^2 + 2 \Sigma}\right\vert_{\Sigma=\frac{h^2}{2}}- (\Phi^1)^2 + \frac{Z^{(0)}_g}{g^2} = 0~.
\end{equation}
Using the determination of $\frac{Z^{(0)}_g}{g^2}$ in terms of the mass-squared for the theory with $h=0$ in eq. \eqref{eq:Ntad}, we get
\begin{equation}\label{eq:VEVPhi1}
(\Phi^1)^2  = \int \frac{\rd^2 p}{(2\pi)^2}\left(\frac{1}{p^2 + m^2_0} -\frac{1}{p^2 + h^2}\right) =\frac{1}{4\pi} \log \frac{h^2}{m^2_0}~.
\end{equation}
Plugging everything back to the vacuum energy density we obtain
\begin{align}
\begin{split}
F_0(h) & = \biggl[ \frac{1}{2}\int \frac{\rd^2 p}{(2\pi)^2} \log[p^2 + 2 \Sigma]  +\left(\Sigma -\frac{h^2}{2}\right)(\Phi^1)^2-  \frac{Z^{(0)}_g}{g^2} \Sigma+ C^{(0)} \biggr]_{\Sigma  =\frac{h^2}{2}}\\
& = \frac{1}{2}\int \frac{\rd^2 p}{(2\pi)^2} \log[p^2 + h^2] -  \frac{h^2}{2}\int \frac{\rd^2 p}{(2\pi)^2} \frac{1}{p^2 + m_0^2}+ C^{(0)}~.
\end{split}
\end{align}
Note that we did not really need the result for $(\Phi^1)^2$ here because the dependence on this VEV canceled when plugging the value of $\Sigma$. 

Taking the difference we get the following LO result for the observable\footnote{The leading free energy $\CF_0$ was computed in \cite{zarembo}, see eq.(2.7), where $\mu_{{\rm there}} = h_{{\rm here}}$. However, \cite{zarembo} missed the non-perturbative term proportional to $m_0^2$, crucial to establish that $\CF_0(h=m_0)=0$.}
\begin{align}
\label{leading-fh}
\begin{split}
\CF_0(h) = F_0(h)-F_0(0) & = \int\frac{\rd^2 p}{(2\pi)^2} \left\{\frac 12\log\left[\frac{p^2 + h^2}{p^2 + m_0^2}\right] -\frac{h^2-m_0^2}{2} \frac{1}{p^2+ m_0^2}\right\} \\
& =  -\frac{h^2}{8\pi}\left[ \log \frac{h^2}{m_0^2}-1 \right] - {m_0^2  \over 8 \pi}~.
\end{split}
\end{align}
Note that we are slightly abusing notation: the symbols $F_0(h)$ and $F_0(0)$ do not denote the same function evaluated at two different arguments, as is clear from the fact that the difference does not vanish for $h=0$. This is because we are considering a different stationary point of the effective potential for $h\neq 0$, so we are actually taking the difference between the free energies of two different states, whose energies cross when $h = m_0$ where indeed the observable vanishes (recall that $m_0$ denotes the physical mass of the bosons at LO). For $h > m_0$ the vacuum with the condensate $\Phi^1\neq 0$ is energetically favored.

\subsubsection{Next-to-leading order vacuum diagrams}
The NLO (next-to-leading order) vacuum energy density for $h=0$ is given by the following diagrams
\begin{align}
\begin{split}
F_1 &= ~~\includegraphics[width=.1\columnwidth , valign=c]{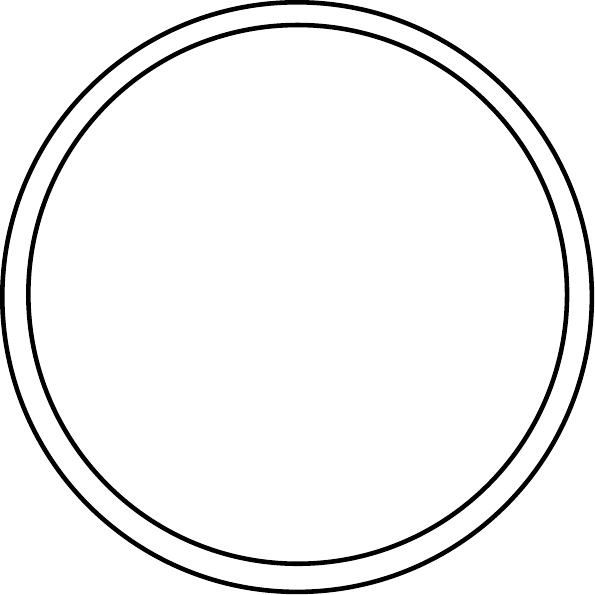}~~+~~\raisebox{2mm}{\includegraphics[width=.1\columnwidth , valign=c]{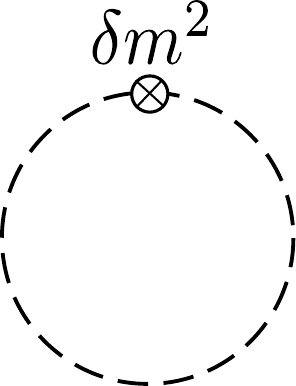}}~~+~~\raisebox{2mm}{\includegraphics[width=.1\columnwidth , valign=c]{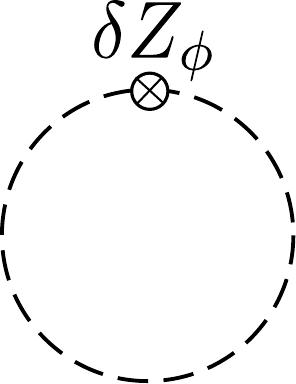}}~~+~~\raisebox{0mm}{\includegraphics[width=.035\columnwidth , valign=c]{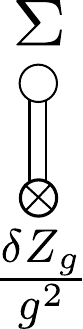}}~~+~~\raisebox{-2mm}{\includegraphics[width=.03\columnwidth , valign=c]{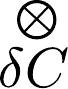}}~~-m_0^2 \,\frac{\partial F_0}{\partial m_0^2} \,r~.
\label{eq:NLOaux1}
\end{split}
\end{align}
 It is a general fact that the various diagrams linear in the NLO correction to the VEVs  $\delta \Sigma$, $\delta \boldsymbol{\Phi}$ drop from the NLO energy density thanks to the LO tadpole condition, so we avoided drawing such diagrams. We will postpone the NLO tadpole condition for $\hat{\sigma}$ to the subsection \ref{sec:NLOtadh0} and leave $\delta Z_g$ as undetermined for the time being. The last contribution comes from the NLO correction to the physical mass-squared
\begin{equation}\label{eq:m2NLO}
m^2 = m^2_0\left(1+\frac{r}{N}\right)~,
\end{equation}
when $F_0$ is re-expressed in terms of the physical mass-squared $m^2$. Note that in all the NLO diagrams we can just use $m^2$ as the mass-squared of the bosons.
 
The first diagram in \eqref{eq:NLOaux1} is a closed loop of the $\hat{\sigma}$ field. The propagator coming from the resummation of the $\hat{\boldsymbol{\phi}}$ bubbles is
\begin{equation}
\langle\hat{\sigma}(p)\hat{\sigma}(-p)\rangle = \frac{1}{-2 B(m^2,p)}~,
\end{equation}
where $B(m^2,p)$ is the ``bubble function''
\begin{align}
\begin{split}
B(m^2,p) &  =\int \frac{\rd^2 q}{(2\pi)^2} \frac{1}{(q+p)^2 + m^2}\frac{1}{q^2 + m^2} \\ & = \frac{1}{4\pi m^2} \frac{\log \left(1+\frac{p^2}{2 m^2}+\sqrt{\frac{p^2}{m^2} \left(1+\frac{p^2}{4 m^2}\right)}\right)}{\sqrt{\frac{p^2}{m^2} \left(1+\frac{p^2}{4 m^2}\right)}}~.
\end{split}
\end{align}
Therefore
\begin{equation}
\includegraphics[width=.1\columnwidth , valign=c]{Figs/Vach0SigBubble.pdf} = \frac 12 \int \frac{\rd^2 p}{(2\pi)^2} \log[-2 B(m^2,p)]~.
\end{equation}
Summing up with the other diagrams, we find
\begin{equation}
F_1 = \int \frac{\rd^2 p}{(2\pi)^2}\frac 12\left\{  \log[-2 B(m^2,p)] + \frac{\delta Z_\phi p^2 +\delta m^2}{p^2 + m^2}\right\} - \frac{\delta Z_g}{g^2}\frac{m^2}{2} + \delta C-m_0^2 \,\frac{\partial F_0}{\partial m_0^2} \,r~.
\end{equation}

The NLO vacuum energy density for $h\neq0$ is given by the following diagrams
\begin{align}
\begin{split}
F_1 (h) &= ~~\includegraphics[width=.11\columnwidth , valign=c]{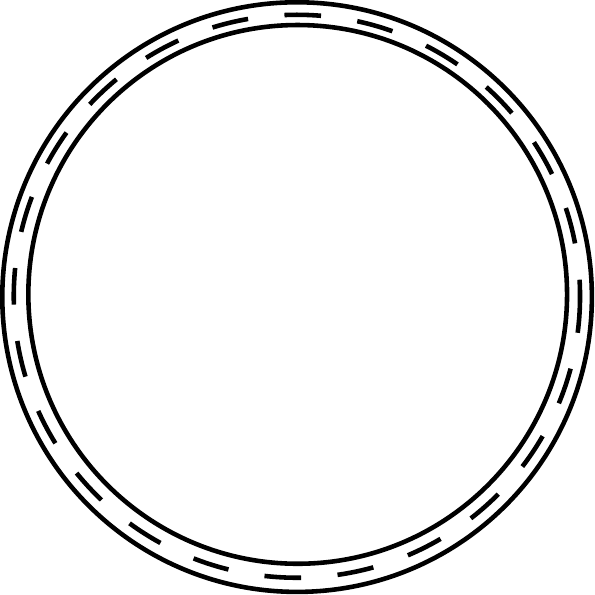}~~-~~\frac{2}{N}~\includegraphics[width=.1\columnwidth , valign=c]{Figs/VacNh0Bubble.pdf}~~+~~\raisebox{2mm}{\includegraphics[width=.1\columnwidth , valign=c]{Figs/Vach0Bubblect.pdf}}~~+~~\raisebox{2mm}{\includegraphics[width=.1\columnwidth , valign=c]{Figs/Vach0Bubblewf.pdf}}\\
&~~+~~\raisebox{0mm}{\includegraphics[width=.035\columnwidth , valign=c]{Figs/Vach0Tadct.pdf}}~~+~~\raisebox{0mm}{\includegraphics[width=.14\columnwidth , valign=c]{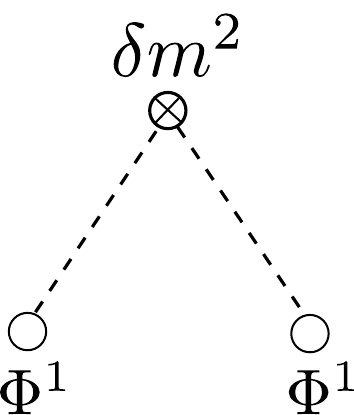}}~~+~~\raisebox{0mm}{\includegraphics[width=.14\columnwidth , valign=c]{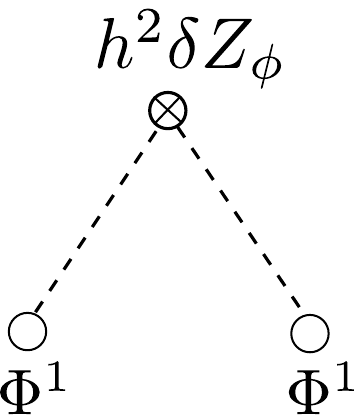}}~~+~~\raisebox{-2mm}{\includegraphics[width=.03\columnwidth , valign=c]{Figs/Vach0ct.pdf}}~~-m_0^2 \,\frac{\partial F_0(h)}{\partial m_0^2} \,r~.
\end{split}
\end{align}
We again used the LO tadpole condition to avoid drawing all diagrams with insertions of $\delta \Sigma$ and $\delta \boldsymbol{\Phi}$. Note that due to the quadratic mixing at this order we receive a contribution from the bubble of the $\hat{\sigma}$-$\hat{\phi}^1$-$\hat{\phi^2}$ propagator (first diagram) and to avoid overcounting we need to subtract the contribution of $\hat{\phi}^1$ and $\hat{\phi^2}$ to the LO closed loop of the bosons (second diagram).

Using a matrix notation, we can express the quadratic action involving $\hat{\sigma}$, $\hat{\phi}^1$ and $\hat{\phi}^2$ as
\begin{align}\label{eq:qact}
\begin{split}
\int \frac{\rd^2 p}{(2\pi)^2}&
\frac{1}{2}\begin{pmatrix}
 \hat{\phi}^1(-p) & \hat{\phi}^2(-p) & \hat{\sigma}(-p) 
\end{pmatrix} K(p)
\begin{pmatrix}
 \hat{\phi}^1(p) \\ \hat{\phi}^2(p) \\ \hat{\sigma}(p) 
\end{pmatrix}~,\\
&K(p) \equiv \begin{pmatrix}
 p^2 &-2 h\,p_\tau  & 2 \Phi^1 \\
 2 h\, p_\tau & p^2 & 0  \\
2 \Phi^1 & 0 &- 2B(h^2,p) 
\end{pmatrix}~.
\end{split}
\end{align} 
Here $p_\tau$ is the momentum in the Euclidean time direction. Therefore we have
\begin{align}
\begin{split}
& \includegraphics[width=.11\columnwidth , valign=c]{Figs/VachSigBubble.pdf}~~-~~\frac{2}{N}~\includegraphics[width=.1\columnwidth , valign=c]{Figs/VacNh0Bubble.pdf}  = \int \frac{\rd^2 p}{(2\pi)^2}\left\{\frac 12 \log[\det K(p)] - \frac{2}{N} \, \frac{N}{2}\log[p^2 + h^2] \right\} \\
&= \int \frac{\rd^2 p}{(2\pi)^2}\frac{1}{2}\left\{\log\left[ - 2 B(h^2,p) - \frac{1}{\pi}\log\frac{h^2}{m^2} \,\frac{p^2}{p^4+ 4 h^2 p_\tau^2}\right] + \log\left[\frac{p^4+ 4 h^2 p_\tau^2}{(p^2 + h^2)^2}\right]\right\}~.
\end{split}
\end{align}
We substituted \eqref{eq:VEVPhi1} inside the determinant. Summing up with the counterterm diagrams, we find
\begin{align}
\begin{split}
F_1 (h) & = \int \frac{\rd^2 p}{(2\pi)^2}\frac{1}{2}\left\{\log\left[ - 2 B(h^2,p) - \frac{1}{\pi}\log\frac{h^2}{m^2} \,\frac{p^2}{p^4+ 4 h^2 p_\tau^2}\right] + \log\left[\frac{p^4+ 4 h^2 p_\tau^2}{(p^2 + h^2)^2}\right]\right.\\
 & \left. + \frac{\delta Z_\phi p^2 +\delta m^2_h}{p^2 + h^2}\right\}+\frac{\delta m^2_h-\delta Z_\phi h^2}{8\pi} \log\frac{h^2}{m^2}- \frac{\delta Z_g}{g^2}\frac{h^2}{2}+ \delta C+ \delta C_h-m_0^2 \,\frac{\partial F_0(h)}{\partial m_0^2} \,r~.
\end{split}
\end{align}
The subscript $h$ in the mass-squared counterterm reminds us that the definition \eqref{eq:ctcollect} of $\delta m^2$ contains a factor of the VEV $\Sigma$, which depends on whether we are at $h=0$ or $h\neq 0$.

Taking the difference $F_1 (h)  -F_1 (0) $ we note that all the contributions involving the propagator counterterms cancel and we are left with
\begin{align}
\begin{split}\label{eq:obs1}
\hspace{-0.3cm}{\cal F}_1 (h)  = &  \int \frac{\rd^2 p}{(2\pi)^2}\frac{1}{2}\left\{\log\left[\frac{B(h^2,p) +\frac{1}{2\pi}\log\frac{h^2}{m^2} \,\frac{p^2}{p^4+ 4 h^2 p_\tau^2}}{B(m^2,p)}\right] + \log\left[\frac{p^4+ 4 h^2 p_\tau^2}{(p^2 + h^2)^2}\right]\right. \\
& \left. + \frac{\delta Z_\sigma (h^2-m^2)}{2}\frac{1}{p^2 +m^2}\right\} - \frac{\delta Z_g}{g^2}\frac{h^2-m^2}{2} + \delta C_h + \frac{m^2 - h^2}{8\pi} \,r~.
\end{split}
\end{align}
Here we used \eqref{leading-fh} to evaluate the terms involving $m_0^2 \frac{\partial}{\partial m_0^2}$. We still have a UV divergent integral and some counterterms that did not cancel in the difference, so we need to fix those to get the finite result for the observable.

\subsubsection{Tadpole condition and propagator correction}\label{sec:NLOtadh0}
The NLO Tadpole condition for $h=0$ is
\begin{align}
\begin{split}
& \raisebox{1.6 mm}{\includegraphics[width=.1\columnwidth , valign=c]{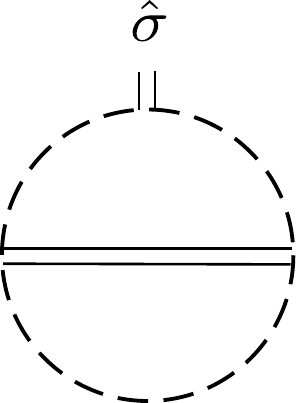}} ~~+~~\raisebox{0 mm}{\includegraphics[width=.1\columnwidth , valign=c]{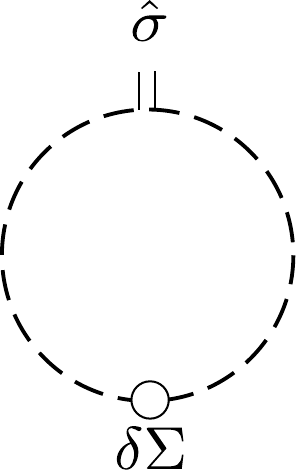}}\\
& \hspace{2cm}+~~\raisebox{0 mm}{\includegraphics[width=.1\columnwidth , valign=c]{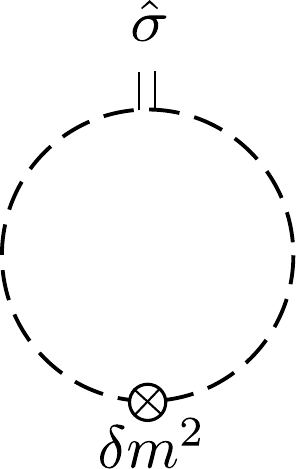}}~~+~~\raisebox{-0.8 mm}{\includegraphics[width=.1\columnwidth , valign=c]{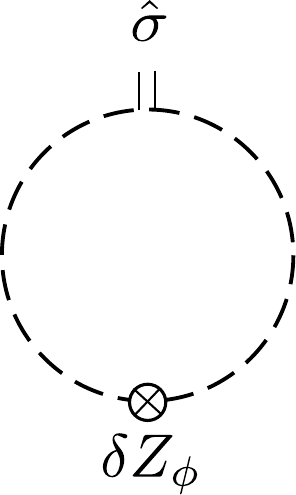}}~~+~~\raisebox{1.6 mm}{\includegraphics[width=.1\columnwidth , valign=c]{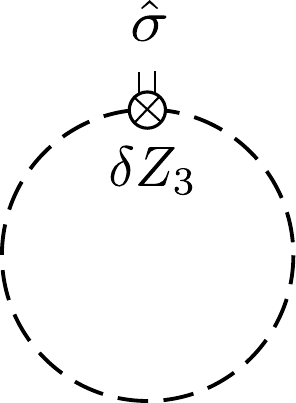}} ~~+~~\raisebox{-1 mm}{\includegraphics[width=.04\columnwidth , valign=c]{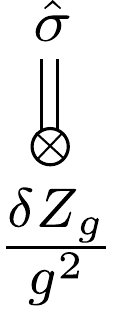}} ~= 0~. 
\end{split}
\end{align}
The two-loop diagram gives
\begin{align}
\begin{split}
&\raisebox{1.6 mm}{\includegraphics[width=.1\columnwidth , valign=c]{Figs/Tadh0Bubble.pdf}} \\ &\hspace{1cm}= \left(-\frac{2}{\sqrt{N}}\right)^3\frac 12 N\int\frac{\rd^2 p}{(2\pi)^2}\int\frac{\rd^2 q}{(2\pi)^2} \frac{1}{-2 B(m^2,p)}\frac{1}{((p+q)^2+m^2)^2}\frac{1}{q^2 +m^2} \\
&\hspace{1cm}= \frac{2}{\sqrt{N}} \int\frac{\rd^2 p}{(2\pi)^2} \left\{\frac{1}{4\pi m^2(p^2 + 4 m^2)B(m^2,p)}+\frac{1}{p^2 + 4 m^2}\right\}~.
\end{split}
\end{align}
Going to the second line we performed the integral in $q$, which gives
\begin{align}
\begin{split}
& \int\frac{\rd^2 q}{(2\pi)^2}\frac{1}{((p+q)^2+m^2)^2}\frac{1}{q^2 +m^2}\\
& = \frac{1}{16 \pi  m^4}\frac{1+\frac{\log \left(1+\frac{p^2}{2 m^2}+\sqrt{\frac{p^2}{m^2} \left(1+\frac{p^2}{4 m^2}\right)}\right)}{\sqrt{\frac{p^2}{m^2} \left(1+\frac{p^2}{4 m^2}\right)}}}{ 1+\frac{p^2}{4 m^2}}  = \frac{1}{4 \pi  m^2} \frac{1}{p^2+4 m^2} + \frac{B(m^2,p)}{p^2 + 4 m^2}~.
\end{split}
\end{align}
Plugging this result and evaluating the one-loop diagrams the NLO tadpole condition can be rewritten as
\begin{align}
\begin{split}
& \frac{\delta Z_g}{g^2} -\frac{\delta Z_\sigma}{2}\int\frac{\rd^2 p}{(2\pi)^2}\frac{1}{p^2 + m^2} \\ & \hspace{-0.5cm}= -2 \int\frac{\rd^2 p}{(2\pi)^2} \left\{\frac{1}{4\pi m^2(p^2 + 4 m^2)B(m^2,p)}+\frac{1}{p^2 + 4 m^2}\right\}-\frac{2\delta\Sigma + \delta m^2 - m^2 \delta Z_\phi}{4\pi m^2}~.
\end{split}
\end{align}
The left-hand side is precisely the combination that appear in the observable in eq. \eqref{eq:obs1} multiplied by $-\frac{h^2-m^2}{2}$, so substituting this relation we find
\begin{align}
\begin{split}\label{eq:obs2}
 \hspace{-0.2cm}{\cal F}_1 (h)  = &  \int \frac{\rd^2 p}{(2\pi)^2}\frac{1}{2}\left\{\log\left[\frac{B(h^2,p) +\frac{1}{2\pi}\log\frac{h^2}{m^2} \,\frac{p^2}{p^4+ 4 h^2 p_\tau^2}}{B(m^2,p)}\right] + \log\left[\frac{p^4+ 4 h^2 p_\tau^2}{(p^2 + h^2)^2}\right]\right. \\
& \hspace{-0.2cm}\left. +2(h^2-m^2)\left( \frac{1}{4\pi m^2(p^2 + 4 m^2)B(m^2,p)}+\frac{1}{p^2 + 4 m^2}\right) \right\} \\ & + (h^2 - m^2)\frac{2\delta\Sigma + \delta m^2 - m^2 \delta Z_\phi}{8\pi m^2}+ \delta C_h + \frac{m^2 - h^2}{8\pi} \,r~.
\end{split}
\end{align}
However, we are still left with a UV divergent integral and some combination of counterterms. To fix this remaining combination we need to consider the correction to the physical mass-squared.

The $1/N$ correction to the inverse propagator of $\widehat{\boldsymbol{\phi}}$ for $h=0$ is given by 
\begin{align}
\begin{split}
& \raisebox{0 mm}{\includegraphics[width=.17\columnwidth , valign=c]{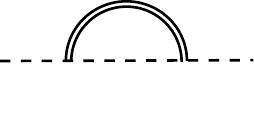}} ~~+~~\raisebox{-2 mm}{\includegraphics[width=.13\columnwidth , valign=c]{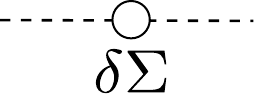}} ~~+~~\raisebox{-2.2 mm}{\includegraphics[width=.13\columnwidth , valign=c]{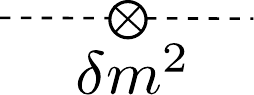}}~~+~~\raisebox{-2.9 mm}{\includegraphics[width=.13\columnwidth , valign=c]{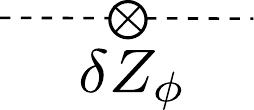}}~~. 
\end{split}
\end{align} 
Summing up this diagrams, and imposing that the physical mass-squared at NLO is given by eq. \eqref{eq:m2NLO} we obtain
\begin{equation}\label{eq:shiftpole}
-\delta Z_\phi m^2 + 2 \delta \Sigma + \delta m^2 = m^2 r -2\left.\int \frac{\rd^2p}{(2\pi)^2} \frac{1}{B(m^2,p)((p+k)^2 + m^2)}\right\vert_{k^2 = -m^2}~.
\end{equation}

Plugging \eqref{eq:shiftpole} in \eqref{eq:obs2}, the residual dependence on the counterterms and on $r$ cancels and we obtain
\ben
 {\cal F}_1 (h)  \! & \!= \! & \! \int\! \frac{\rd^2 p}{(2\pi)^2}\frac{1}{2}\left\{\log\!\left[\frac{B(h^2,p) +\frac{1}{2\pi}\log\frac{h^2}{m^2} \,\frac{p^2}{p^4+ 4 h^2 p_\tau^2}}{B(m^2,p)}\right] \! + \log\left[\frac{p^4+ 4 h^2 p_\tau^2}{(p^2 + h^2)^2}\right]  +  2(h^2-m^2) \right. \nonumber  \\
&& \left. \hspace{-0.8cm}  \times\left[ \frac{1}{4\pi m^2B(m^2,p)}\left(\frac{1}{p^2 + 4 m^2}-\left.\frac{1}{(p+k)^2 + m^2}\right\vert_{k^2 =-m^2}\right)+\frac{1}{p^2 + 4 m^2}\right] \right\}+ \delta C_h~.
\een
The integral appearing in this final expression is UV finite, confirming that $\delta C_h$ is a finite shift. In order to fix the finite shift $\delta C_h$ we impose as a further renormalization condition that also at NLO the observable vanishes for $h = m$. This ensures that after the inclusion of $1/N$ corrections it remains true that the state with a condensate $\Phi_1 \neq 0$ becomes favored precisely in the range $h\geq m$ (recall the comments below eq. \eqref{leading-fh}). By performing the integral for $h=m$, one can check that this condition is satisfied by fixing $\delta C_h = \frac{h^2}{4\pi}$.

To write more explicitly the $k$-dependent part, it is convenient to choose $k$ in the Euclidean time direction and simply plug $k=(\ri m, 0)$, so that
\begin{equation}
\frac{1}{(p+k)^2 + m^2} = \frac{1}{p^2 + 2 \ri m p_\tau} = \frac{p^2 - 2 \ri m p_\tau}{p^4 + 4 m^2 p_\tau^2}~.
\end{equation}
We can then drop the imaginary part because it is odd in $p_\tau$. Therefore we obtain the final result
\begin{align}
\begin{split}\label{eq:obsfinal}
& \hspace{-0.5cm}\CF_1 (h) = \int \frac{\rd^2 p}{(2\pi)^2}\frac{1}{2}\left\{\log\left[\frac{B(h^2,p) +\frac{1}{2\pi}\log\frac{h^2}{m^2} \,\frac{p^2}{p^4+ 4 h^2 p_\tau^2}}{B(m^2,p)}\right] + \log\left[\frac{p^4+ 4 h^2 p_\tau^2}{(p^2 + h^2)^2}\right]\right. \\
&\left. +2(h^2-m^2)\left[ \frac{1}{4\pi m^2B(m^2,p)}\left(\frac{1}{p^2 + 4 m^2}-\frac{p^2}{p^4 + 4m^2 p_\tau^2}\right)+\frac{1}{p^2 + 4 m^2}\right] \right\} + \frac{h^2}{4\pi}~.
\end{split}
\end{align}
 
 \subsection{The $1/N$ expansion from the Bethe ansatz}
 
As discussed in section \ref{sec-fe-int}, since the NLSM is integrable, the free energy $\CF(h)$ can be calculated exactly by solving the integral equation (\ref{intetwo}).
One could think that the $1/N$ expansion of the free energy can be obtained by simply expanding the kernel in a power series in $1/N$. However, 
this turns out to be subtle, since the kernel is singular in the limit $\Delta \rightarrow 0$, see appendix \ref{app:analytic}. The $1/N$ expansion of the integral equation 
(\ref{intetwo}) for the NLSM was studied in \cite{zarembo} at the first non-trivial order. The calculation of higher order corrections in similar models introduces additional subtleties, as shown in \cite{ksz}, but in our analysis we will restrict to the next-to-leading order term. 
 
It turns out that the $1/N$ expansion of the kernel requires the introduction of distributions. The expansion of the kernel reads, 
\be
K(\theta)=\delta(\theta)+ \sum_{k \ge 1} \Delta^k K_k(\theta), 
\label{eq:kernelExp}
\ee
where $\delta(\theta)$ is Dirac's delta function. The first two terms in the expansion read
\be
\ba
K_1(\theta)&=  -{\rd \over \rd \theta}\CL_1(\theta), \\
K_2(\theta)&= -2 \pi^2 \delta'' (\theta)-{\rd \over \rd \theta} \CL_2 (\theta). 
\ea
\ee
In these expressions, 
\be
\CL_1(\theta)= \frac{1}{\theta}+{1 \over \sinh(\theta)},
\ee
and
\be
\CL_2(\theta)={\ri  \over 4 \pi }\left\{ \psi ^{(1)}\left(\frac{\ri \theta+\pi
   }{2 \pi }\right)- \psi ^{(1)}\left(\frac{\pi -\ri \theta}{2 \pi }\right)-\psi
   ^{(1)}\left(\frac{\ri \theta}{2 \pi }\right)+\psi ^{(1)}\left(-\frac{\ri \theta}{2 \pi }\right) \right\},
   \ee
where $\psi^{(1)}$ is the first derivative of the digamma function. The expansion (\ref{eq:kernelExp}) has to be understood in the following sense: when acting on a test function $f(\theta)$ 
which is differentiable and vanishes at the boundaries, 
one has 
\be
\ba
\int_{-B}^B K(\theta- \theta') f(\theta') \rd \theta' &= f(\theta) -\Delta \left( {\rm P}  \int_{-B}^B \CL_1 (\theta- \theta') f'(\theta') \rd \theta'  \right)\\
& + \Delta^2 \int_{-B}^B K_2 (\theta- \theta') f(\theta') \rd \theta' + \CO(\Delta^3). 
\ea
\ee
where ${\rm P}$ means the principal value.

We can now solve the integral equation by using a large $N$ ansatz for the distribution
\be
\epsilon (\theta)={1\over \Delta}  \sum_{k \ge 0} \Delta^k \epsilon_k (\theta), 
\ee
and for the endpoint
\be
B= \sum_{k \ge 0} \Delta^k B_k. 
\ee
The leading term in the integral equation (\ref{intetwo}) is then given by 
\be
\label{pre-largen}
{\rm P} \int_{-B_0}^{B_0} \CL_1(\theta- \theta') \epsilon'_0(\theta') \rd \theta'= h- m \cosh(\theta). 
\ee
By splitting the kernel $\CL_1(\theta)$ as 
\be
\CL_1(\theta) = {2 \over \theta} + \CL_1^{\rm r} (\theta)
\ee
where 
\be
 \CL_1^{\rm r} (\theta)= {1 \over \sinh(\theta)}-{1\over \theta} 
 \ee
 is regular at $\theta=0$, we can write (\ref{pre-largen}) as
\be 
\label{leading-sm}
{\rm P} \int_{-B_0}^{B_0} {2 \epsilon'_0(\theta') \over  \theta- \theta'} \rd \theta' + \int_{-B_0}^{B_0} \CL^{\rm r}_1(\theta- \theta')
 \epsilon'_0(\theta') \rd \theta'= h - m \cosh(\theta). 
 \ee
After integrating by parts and an integration w.r.t. $\theta$, one can also write (\ref{pre-largen}) as
\be
\label{largen}
{\rm P} \int_{-B_0}^{B_0} \CL_1(\theta- \theta') \epsilon_0(\theta') \rd \theta'= h \theta- m \sinh(\theta), 
\ee
which is the form used in \cite{zarembo}. 
 
We conclude that the large $N$ limit of the distribution appearing in the TBA equation (\ref{intetwo}), $\epsilon_0(\theta)$, can be obtained as a solution of the singular integral equations (\ref{pre-largen}) or (\ref{largen}). (\ref{largen}) is very similar to 
 the integral equation that one would find for the 
 density of states of a large $N$ matrix integral. The singular part of $\CL_1 (\theta)$ is what one would find for a conventional, Hermitian 
 matrix model. The additional term $\CL^{\rm r}_1(\theta)$ would correspond to a non-conventional eigenvalue interaction. We note that the integral equation (\ref{largen}) does not determine by itself the value of $B_0$ as a function of $h$. 
 It was argued in \cite{zarembo} that this value is fixed by requiring the following behavior near the edges of the distribution:
 \be
 \label{bc}
 \epsilon_0(\theta) \sim (B_0^2-\theta^2)^{3/2}, \qquad \theta \rightarrow \pm B_0. 
 \ee
When $\CL^{\rm r}_1(\theta)=0$, as it happens in the PCF model studied in the next section, 
one can show that the condition (\ref{bc}) follows from (\ref{leading-sm}), by requiring $\epsilon'_0(\theta)$ to be regular 
at $\pm B_0$. 

The solution to the singular integral equation (\ref{largen}) is not known explicitly. 
However, there is both analytic and numerical evidence that 
 the leading order free energy obtained from this solution, 
 \be
\CF_0(h)= -{m \over 2 \pi}  \int_{-B_0}^{B_0} \rd \theta \, \cosh(\theta) \, \epsilon_0(\theta), 
 \ee
 agrees exactly with (\ref{leading-fh}). By assuming this analytic value for $ \CF_0(h)$ one can deduce the value of $B_0$  \cite{zarembo}:
 \be
  B_0(h)= {\sqrt{ \log\left({h \over m} \right)\left( \log\left({h \over m} \right)  + 1\right)}} + \sinh^{-1} \left[ \sqrt{\log\left({h \over m} \right)} \right]. 
  \ee

 Let us now write down the equation for the next-to-leading correction $\epsilon_1(\theta)$. By using (\ref{bc}), it can be seen that $B_0$ does not get corrected at that order, and one finds the equation
\be
 {\rm P} \int_{-B_0}^{B_0} \CL_1(\theta-\theta') \epsilon_1(\theta') \rd \theta'=- 2 \pi^2 \epsilon'_0(\theta)- \int_{-B_0}^{B_0} \CL_2(\theta-\theta') \epsilon_0(\theta') \rd \theta'. 
  \ee
This equation can be solved numerically to obtain the next-to-leading correction to the free energy, given by 
\be
\CF_1(h)= -{m \over 2 \pi}  \int_{-B_0}^{B_0} \rd \theta \, \cosh(\theta) \, \epsilon_1(\theta). 
\label{eq:F1NLSMsing}
 \ee
We checked, for some values of $h$, that $\CF_1(h)$ computed as above agrees with \eqref{eq:obsfinal}. 
The numerical resolution of the singular integral equation is quite time consuming: with few hours of computation we reached an agreement with a relative error of order $10^{-7}$. 
Proceeding to higher orders in $\Delta$ in the expansion of the kernel \eqref{eq:kernelExp} and solving singular integral equations to compute $\CF_k$ for $k > 1$ becomes very challenging.  
We have been able to extract more coefficients of the series in $\Delta$ by following an indirect procedure: we first solve numerically with high precision the TBA for fixed $h_0$ and different values of $\Delta$, and from this sequence of $\CF(h_0)$ we compute the value of the asymptotic expansion at each order, using Richardson transforms to accelerate the series. This allowed us to compute, for a given $h$, the functions $\CF_k(h)$ up to $k=6$. These first coefficients do not present a factorial growth as it would be expected from the properties of the kernel discussed in appendix \ref{app:analytic}. However, they are not enough to allow us to make claims on the nature of the $1/N$ series.

\subsection{Trans-series expansion and comparison with perturbation theory} 

We can now compare the exact results for ${\cal F}_0$ and ${\cal F}_1$ found in the previous subsections with perturbation theory. Up to order $\Delta$, \eqref{eq:LNPTC6} gives 
\be
h = \re^{\frac{1}{\alpha}} m \big(1+\Delta \log \alpha \big)+ {\cal O}(\Delta^2) \,.
\label{eq:alphaNLSM}
\ee

Given \eqref{leading-fh} and the definitions \eqref{eq:FNLSM} and \eqref{eq:TransPhi}, it is straightforward to get
\be
\varphi_0^{(0)}(\alpha)= - {1\over \alpha}+{1\over 2} \,, \qquad\varphi_0^{(1)}(\alpha) = \frac{1}{2}\,, \qquad \varphi_0^{(\ell)}(\alpha) =0 \,, \;\; \ell \geq 2 \,.
\label{eq:varphiLO_nlsm}
\ee
The perturbative series $\varphi_0^{(0)}$ agrees with \eqref{eq:phi0NLSM}, as it should, but in addition we see a non-perturbative single trans-series term, which 
cannot be captured in perturbation theory. The mismatch between the exact free energy and its perturbative calculation 
is due to the $h$-independent term proportional to $m^2$.
It is given by the non-perturbative free energy at $h=0$, 
evaluated at the non-trivial large $N$ point, which has been calculated in \cite{bcr} and has the value
\be
\label{subt}
\Delta \CF (0)={m^2 \over 8 \pi} + \CO(\Delta^2). 
\ee
This suggests that the difference between perturbation theory and the $1/N$ expansion is only due to 
the difference between a perturbative and a non-perturbative evaluation at $h=0$. We will 
give further evidence of this when we consider the next-to-leading term in the $1/N$ expansion. 

\begin{figure}\begin{center}
\includegraphics[height=7.5cm]{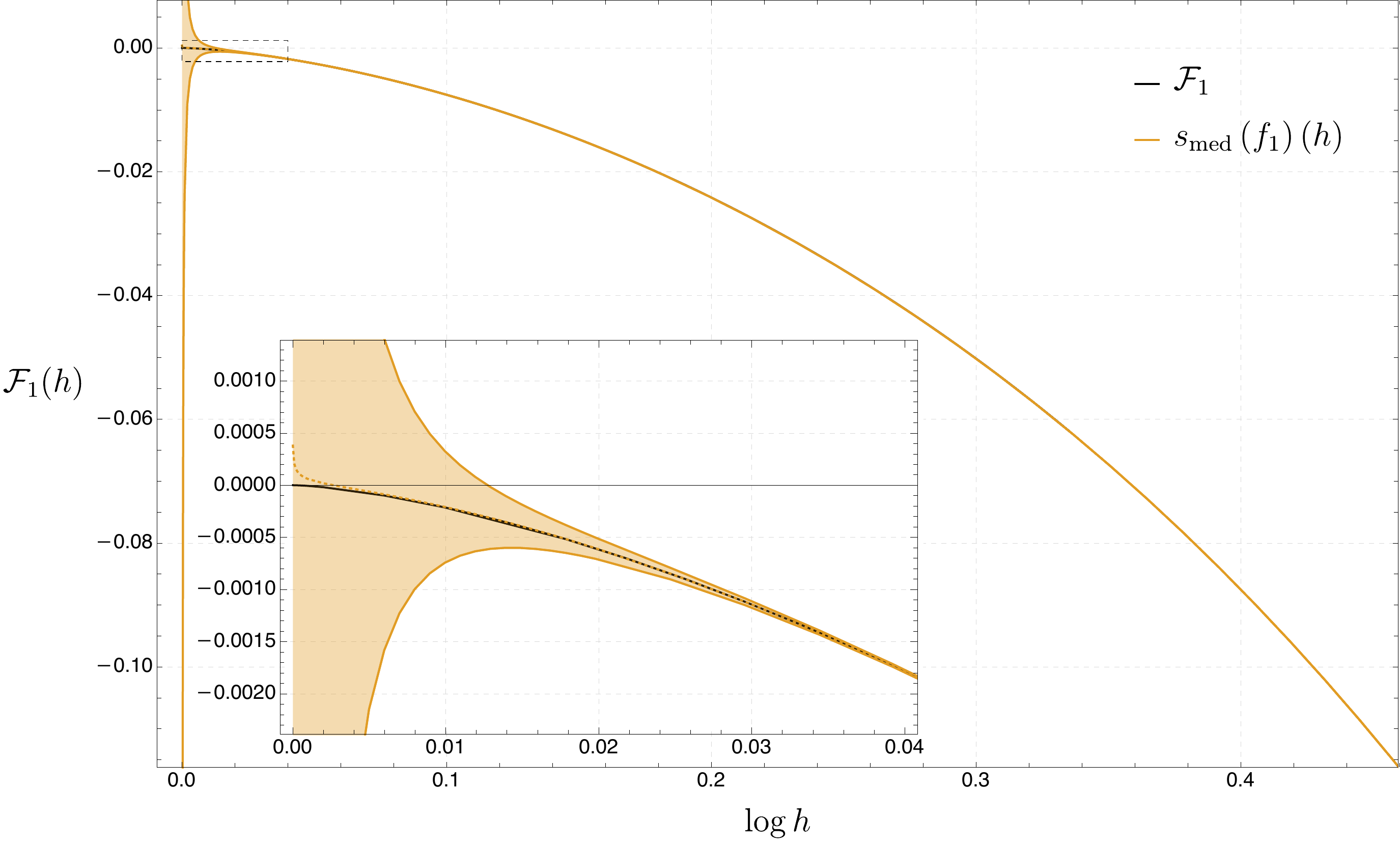}
\end{center}
\caption{Comparison between the median resummation of the perturbative series and the exact result. The orange curve is obtained computing the $+$ and $-$ lateral Pad\'e-Borel resummations of the first 41 terms of the perturbative series and taking their average, while the error is estimated from the convergence of the resummation. The black curve (mostly covered by the orange one) is the numerical evaluation of \eqref{eq:obsfinal}.  The mass $m$ is set to 1, so $h=1$ is maximally strong coupling.}
\label{fig:NLSM}
\end{figure}

A direct expansion in $\alpha$ of the NLO term ${\cal F}_1$ in \eqref{eq:obsfinal} is challenging. Luckily enough, we will verify that it has the same structure of ${\cal F}_0$, namely its trans-series expansion is composed of only two terms $\varphi_1^{(0)}$ and $\varphi_1^{(1)}$. The coefficient  $\varphi_1^{(0)}$ is the perturbative result and its first terms can be read from \eqref{sigma1}. 
This series is factorially divergent, and its Borel transform has a Borel singularity in the positive real axis (i.e. an IR renormalon), which 
was analyzed in detail in \cite{mmr}. We can then use lateral Borel resummations $s_\pm (\varphi_1^{(0)})(\alpha)$, i.e. integrating the Borel transform slightly above or below the real axis avoiding in this way the singularities, indicated respectively by $s_+$ and $s_-$. They lead to an imaginary piece. 
Let us denote by $s_\pm (\varphi_1^{(0)})$ the function obtained from $\varphi_1^{(0)}$ by using the lateral Borel 
resummations of the series (\ref{sigma1}). The results of \cite{mmr} imply that 
\be
 {\rm Im} \left( s_\pm \left(\varphi_1^{(0)} \right)\right)=\mp {\pi \over 2}\re^{-\frac{2}{\alpha}}, 
 \ee
 which is the imaginary contribution of the IR renormalon unveiled in \cite{mmr}. A detailed numerical comparison 
 indicates that the exact $1/N$ result \eqref{eq:obsfinal} agrees precisely with the {\it real} part of the lateral Borel resummations. Defining 
 $f_1(\alpha) = -h^2/(4\pi) ( - \log \alpha+ \varphi_1^{(0)} (\alpha))$, we have
\be
 \CF_1(h)= {\rm Re}  \left( s_\pm \left(f_1 \right) (\alpha)\right) \,,
 \label{eq:F1NLSMexp}
 \ee
where in \eqref{eq:F1NLSMexp} $\alpha = 1/\log(h/m)$. We can use the so-called median resummation, which in this case is simply
 \be
 s_{\rm med}= {s_+ + s_- \over 2} 
 \ee
 to write the above result as
 \be
\CF_1(h)=  s_{\rm med} \left(f_1\right)(h)\,.
 \ee
 These results indicate that the trans-series expansion of ${\cal F}_1$ is rather trivial, namely we have
 \be
 \varphi_1^{(1)}(\alpha, C_\pm)  =  C_\pm \frac{\pi}{2}\,,  \qquad  \varphi_1^{(\ell)} =0 \,, \;\; \ell\geq 2 \,,
 \ee
 where $C_\pm = \pm \ri$. In fig. \ref{fig:NLSM} we illustrate the agreement between the median resummation and the numerical evaluation of the exact result.
In order to show the high precision of the resummation and the agreement with the exact value, in table \ref{tab:nlsm_F} we report  the comparison between the numerical evaluation of $\CF_1$ in \eqref{eq:obsfinal} and the median resummation of $f_1$ for a fixed value of $h$, taken to be $h=3$.
\begin{table}[t]
  \centering
  \begin{tabular}{c|c}
    \hline
    $\CF_1(h=3)$ & $-1.00252688160157404\text{\dots}$ \\\hline
     $ s_{\rm med} \left(f_1 \right)(h=3)$ & $-1.00252688160157404(4)$ \\\hline
  \end{tabular}
  \caption{Comparison between the exact result and the median resummation of the perturbative series for $h=3$ and $m$ set to 1. The error in the Pad\'e-Borel resummations of the first 41 terms of the perturbative series  is estimated from the convergence of the resummation.}
  \label{tab:nlsm_F}
\end{table}

This is similar to the result found in \cite{fkw1,fkw2} for the PCF  model at large $N$, in which the exact answer is the real part of a Borel-resummed series. One additional insight of the result above is that we have an explicit diagrammatic interpretation of the underlying perturbative series, in terms of the ring diagrams considered in \cite{mmr}.  


\section{The principal chiral field model}
\label{sec:PCF}

In this section we present our results for the PCF model. In the first subsection we review some general aspects of the model. Then, we present the large $N$ exact solution for $\CF_0(h)$. In the final subsection we ``decode" this exact solution in terms of an explicit, resurgent trans-series: we find a trans-series extension of the perturbative expansion, and we show that the exact solution can be obtained from this trans-series by Borel resummation. 

\subsection{General aspects}

The PCF model is a quantum field theory for maps $\Sigma: \IR^2 \rightarrow SU(N)$, with Lagrangian density
\be
\CL= {1\over  g_0^2} \tr \left(\partial_\mu \Sigma\,  \partial^\mu \Sigma^\dagger\right).
\ee
In our convention (\ref{betaf})  we have
\be
\beta_0={1\over 16  \pi \overline \Delta},  \qquad \xi=\frac{1}{2}. 
\ee
The PCF has a global symmetry $SU(N)_L \times SU(N)_R$, therefore there are 
conserved charges that can be used to construct the free energy $F(h)$. 
We will consider a vectorial symmetry $SU(N)_V \subset SU(N)_L \times SU(N)_R$, and the corresponding charge will be denoted by $Q$. 
Its eigenvalues in the fundamental representation will be denoted by $\boldsymbol{q}=(q_1, \cdots, q_N)$. Two choices of charges have been studied in the literature. In \cite{fkw1, fkw2} one takes
\be
\label{fkw-charges}
q_k=r_k-r_{k-1}, \qquad r_k = {\sin(\pi k /N) \over \sin(\pi/N)}. 
\ee
This leads to an explicitly solvable large $N$ limit. There is a different setting, considered originally in \cite{pcf} to determine the mass gap 
of the theory, in which one chooses the charge
\be
\label{or-charges}
 \boldsymbol{q}= \left({1\over 2}, -{1 \over 2(N-1)}, \cdots, -{1\over 2(N-1)} \right). 
 \ee
We will consider in this paper the setting \eqref{or-charges}.  
As we will see, the large $N$ limit is also solvable in this case. In addition, the resulting large $N$ free energy leads to an infinite series of IR renormalon contributions which can be calculated explicitly.  
The relevant kernel for the choice of charges (\ref{or-charges}) is reported in \eqref{eq:PCF1}.
In the PCF model it is convenient to define the 't Hooft  coupling
\be
\alpha \equiv \alpha \Big(\mu=\sqrt{\frac{2\pi}{\re}} h\Big)\,,
\ee
where $\alpha(\mu)$ is the TBA coupling defined in \eqref{eq:LNPTC6}.

The free energy $\CF(h)$ can be computed in perturbation theory. The one-loop result was presented in \cite{pcf} for 
an arbitrary choice of charges, and it is given by  
\be
\label{foneloop}
\CF(h)=-{4 h^2 \over \overline g^2} \sum_{j=1}^N q_j^2-{h^2 \over 2 \pi} \sum_{1\le i<j\le N} (q_i -q_j)^2 \left[ \log |q_i-q_j| -{1\over 2} \right]+ \CO(\overline g^2), 
\ee
where $\overline g^2$ is the $\overline{{\rm MS}}$ coupling defined by 
\be
\label{RG-evol}
\log\left( {\mu \over h} \right)= -\int_{g}^{{\overline g}} {\rd x \over \beta^{\overline{{\rm MS}}}(x)}. 
\ee
It follows from (\ref{foneloop}) that in terms of the 't Hooft coupling $\alpha$, we have
\be
 \CF_0(h) \sim-{h^2 \over 8 \pi} \left\{ {1\over  \alpha} - {1\over 2}+ \CO( \alpha) \right\}, 
\ee
where $\CF_0$ is the leading $1/N$ term of $\CF$ appearing in \eqref{eq:FPCF}.

The free energy $\CF(h)$ can also be computed from the TBA equations (\ref{intetwo}), (\ref{fh-bethe}). 
The TBA solution can be used to extract the perturbative expansion of $\CF_0(h)$ up to very high orders in the 
coupling constant and at finite $N$, by using the methods in \cite{volin,volin-thesis}. 
This was done in \cite{mr-ren} for the normalized energy density, but the results in that paper can be easily translated into an expansion in $\alpha$ 
for $\CF_0(h)$, and one finds
\be
 \CF_0(h) \sim-{h^2 \over 8 \pi}  \left\{ {1\over  \alpha} - {1\over 2}-{\alpha \over 4} -{5 \alpha^2 \over 16} -{53 \alpha^3 \over 96} +\CO(\alpha^4)\right\}. 
\ee

\subsection{Exact solution at large $N$}

As in the case of the NLSM, the kernel (\ref{int-kernel}) given by (\ref{eq:PCF1}) has a $1/N$ expression which involves distributions:
\be
K(\theta)= \delta(\theta) + \sum_{k \ge 1} \overline \Delta^k K_k (\theta)\,,
\ee
where the first term is simply
\be
K_1(\theta)=-{\rd \over \rd \theta} \CL_1(\theta), \qquad \CL_1 (\theta)= {2\over \theta}\,.
\ee
The integral equation at leading order in $1/N$, (\ref{largen}), reads in this case
\be 
\label{e0ie}
{\rm P} \int_{-B}^{B} {2 \epsilon_0(\theta') \over  \theta- \theta'} \rd \theta'= h \theta- m \sinh(\theta). 
 \ee
 Here, we have denoted $B_0$ by $B$, since we will not consider subleading corrections to the value of the endpoint. 
 Due to the simplicity of the leading kernel, (\ref{e0ie}) is the equation for the density of eigenvalues of a Hermitian matrix model with a potential 
 \be
 \label{mmpot}
 V(x)= {h x^2 \over 2}- m \cosh(x). 
 \ee
 Since the support of $\epsilon_0(\theta)$ is the full interval $[-B, B]$, we are considering a 
 so-called one-cut solution. This solution can be obtained by using standard matrix model techniques, 
 see e.g. \cite{mmbook}. The density is given by 
 \be
 \epsilon_0(\theta)= {1\over 2 \pi} {\sqrt{B^2 -\theta^2}} M(\theta), 
 \ee
 where the function $M(\theta)$ can be written as a contour integral around $z=0$
 \be
 M(\theta)= \oint_0 {V'(1/z) \over 1- z\theta} {1\over {\sqrt{1-B^2 z^2}}} {\rd z \over 2 \pi \ri}. 
 \ee
To obtain the explicit solution, it turns out to be useful to write $V(x)$ in (\ref{mmpot}) as a power series around $x=0$, 
\be
V(x)= \sum_{k \ge 0} g_{k} x^{2k}, 
\ee
where, in our case, 
\be
g_1={h-m\over 2}, \qquad g_k = -{m \over (2k)!}, \quad k \ge 2.
\ee
Then, by using the expansion 
\be
 {1\over {\sqrt{1-B^2 z^2}}}=\sum_{k=0}^{\infty} {2 k \choose k} \left( {B^2 \over 4} \right)^k  z^{2k}, 
 \ee
 we find the expression
 \be
 \epsilon_0(\theta)= {1\over 2 \pi} {\sqrt{B^2-\theta^2}} \left( h- 2m \sum_{r,k \ge 0} {2k \choose k} {r+ k+1 \over (2(r+k+1))!} \left( {B^2 \over 4}\right)^k \theta^{2r} \right). 
 \ee
 The value of $B$ is determined by the condition (\ref{bc}), as in \cite{zarembo}, and one finds
 \be
 h- 2 m \sum_{r,k \ge 0} {2k \choose k} {r+ k+1 \over (2(r+k+1))!} 4^{-k} B^{2(k+r)}=0. 
 \ee
This series defines an entire function of $B$ which can be written down in closed form in terms of the Bessel functions $I_{1,2}(z)$:
 \be
 \label{b0h}
 {2 \over B} I_1(B)+ I_2(B) ={h \over m}\,.
 \ee
This gives the relation between $B$ and $h$.

To obtain the large $N$ free energy we just have to calculate the integral
\be
\CF_0(h)= -{m \over 2 \pi} \int_{-B}^{B} \epsilon_0(\theta) \cosh(\theta) \rd \theta. 
\ee
By expanding the $\cosh$, we find 
\be
\CF_0(h)= -{m \over 2 \pi} \sum_{t \ge 0} { \mathfrak{M}_t\over (2t)!}, 
\ee
where
\be
\mathfrak{M}_t=\int_{-B}^{B} \epsilon_0(\theta) \theta^{2t} \rd \theta
\ee
are the moments of $\epsilon_0(\theta)$. As it is well-known from the matrix model literature, they 
can be calculated by using the expansion at infinity of the resolvent 
\be
\omega_0(\theta)={1\over 2} \left( V'(\theta) - {\sqrt{\theta^2- B^2}} M(\theta)  \right)= \sum_{t \ge 0}\mathfrak{M}_t \theta^{-2t-1}. 
\ee
By doing this, one finds
\be
\ba
\CF_0(h)=-{m^2\over 4 \pi} \biggl\{ &{ h \over m} B I_1(B)\\
& - B^2 \sum_{k,r,t \ge 0}  {2k \choose k} { (2(r+t))! \over (2(r+k+1))!} { r+ k+1 \over r+t+1}
{1\over ((r+t)!)^2 (2t)!} \left( {B^2 \over 4} \right)^{k+r+t} \biggr\}. 
\ea
\ee
Note that, from the point of view of quantum field theory, this is a strong coupling expansion, since small $B$ corresponds to small $h/m$. 
Fortunately, this expression can be summed up in closed form in terms of a generalized hypergeometric function, and we obtain in the end
\be
\label{f0b}
\CF_0(h)= -{h^2\over 4 \pi} \left\{ {B^2 I_1(B) \over 2 I_1(B ) +B  I_2(B) }- 
{1\over 2} {B^4 \over  \left( 2 I_1(B) +B  I_2(B) \right)^2} {}_1 F_2\left( {1\over 2} ; 1, 2; B^2 \right)  \right\}.
\ee
From these expressions it is also possible to obtain exact formulae for the large $N$ limits of the density of particles and of the energy density, defined by 
\be
\overline \Delta \rho=\rho_0 + \CO(\overline \Delta), \qquad \overline \Delta e=e_0 + \CO(\overline \Delta). 
\ee
One finds, 
\be
\label{e0rho0}
\ba
{\rho_0 \over m}= {B \over 4 \pi} I_1(B),\qquad 
{e_0 \over m^2}={B^2 \over 8 \pi} {}_1 F_2 \left( {1\over 2};1,2;B^2\right). 
\ea
\ee

\subsection{Trans-series expansion}

We can now address the question of what is the relation between the exact large $N$ result (\ref{f0b}), and the 
trans-series expansion of the free energy, including exponentially small corrections. It turns out that all the special functions 
appearing in (\ref{f0b}) and (\ref{b0h}) have simple trans-series expansions which can be used to obtain a trans-series expansion of $\CF_0(h)$. 
Let us start with (\ref{b0h}). By using the trans-series asymptotics of the Bessel functions, we have the following equality, valid for $B>0$:
\be
{2 \over B} I_1(B)+ I_2(B)= {\re^B \over {\sqrt{2 \pi B}}} \left( s_\pm (\gamma^{(0)})(B)+ C_\pm \re^{-2B} s(\gamma^{(1)})(B) \right). 
\ee
Here, 
\be
\gamma^{(0)}(B)=1+ {1\over 8 B}+ {9 \over 128 B^2}+ \cdots, \qquad \gamma^{(1)}(B)= \gamma^{(0)}(-B)
\ee
are Gevrey-1 series, $s$ with no subscripts $\pm$ denotes the standard Borel resummation, available when there are no singularities in the positive real axis of the Borel plane, and 
\be
C_\pm= \mp \ri. 
\ee
As usual in Borel--\'Ecalle resummation, the value of $C_\pm$ is correlated with the choice of lateral resummation. We also have
 \be
 \label{bi1}
 B I_1(B)\sim {\sqrt{ B \over 2 \pi}} \re^B \left( \nu^{(0)}(B) - C_\pm \re^{-2B} \nu^{(1)}(B) \right), 
 \ee
 where
 \be
 \nu^{(0)}(B)= 1-{3 \over 8B} -{15 \over 128 B^2}+ \cdots, \qquad \nu^{(1)}(B)= \nu^{(0)}(-B). 
 \ee
  Finally, we have the following formula for the generalized hypergeometric function, 
 \be
 {B^2 \over 2} {}_1 F_2\left( {1\over 2} ; 1, 2; B_0^2 \right) ={ \re^{2B} \over 4 \pi} \left( s_\pm ( f^{(0)})(B)+ C^{\pm} \re^{-2B} 4B s_\pm (f^{(1)})(B)+ 
 \re^{-4B}  s(f^{(2)})(B) \right), 
 \ee
 where
 \be
 f^{(0)}(B)=1+{1\over  4B}+ \cdots, \qquad f^{(2)}(B)= f^{(0)}(-B), \qquad f^{(1)}(B)= 1-{1\over 8 B^2}+ \cdots
 \ee

 It follows from (\ref{f0b}) that $\CF_0(h)$ has a trans-series expansion in terms of the small parameters $1/B$, $\re^{-2B}$. We are 
 however interested in obtaining the trans-series expansion in terms of the 't Hooft coupling $\alpha$ introduced in (\ref{eq:LNPTC6}), which makes contact with conventional perturbation theory. To do this, we need the relation 
between $\alpha$ and $B$, which is obtained by combining (\ref{eq:LNPTC6}) and  (\ref{b0h}). This relation can be written as a trans-series equation,  
\be
B-{1\over 2} \log(B) -{1\over 2}+ \log \gamma^{(0)}(B)+ 
\log \left(1+ C_\pm  \re^{-2B} {\gamma^{(1)}(B) \over \gamma^{(0)}(B)} \right) = {1\over  \alpha}+ {1\over 2} \log ( \alpha), 
\ee
and it has a trans-series solution of the form 
\be
B= {1\over \alpha} \mathfrak{B}^{(0)}(\alpha)+ \sum_{\ell \ge 1} C_\pm ^\ell \re^{-{2 \ell /  \alpha}} \mathfrak{B}^{(\ell)}(\alpha). 
\ee
The leading term in this trans-series is given by 
\be
b=  {1\over \alpha} \mathfrak{B}^{(0)}(\alpha)={1\over \alpha}+ {1\over 2}+{\alpha\over 8} -{11 \alpha^3\over 384} -{35 \alpha^4 \over 768}+ \cdots
\ee
We can also compute the first exponential corrections. This is better done in terms of $b$. We find, for the very first terms, 
\be
\ba
 \re^{-{2  / \alpha}} \mathfrak{B}^{(1)}( \alpha)&= -\re^{-2  b} \left(1+ {1 \over 4  b} + {9 \over 32   b^2} + \cdots\right),\\
  \re^{-{4 /  \alpha}} \mathfrak{B}^{(2)}( \alpha)&= -{3 \over 2} \re^{-4   b} \left(1+ {2 \over 3  b} + {3 \over 4   b^2} + \cdots\right).
 \ea
 \ee
We obtain in this way the following trans-series structure for $\CF_0(h)$:
\be
\label{ts-F0}
\ba
\CF_0(h)\sim  -{h^2 \over 8 \pi} &\biggl\{ {1\over  \alpha}- {1\over 2} - { \alpha\over 4} -{5 \alpha^2 \over 16} -{53 \alpha^3 \over 96}-\frac{487 \alpha ^4}{384}-\frac{13789 \alpha ^5}{3840}-\frac{185143 \alpha ^6}{15360}+\CO\left(\alpha
   ^7\right)\\
&-{4 C_\pm \over \re   \alpha^2} \re^{-2 /\alpha} \left( 1 + \alpha + {\alpha^2 \over 4} -\frac{\alpha ^3}{16}+\frac{\alpha ^4}{96}-\frac{31 \alpha ^5}{384}-\frac{23 \alpha ^6}{1280}+\CO\left(\alpha ^7\right)  \right)\\
&+{2 C^2_\pm\over \re^2  \alpha} \re^{-4 / \alpha} \left( 1-\frac{\alpha }{4}+\frac{3 \alpha ^2}{8}-\frac{\alpha ^3}{2}+\frac{4 \alpha ^4}{3}-\frac{181 \alpha ^5}{64}+\frac{3227 \alpha
   ^6}{320}+\CO\left(\alpha ^7\right)\right) \\
   &+ \CO \left( \re^{-6 / \alpha}  \right) \biggr\}, 
\ea
\ee
where $C_\pm^2=-1$. To make this completely explicit, we can write (\ref{ts-F0}) with the notations introduced in \eqref{eq:TransPhi} as a trans-series in $\alpha$, $\re^{-2/\alpha}$. We have
\be
\label{Phits}
\CF_0(h)\sim  -{h^2 \over 8 \pi} \Phi(\alpha,C_\pm) =  -{h^2 \over 8 \pi} \sum_{\ell \ge 0} C_\pm ^\ell \re^{-2 \ell /\alpha} \varphi^{(\ell)} (\alpha)\,, 
\ee
where we have factorized $C_\pm$ and dropped the unnecessary subscript $0$: 
\be
\label{Phits_pcf}
 \Phi(\alpha,C_\pm)  \equiv  \Phi_0(\alpha,C_\pm) \,, \qquad
\varphi_0^{(0)} (\alpha) \equiv  \varphi^{(0)} (\alpha) \,, \qquad 
\varphi_0^{(\ell)} (\alpha,C_\pm ) \equiv C_\pm^{(\ell)} \varphi^{(\ell)} (\alpha)\,, \;\; \ell \geq 1\,.  
\ee
The series $ \varphi^{(\ell)}(\alpha)$ can be read from (\ref{ts-F0}),  
\be
\varphi^{(0)}(\alpha)= \frac{1}{\alpha}-{1 \over 2} +\cdots, \quad\; \varphi^{(1)}(\alpha)= -{4 \over \re \alpha^2 }\left( 1 + \alpha +\cdots\right), \quad  \;
\varphi^{(2)}(\alpha)= {2 \over \re^2 \alpha}\left( 1 -{ \alpha \over 4} +\cdots\right).
\ee
Then, we have the following equality:
\be
\ba
\label{phi-med}
\CF_0(h)&=-{h^2 \over 8 \pi}s_\pm \left( \Phi \right)(\alpha;C_\pm)=-{h^2 \over 8 \pi} 
\sum_{\ell \ge 0}  C_\pm ^\ell \re^{-2 \ell /\alpha} s_\pm \big( \varphi^{(\ell)} \big) (\alpha)\,.  
\ea 
\ee
There are many aspects of the above result which are worth commenting in detail, both at the physical and the mathematical level. 

From the physics point of view, let us note that the trans-series (\ref{Phits}) 
has an {\it infinite} number of exponentially small corrections, corresponding to an infinite number of IR renormalon singularities. 
This is in contrast to the NLO term in the $1/N$ expansion of the NLSM, studied above, and also to the planar solution of 
the PCF  model \cite{fkw1,fkw2} with the choice of charges (\ref{fkw-charges}). 
However, all corrections are built up of a finite number of trans-series in the variable $B$, appearing in (\ref{f0b}). 
A similar phenomenon appears in the trans-series solution of certain Riccati ordinary differential equations \cite{bssv,ss}, in which all the exponential corrections in the trans-series are obtained from a finite number of building blocks. 

From a more formal point of view, one can ask whether the exponentially small corrections are determined by 
the perturbative series. It turns out that, in this case, $\Phi (\alpha;C)$ satisfies the same resurgent equations that the trans-series 
solution to Painlev\'e II described in detail in \cite{mmnp}. Namely, we conjecture the following equality of laterally resummed trans-series
\be
\label{med-res-p}
s_+\left(\Phi \right) (\alpha;C)= s_-\left(\Phi \right) (\alpha;C+\mathsf{S}), 
\ee
where 
\be
\mathsf{S}=2 \ri
\ee
and $C$ is now an {\it arbitrary} complex parameter. As shown in \cite{mmnp}, (\ref{med-res-p}) leads to the following relationships 
\be
\label{stokes}
s_+(\Phi^{(\ell)})(\alpha)- s_-(\Phi^{(\ell)})(\alpha)= \sum_{k\ge 1} {\ell + k \choose \ell} \mathsf{S}^k s_-(\Phi^{(\ell+k)})(\alpha), \qquad \ell\ge 0. 
\ee
where
\be
\Phi^{(\ell)} (\alpha) \equiv \re^{-2 \ell /\alpha} \varphi^{(\ell)} (\alpha)\,. 
\ee
We have explicitly verified some of these equations numerically, by including up to the fourth term in the trans-series. (\ref{stokes}) gives 
the so-called Stokes automorphism of $\Phi^{(\ell)}$ across the positive real axis, and 
it expresses it as an infinite linear combination of the higher order terms in the trans-series, $\Phi^{(\ell+k)}(\alpha)$. 
The coefficients in the r.h.s. of (\ref{stokes}) are called Stokes constants 
and are explicitly known. Equivalently, (\ref{stokes}) says that the Borel transform of $\Phi^{(\ell)}(\alpha)$, $\widehat \Phi^{(\ell)}(\zeta)$, 
has singularities at $\zeta= 2 k$, $k\in \IZ_{>0}$. By expanding $\widehat \Phi^{(\ell)} (\zeta)$ around the $k$-th singularity, 
one can obtain $\Phi^{(k+\ell)}(\alpha)$. In particular, when applied to $\ell=0$, 
(\ref{stokes}) implies that {\it all} the higher order terms in the trans-series can be obtained from the perturbative series.
\begin{figure}[t]		
\centering			
\includegraphics[scale=.45]{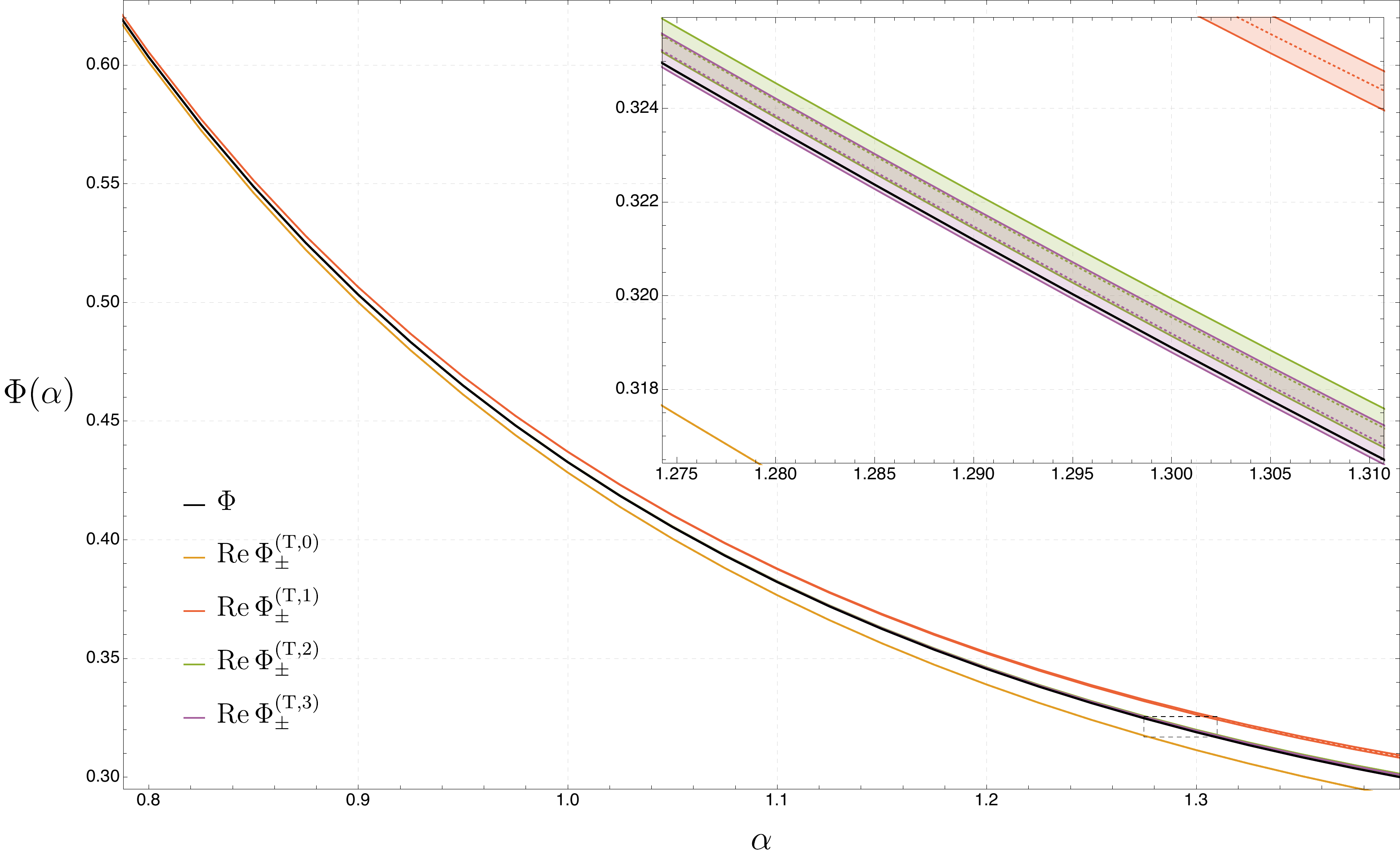} 
	\caption{Leading rescaled free energy $\Phi$ as a function of the coupling $\alpha$. The black line corresponds to the exact result, the 
	orange, red, green and blue lines to the approximate result given by Borel resumming the first series in the trans-series, as in the legend. The box to the bottom-right is a zoom
	of the region marked by the dashed black rectangle above. The dashed lines are the central values, while the shaded area corresponds to the error associated to the Borel resummation, as explained 
	in the main text.}
	\label{fig:PCF_alpha}
\end{figure}
It follows from (\ref{stokes}) that the Borel resummed trans-series 
 \be
 \label{med-res}
 s_+\left(\Phi\right) (\alpha;C-\mathsf{S}/2)=  s_-\left(\Phi \right)(\alpha;C+\mathsf{S}/2)
 \ee
 is real for any real $C$ \cite{mmnp}. This is sometimes called a {\it median resummation} (see also \cite{as13}). Since $C_\pm=\mp\ri = \mp \mathsf{S}/2$, (\ref{phi-med}) is a median resummation 
corresponding to $C=0$. 

For illustration we compare in fig. \ref{fig:PCF_alpha} the exact value of the leading rescaled free energy $\Phi \equiv s_\pm \left( \Phi \right)(C_\pm)$ with the real part of its approximations given by the truncated trans-series 
\be
\Phi^{({\rm T},n)}_\pm \equiv  \sum_{\ell=0}^n  C_\pm^\ell s_\pm(\Phi^{(\ell)}) \,. 
\label{eq:Phin+}
\ee
The term $\Phi^{({\rm T},0)}_\pm$ corresponds to the lateral Borel resummation of the perturbative series, $\Phi^{({\rm T},1)}_\pm$ to the lateral Borel resummation of the perturbative series plus the first trans-series, and so on. Notice that $\Phi^{({\rm T},n)}_+$ and  $\Phi^{({\rm T},n)}_-$ are complex conjugate. The Borel resummation of the asymptotic series entering $\Phi^{(\ell)}$ with $\ell\leq 2$ has been performed using 89 perturbative coefficients and by reconstructing the Borel function using a diagonal $[44/44]$ Pad\'e approximant, while for $\ell=3$ we have used 29 coefficients and reconstructed the Borel function with the diagonal $[14/14]$ Pad\'e approximant. The dashed lines represent the central values, with an error band given by the uncertainty in the numerical reconstruction of the Borel function. The dominant source of uncertainty comes from the convergence of the Pad\'e approximation, estimated by taking the difference between the two highest diagonal approximants: ${\rm P}[44/44]-{\rm P}[43/43]$. Other sub-dominant contributions -such as the one obtained by introducing one or more dummy variables (e.g. a Borel-Le Roy parameter) and minimizing with respect to them- have been neglected (see e.g. section 4 in \cite{Serone:2018gjo} for an overview of possible numerical recipes to estimate the error when using Borel resummation techniques). We see how the exact result is approached when considering more and more terms in the trans-series expansion.
In table \ref{tab:pcf_Phi} we show more in detail the cancellations happening between the different orders of the trans-series. 
We report, for fixed $\alpha=\frac{1}{5}$, the difference between the exact value and the real part of the truncated trans-series, and its imaginary part. It is interesting to notice how both values approach zero with a change of magnitude happening every two orders, a consequence of the fact that the $\Phi^{(\ell)}$'s satisfy the relationships (\ref{stokes}).

\begin{table}[t]
  \centering
  \begin{tabular}{c|c c}
    \hline
               & $\Phi(\frac{1}{5})-{\rm Re}\,\Phi^{({\rm T},n)}_\pm(\frac{1}{5})$                       & ${\rm Im}\,\Phi^{({\rm T},n)}_\pm(\frac{1}{5})$                      \\\hline
    $n=0$ & $2.68541385336\text{\dots} \cdot 10^{-9}$ & $\pm0.0020200401313\text{\dots}$ \\\hline
     $n=1$ & $-2.685413850(1)\cdot10^{-9}$ & $\mp9.162556(5)\cdot10^{-14}$ \\\hline
     $n=2$ & $-3.9(1)\cdot10^{-18}$ & $\pm4.581282(5)\cdot10^{-14}$  \\\hline
      $n=3$ & $1.4(1)\cdot10^{-18}$ & $\mp3(5)\cdot10^{-20}$  \\\hline
  \end{tabular}
  \caption{First orders of the truncated trans-series. In the first column the difference between the exact value and its real part and in the second column its imaginary part. The uncertainty correspond to the error associated to the Borel resummation, as explained in the main text.}
  \label{tab:pcf_Phi}
\end{table}

\begin{figure}[t]		
\centering			
\includegraphics[scale=.45]{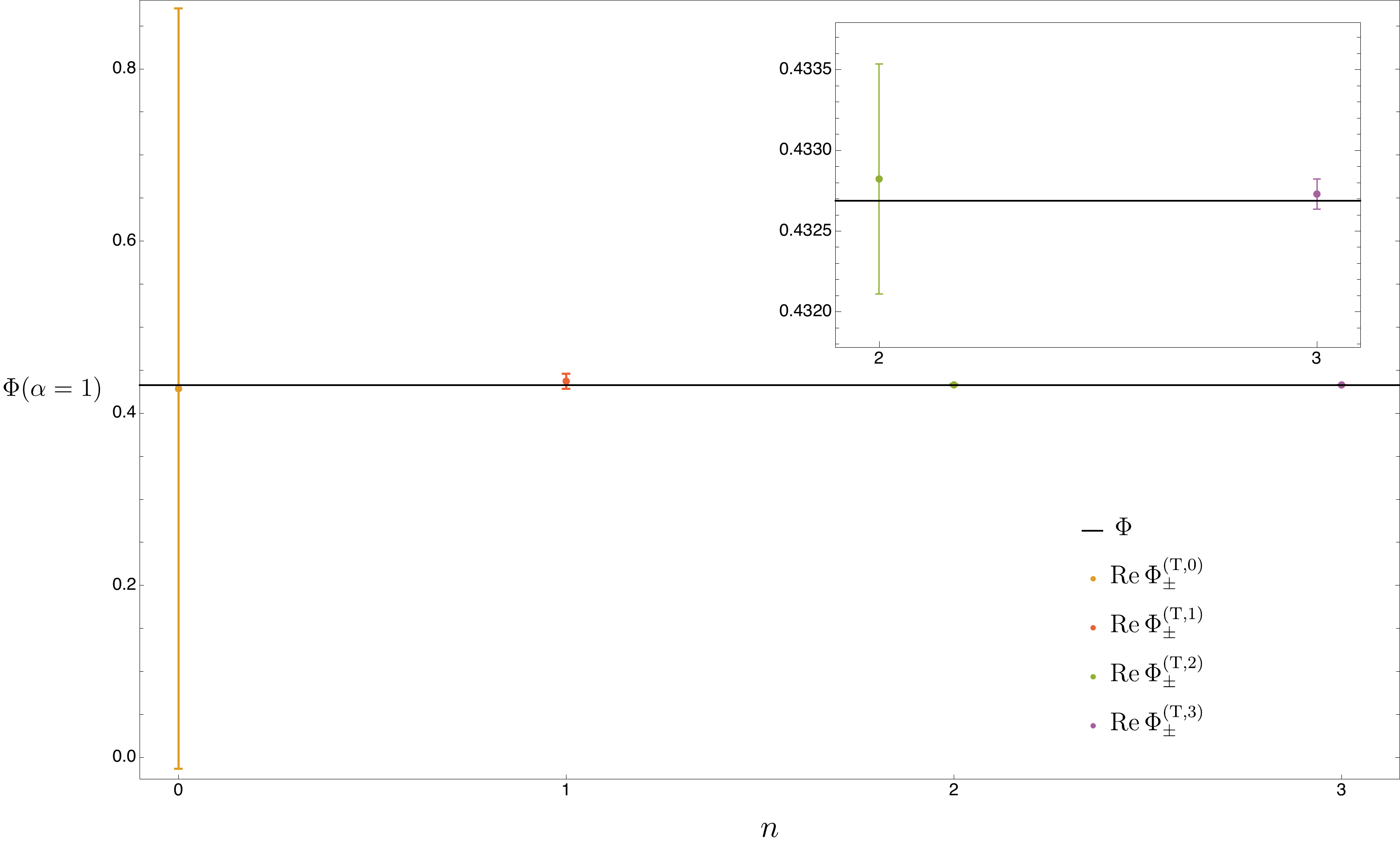} 
	\caption{The real part of of the resummed truncated trans-series $\Phi^{({\rm T},n)}$ defined as in \eqref{eq:Phin+} as a function of $n$. The black horizontal line is the exact value.
	The box to the top-right is a zoom of the last two points and their error bars. The error bars are given by the left-over ambiguity in the truncated trans-series.}
	\label{fig:PCF_n}
\end{figure}

In order to better appreciate how the exact result is approached when taking more and more terms in the trans-series, we show in fig. \ref{fig:PCF_n}  the resummation of $\Phi$ for a fixed $\alpha=1$, and as a function of the number of trans-series terms $n$ included in the resummation. In this plot the error interval is given by the left-over ambiguity in the trans-series, namely by the imaginary part of $s_\pm(\Phi^{(n)})$ at each order. The numerical error associated to the Borel resummation is sub-leading and has been neglected. Note how quick is the convergence to the exact result and how the choice of the uncertainty gives a reliable estimate of the error.

All the statements above were made for the observable $\CF_0(h)$, but a similar trans-series expansion can be made for the normalized 
energy density using (\ref{e0rho0}). In this expansion it is convenient to introduce the coupling $a$ defined by 
\be
\label{rhoalpha}
 \log \left({4 \sqrt{2} \pi^{3/2} \rho_0 \over \re\, m} \right)={1\over a}-{1\over 2}\log(a)\,,
\ee
where the numerical factor inside the $\log$ has been chosen to match this definition with that in \cite{mr-ren}. Note that the coupling $a$ here was denoted by $\alpha$ there.
The resulting trans-series is 
\be
\label{nen-series}
\ba
{e_0 \over 2 \pi \rho_0^2}&\sim a+ {a^2 \over 2} + {a^3 \over 4}+  {5 a^4 \over 16}+\frac{53 a ^5}{96}+ \cdots\\
&+ {4 C_\pm  \over \re} \re^{-2/a} \left(1+ a+ {a^2 \over 4}- { a^3  \over 16}+ {7a^4 \over 96}+ \cdots \right) \\
& +{2 C_\pm^2 \over \re^2 }\re^{-4/a} \left( a + { a^2 \over 4} - {7 a^3 \over 8}+ { a^4}+ \cdots\right)\\
& + \CO(\re^{-6/a}). 
\ea
\ee
The first line reproduces the result obtained in \cite{mr-ren}. The trans-series appearing here, in terms of $a$, has the same formal properties of its close cousin (\ref{ts-F0}), like for example (\ref{med-res-p}). 

Our main conclusion is that, in this example, the exact large $N$ free energy of the PCF model can be obtained by a median Borel resummation of a 
non-trivial resurgent trans-series, and therefore provides a beautiful success for the program of resurgence in an asymptotically free quantum field theory. Mathematically, this works because the building blocks of the exact solution (\ref{f0b}) are special functions with known trans-series representations.  From the physical point of view it is however gratifying to have a non-trivial example 
of resurgence at work with infinitely many non-trivial 
IR renormalons, yet analytically tractable. 

In a recent {\it tour de force}, Abbott and collaborators \cite{abbh1,abbh2} were able to obtain detailed information on 
the trans-series expansion for the normalized energy density in the $O(4)$ sigma model, which is nothing but the PCF model
we are studying at $N=2$. By extrapolating numerical results to analytic results, they obtained an expression very similar to (\ref{nen-series}), and they gave evidence that the exact answer can be recovered by median resummation of the trans-series. 
Our results are an analytic version, at large $N$, of their result for $N=2$.

\section{The Gross--Neveu model}
\label{sec:GN}

In this section we present our results for the GN model. In subsection \ref{sec:GNgen} we review some general aspects of the model. Then, in subsection \ref{sec:GNTsE} we present the trans-series expansion for $\CF_0(h)$ and $\CF_1(h)$ and show that these trans-series cannot be reconstructed using resurgence. Finally in subsection \ref{sec:GNExpN} we numerically study the $1/N$ series expansion of $\CF$ up to high order, 
we analytically continue the series beyond its radius of convergence, and see how this continuation matches with known dualities between GN models at low $N$ and sine-Gordon theories.

\subsection{General aspects}
\label{sec:GNgen}

The Lagrangian density describing the Gross--Neveu (GN) model \cite{gross-neveu} is 
\be
\CL= \frac{\ri}{2} \overline{\boldsymbol{\chi}} \cdot \slashed{\partial} \boldsymbol{\chi}+ {g^2\over 8} \left(\overline{\boldsymbol{\chi}} \cdot \boldsymbol{\chi}  \right)^2\,,
\label{eq:GNLag}
\ee
where $\boldsymbol{\chi}=(\chi^1,\ldots, \chi^N)$ is a set of $N$ Majorana fermions in two dimensions. 
As is well-known, a non-perturbatively generated mass gap and spontaneous breaking of a $\IZ_2$ chiral symmetry occur in this theory.
For $N>4$ the lightest particle in the spectrum is the fundamental fermion in the Lagrangian \eqref{eq:GNLag}.
In our conventions (\ref{betaf}) 
we have
\be
\beta_0={1\over 4  \pi  \Delta}, \qquad   \xi = -\Delta. 
\label{eq:betaxiGN}
\ee
The GN model has a $O(N)$ global symmetry, under which the $N$ fermions transform as vectors, with conserved currents given by 
\be
J_\mu^{IJ} = \bar \chi^I \gamma^\mu \chi^J\,.
\ee
We couple the fermions to a chemical potential $h$ associated to the $U(1)\subset O(N)$ charge $Q^{12}$.
In the GN model it is convenient to define the 't Hooft  coupling
\be
\alpha \equiv \alpha (\mu=2h)\,,
\ee
where $\alpha(\mu)$ is the TBA coupling defined in \eqref{eq:LNPTC6}. 

The first two terms $\CF_0$ and $\CF_1$ in the $1/N$ expansion \eqref{eq:FGN} have been analytically computed in \cite{fnw1,fnw2} using both QFT and TBA techniques.
They read 
\ben
{\cal F}_0(h) & = & -\frac{h^2}{2\pi} \Big(\tanh B_0 -\frac{B_0}{\cosh^2 B_0} \Big)\,,  \nonumber \\
{\cal F}_1(h) & = & -\frac{h^2}{2\pi} \frac{2}{\cosh^2 B_0} (\sinh^2 B_0 +  B_0^2- B_0 {\rm Shi}(2B_0))\,,
\label{eq:LNPTC17}
\een
where
\be
B_0= \cosh^{-1} \left({h \over m} \right)\,, 
\ee
and Shi$(x)$ is the hyperbolic sin integral function
\be
 {\rm Shi}(x)  = \int_0^x \frac{\sinh t}{t} \rd t \,.
\ee

\subsection{Trans-series expansion}
\label{sec:GNTsE}

\begin{figure}[t]		
\centering			
\includegraphics[scale=.27]{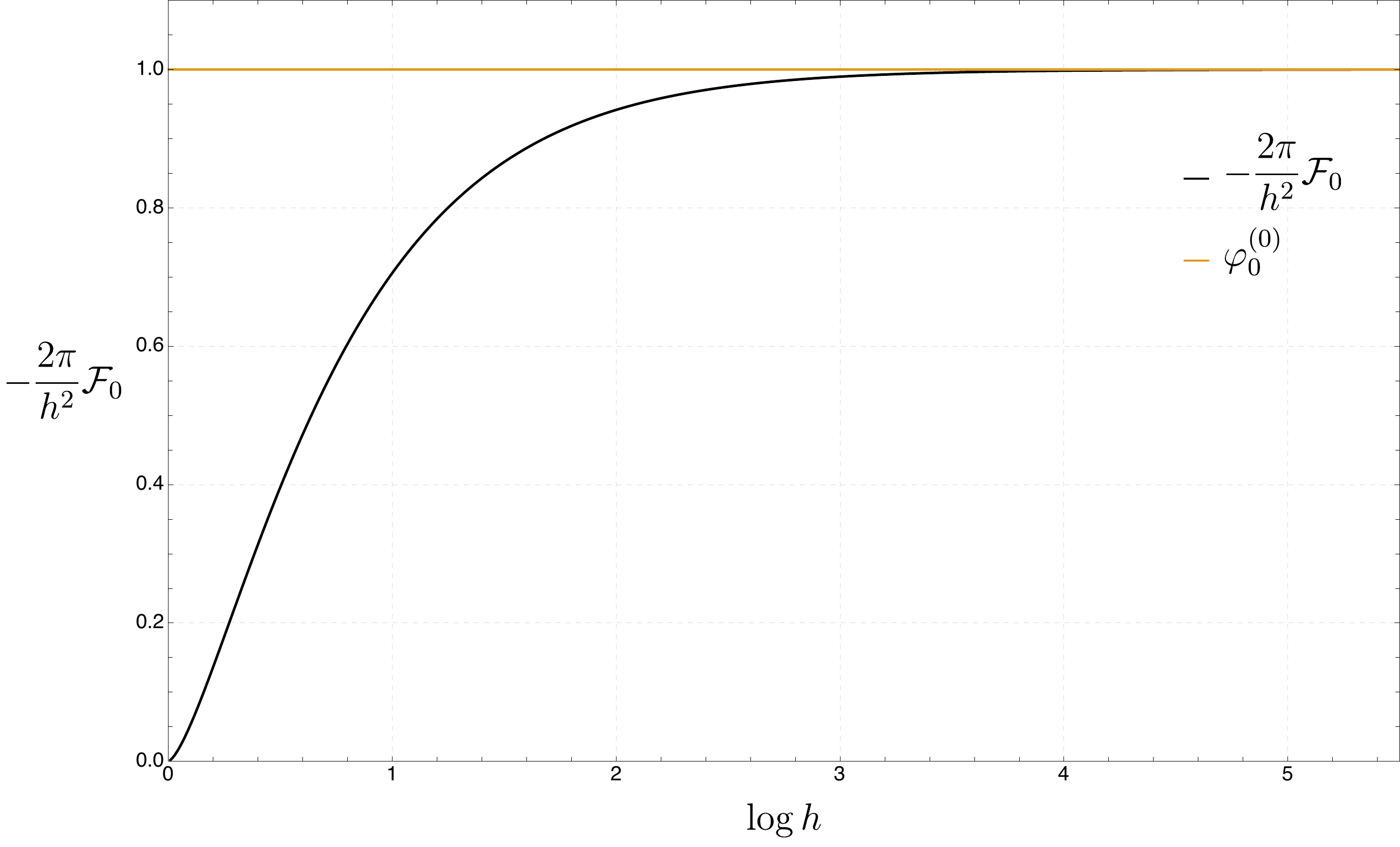} 
\includegraphics[scale=.27]{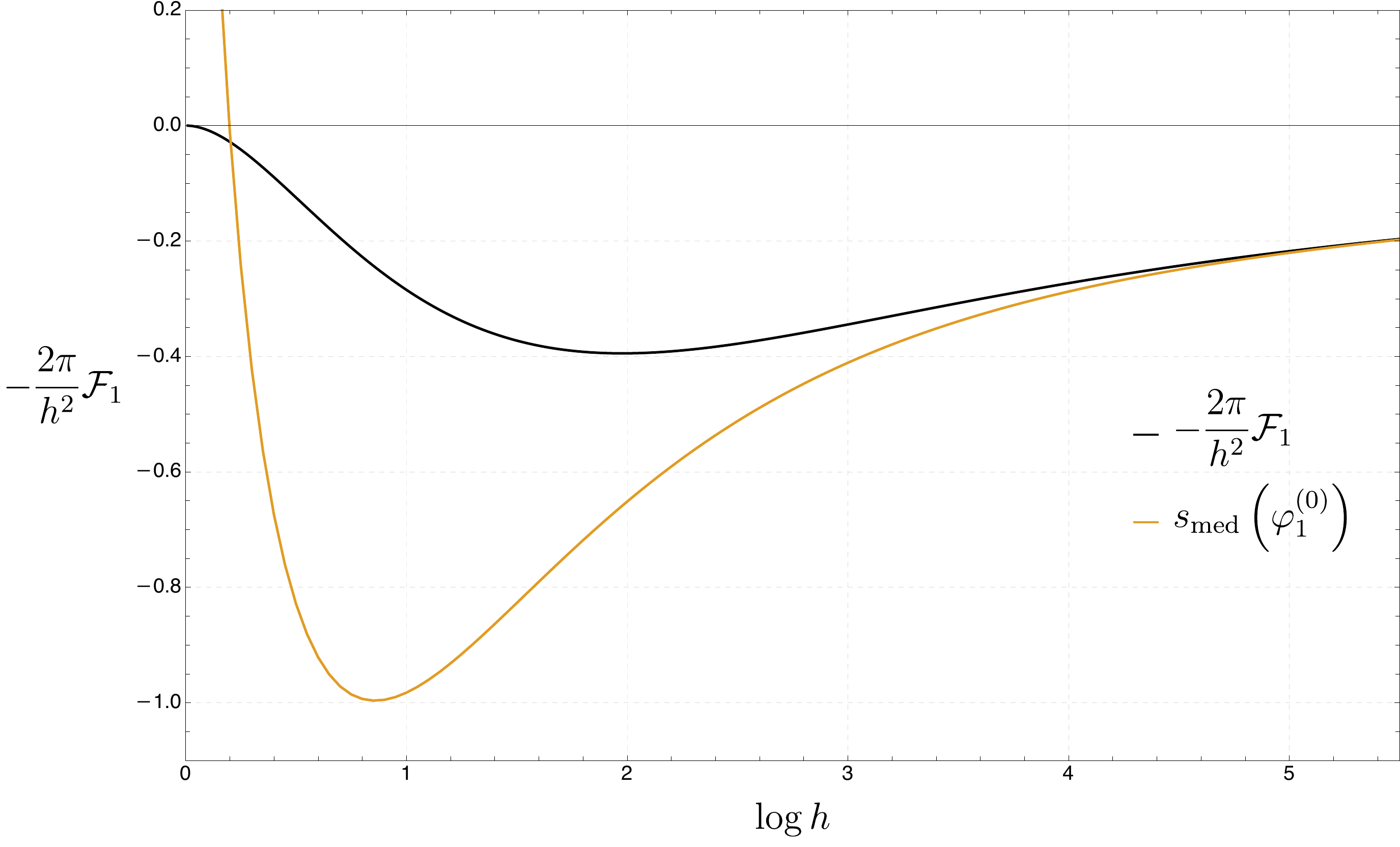} 
	\caption{Free energy coefficients $\CF_{0,1}$ rescaled by $-2\pi/h^2$ as a function of the coupling of the external field $h$ with $m$ set to 1. The orange and black lines correspond to the perturbative and exact results, respectively, for both the leading (left) and next-to-leading (right) orders in $\Delta$.}
	\label{fig:GN_alpha}
\end{figure}

Let us now work out the explicit form of the first terms of the trans-series $\Phi_k(\alpha)$ defined in (\ref{eq:FGN}), for $k=0,1$, using \eqref{eq:LNPTC17} and \eqref{eq:LNPTC6} relating $\alpha$ and $h$.
 This map depends on $\Delta$ through the $\xi$ factor in \eqref{eq:betaxiGN}. Up to order $\Delta$ we get
\be
h = \re^{\frac{1}{\alpha}}\frac{m}{2} \Big(1- \frac{ \Delta}{2} \log (\alpha^2) \Big)+ {\cal O}(\Delta^2) \,.
\label{eq:LNPTC11}
\ee
For $k=0$, the first few terms of the series $\varphi_0^{(\ell)}$ defined in (\ref{eq:TransPhi}) read
\be
\varphi_0^{(0)}(\alpha) = 1\,, \quad\varphi_0^{(1)}(\alpha) = -2-\frac{4}{\alpha}\,, \quad\varphi_0^{(2)}(\alpha) = 2\,, \quad \varphi_0^{(3)}(\alpha) = 2\,, \quad
\varphi_0^{(4)}(\alpha) = \frac{10}{3} \,.
\label{eq:LNPTC16}
\ee
Note that the perturbative expansion $\varphi_0^{(0)}$ is trivial and all trans-series terms $\varphi_0^{(\ell)}$ with $\ell\geq 1$ are truncated, yet non-vanishing.
We see that here is no way to reconstruct the non-perturbative terms  $\varphi_0^{(\ell)}(\alpha)$ with $\ell\geq 1$ from $\varphi_0^{(0)}$.
Since the leading order is somewhat trivial, it is useful to go through the next-to-leading order $\Phi_1$, where each term $\varphi_1^{(\ell)}(\alpha,C_\pm)$ has a non-trivial asymptotic expansion in $\alpha$. After some algebra, the first terms read
 \ben
\varphi_1^{(0)}(\alpha) \!\!& =  & \!\!  -\Big(\alpha+\alpha^2 + \frac{3}{2} \alpha^3 + 3 \alpha^4 + \frac{15}{2} \alpha^5 + \frac{45}{2} \alpha^6 + {\cal O}(\alpha^7)\Big)\,, \nonumber   \\
\varphi_1^{(1)}(\alpha, C_\pm) \!\!& =  &  C_\pm \frac{4 \pi}{\alpha} +\frac{8}{\alpha^2} - \frac{4}{\alpha} \log (\alpha^2 ) - 4  + \alpha \Big(2+ \alpha + \alpha^2 + \frac 32  \alpha^3 + 3 \alpha^4 + {\cal O}(\alpha^5)\Big)\,, \nonumber  \\
\varphi_1^{(2)}(\alpha, C_\pm) \!\!&=&  -C_\pm 4 \pi -\frac{16}{\alpha} +4 \log (\alpha^2 )+ \alpha \Big( 2 +  \frac{1}{2} \alpha + 3 \alpha^2-\frac{3}{4} \alpha^4 + {\cal O}(\alpha^5) \Big)  \,.\label{eq:LNPTC26}
\een
with $C_\pm=\pm \ri$. The perturbative series $\varphi_1^{(0)}$ turns out to be equal to 
\be
-2 \sum_{n=1}^{\infty} \Gamma(n+1) \Big(\frac{\alpha}{2}\Big)^n  \,.
\label{eq:LNPTC27}
\ee
This series is non-Borel resummable, but can be studied analytically.
Its Borel transform is 
\be
B_1(t)  = \frac{2t}{t-2} \,.
\label{eq:LNPTC28}
\ee
The simple pole at $t=2$ hinders Borel summability. We can deform the contour to avoid the pole, passing either above (${\cal C}_+$) or below (${\cal C}_-$) it. 
The Borel resummation of this series gives then 
\be
s_\pm \big(\varphi_1^{(0)}\big)(\alpha) =\frac{2}{\alpha}\int_{{\cal C}_\pm} \! \rd t \, \re^{-\frac{t}{\alpha}} \frac{t}{t-2} =
  \frac{2}{\alpha} {\rm P} \Big(\int_0^\infty \! \rd t \, \re^{-\frac{t}{\alpha}} \frac{t}{t-2}\Big)  \mp \re^{-\frac{2}{\alpha}}\ \frac{4 \ri \pi}{\alpha}  \,.
\label{eq:LNPTC29}
\ee
Nicely enough, the ambiguity in the imaginary part in \eqref{eq:LNPTC29} is exactly the same appearing in $\varphi_1^{(1)}(\alpha, C_\pm)$. The two contributions cancel each other if we choose the contour ${\cal C}_\pm$ for $C_\pm$, respectively. The terms in parenthesis in the series  \eqref{eq:LNPTC26} for $\varphi_1^{(1)}(\alpha, C_\pm)$ form the same asymptotic series \eqref{eq:LNPTC27}.  
Their Borel resummation gives 
\be
s_\pm \big(\varphi_1^{(1)}\big)(\alpha, C_\pm) \supset \int_{{\cal C}_\pm} \! \rd t \, \re^{-\frac{t}{\alpha}} \frac{4}{2-t}  
 = {\rm P} \Big(\int_0^\infty \! \rd t \, \re^{-\frac{t}{\alpha}} \frac{4}{2-t}\Big)  \pm \re^{-\frac{2}{\alpha}} 4 \ri \pi  \,,
\label{eq:LNPTC31}
\ee
where for simplicity we have not reported the first four terms of  $\varphi_1^{(1)}$ appearing in \eqref{eq:LNPTC26} (that's why the $\supset$ sign instead of the equality sign).
Again, if we pick up the contour ${\cal C}_\pm$, the imaginary part in \eqref{eq:LNPTC31} cancels respectively the imaginary terms proportional to $-C_\pm$ appearing in the second trans-series $\varphi_1^{(2)}$.
In the spirit of resurgence, imaginary parts nicely match between one series and the next, but we see a plethora of real non-perturbative terms which cannot be detected.
The asymptotic series $\varphi_1^{(1)}$  and those with $\ell \geq 2$ cannot be reconstructed from the knowledge of $\varphi_1^{(0)}$  only.
In order to quantify and illustrate the phenomenon, in fig. \ref{fig:GN_alpha} we compare 
the exact  free energies $\CF_{0,1}$, rescaled by $h^2$, with the median Borel resummation of their perturbative series expressed in terms of $h$ and $m$,  i.e $s_{\rm med}(\varphi_{0,1}^{(0)})(1/\log(2h/m))$.\footnote{Since $\varphi_{0}^{(0)}(\alpha)=1$, there is no need of resummation at LO and this order does not contribute at the next one, differently to what happens in the NLSM.} We restrict to the perturbative series
because higher trans-series terms can not be  obtained from a resurgence analysis. As expected, for both $k =0,1$, the perturbative and full results are in good agreement at large $h$, i.e. weak coupling, but they significantly differ at strong coupling, when the terms with $\ell \geq 1$ are no longer negligible.

All the above analysis could be repeated for the energy density $e(\rho)$, Legendre transform of $\CF(h)$, expanded in $\Delta$ and written as a trans-series as 
\be
e(\rho) 
\sim \frac{\pi}{2}\rho^2  \sum_{k\ge 0} \widetilde \Phi_{k}(a,C_\pm) {\Delta}^{k} \,,
\ee
where
\be
\widetilde \Phi_{k}(a,C_\pm)   = \sum_{\ell=0}^\infty e^{-\frac{2\ell}{a}} \widetilde \varphi_k^{(\ell)}(a,C_\pm)\,.
\ee
The asymptotic expansion in this case is more conveniently written in terms of the coupling 
\be
a\equiv \alpha(\mu = 2\pi \rho)\,.
\label{eq:LNPTC8}
\ee
For $\widetilde \varphi_0$ we get
\ben
\widetilde \varphi_0^{(0)} (a) =   1\,, \quad \widetilde \varphi_0^{(1)} (a) =   2+\frac{4}{a} \,, \quad \widetilde \varphi_0^{(2)} (a) =   2\,,\quad  \widetilde \varphi_0^{(3)} (a) =  -2\,,\quad
\widetilde \varphi_0^{(4)} (a) =  \frac{10}{3}\,,
\label{eq:LNPTC15}
\een
while for $\widetilde \varphi_1$ we have
\ben
\widetilde \varphi_1^{(0)}(a) &=  & \!\! a+ a^2 + \frac{3}{2}a^3 + 3a^4 + \frac{15}{2}a^5 + \frac{45}{2}a^6 + {\cal O}(a^7) \,, \nonumber   \\
\widetilde \varphi_1^{(1)}(a) &=  &  - C_\pm \frac{4\pi}{a} -\frac{8}{a^2} + \frac{4}{a} \log (a^2 ) + 4  +a\Big(2+a+a^2 + \frac 32 a^3 + 3a^4 + {\cal O}(a^5)\Big) \,,  \nonumber  \\
\widetilde \varphi_1^{(2)}(a) &=  & - C_\pm 4 \pi -\frac{16}{a} +4 \log (a^2 ) -a\Big( 2 +  \frac{1}{2}a+ 3a^2-\frac{3}{4}a^4 + {\cal O}(a^5) \Big)\,.  \label{eq:LNPTC25}
\een
We see that the expansions of $\widetilde \Phi_0$ and $\widetilde \Phi_1$ in terms of $a$ are very similar to those of $\Phi_0$ and $\Phi_1$ in terms of $\alpha$.
The perturbative series $\widetilde \varphi_1^{(0)}$ in \eqref{eq:LNPTC25} agrees with the perturbative expansion found in eq.(A.12) of \cite{mr-ren} using the techniques of \cite{volin, volin-thesis}, while
 $\widetilde \varphi_1^{(\ell)}$, with $\ell>0$, are non-perturbative terms that could not be captured in that analysis.
All the considerations made above about imaginary part cancellations and impossibility of recovering the non-perturbative terms from the perturbative series of $\CF(h)$ 
apply also for $e(\rho)$ and will not be repeated.

\subsection{Higher orders in the $1/N$ expansion}  

\label{sec:GNExpN}

In contrast to the NLSM and PCF models, the kernel in the GN model is analytic at $N=\infty$. This implies that the TBA solution can easily be expanded in powers of $1/N$, with each term a regular function of $\theta$, making it possible to solve the 
integral equations \eqref{intetwo} in a systematic $1/N$ expansion:
\be
K(\theta)= \sum_{k \ge 1} K_k(\theta) \Delta^k \,, \quad \quad 
\epsilon(\theta)=\sum_{k \ge 0} \epsilon_k (\theta) \Delta^k, \quad \quad B= \sum_{k \ge 0} B_k \Delta^k\,.  
\label{expepsB_gn}
\ee
The kernel coefficients $K_k(\theta)$ are trivially derived from the GN kernel reported in (\ref{eq:AEGN1}).  The $B_k$ with $k \geq 1$ can be expressed in terms of the values of the $\epsilon_m (B_0)$ with $m\leq k$ and their derivatives by solving recursively the condition (\ref{eq:Bcondition}) at each order in $\Delta$. For instance, for the first few orders we have
\be
\epsilon_0(B_0)+\Delta \bigg(\epsilon_1(B_0)+\partial_\theta \epsilon_0 \big\vert_{B_0} \  B_1 \bigg)+\Delta^2 \bigg(\epsilon_2(B_0)+\partial_\theta \epsilon_1 \big\vert_{B_0} \  B_1 + \partial^2_\theta \epsilon_0 \big\vert_{B_0} \  \frac{B_1^2}{2} +\partial_\theta \epsilon_0 \big\vert_{B_0} \  B_2 \bigg)+ \dots = 0\,.
\label{epsexpB_gn}
\ee
Plugging (\ref{expepsB_gn}) in (\ref{intetwo}) we have
\ben
\label{expansioneps}
\epsilon(\theta) \!\! & =&\!\! \ h-m \cosh \theta + \sum_{k \ge 1} \Delta^k \int_{-B_0}^{B_0} \rd\theta'\,  \Big(\sum^{k-1}_{n=0}  K_{k-n}(\theta-\theta') \epsilon_n(\theta') \Big) \\
\!\!&+& \!\! \sum_{p \ge 1} \frac{1}{p!}\Big(\sum_{q \ge 1} B_q \Delta^q \Big)^p \partial^{p-1}_{\theta'} \bigg(\sum_{k \ge 1} \Delta^k \sum^{k-1}_{n=0}  \left(K_{k-n}(\theta-\theta')+K_{k-n}(\theta+\theta')\right) \epsilon_n(\theta') \bigg)\bigg|_{B_0}\,, \nonumber
\een
where we used the fact that $K_n(\theta)$ and $\epsilon_n(\theta)$ are even functions. Solving the equation at each order in $\Delta$, we can compute iteratively all the $\epsilon_k(\theta)$ knowing the values of $\epsilon_m(\theta)$ for $m < k$ and the $B_q$ with $q < k-1$.
Finally, in order to compute $\CF_k(h)$ it is enough to expand (\ref{fh-bethe}) in $\Delta$:
\ben
\label{expansionF}
\CF(h)\!\!\!\!&=&\!\!\!\!-\frac{m}{2\pi}\int_{-B}^B \rd \theta  \, \cosh \theta \ \Big(\sum_{k \ge 0} \epsilon_k (\theta) \Delta^k\Big)\\
\!\!\!\!&=&\!\! \!\! -\frac{m}{2\pi}\bigg(  \sum_{k \ge 0} \Delta^k \!\! \int_{-B_0}^{B_0} \!\!\rd \theta \cosh \theta\  \epsilon_k (\theta) +2 \sum_{p \ge 1} \frac{1}{p!}\Big(\sum_{q \ge 1} B_q \Delta^q \Big)^p \sum_{n \ge 0} \Delta^n \partial^{p-1}_\theta \Big(\cosh \theta \  \epsilon_n (\theta) \Big)\bigg|_{B_0} \bigg)\,, \nonumber
\een
where we used once again the fact that $\epsilon_n(\theta)$ is an even function.

In order to make more explicit the procedure and show how the iterative process starts, let's rederive $\CF_0(h)$ and  $\CF_1(h)$ given in \eqref{eq:LNPTC17}. At order $\Delta^0$ we simply have 
\be
K_0=0\,, \quad \quad\epsilon_0(\theta, h)= h-m \cosh(\theta)\,, \qquad B_0= \cosh^{-1} \left({h \over m} \right)\,.
\ee
The leading order free energy reads
\be
\CF_0(h(B_0))=-\frac{m}{2\pi} \int_{-B_0}^{B_0} \rd \theta \cosh(\theta) \ \epsilon_0(\theta, h(B_0))=-\frac{h^2}{2\pi} \bigg( \tanh(B_0)-{B_0 \over \cosh^2(B_0)}\bigg)\,,
\ee
as already reported in the first equation of (\ref{eq:LNPTC17}).

At order $k=1$ we have
\be
K_1(\theta)= {1\over \theta^2} -{\cosh(\theta) \over \sinh^2(\theta)}\,,
\ee
and 
\be
\epsilon_1(\theta,h)= \int_{-B_0}^{B_0} K^{(1)}(\theta-\theta') \ \epsilon_0(\theta', h)\rd \theta'\,.
\ee
Performing the integral we get
\be
\ba
\epsilon_1(\theta, h(B_0))=&\ m \sinh (\theta)( \text{Chi}(B_0+\theta )-\text{Chi}(B_0-\theta )-\log (\sinh (B_0+\theta) \text{csch}(B_0-\theta)))\\
+&\ m \cosh (\theta) (2 B_0-\text{Shi}(B_0-\theta )-\text{Shi}(B_0+\theta ))\,,
\label{eps1_gn}
\ea
\ee
and from it 
\be
B_1(B_0)= \ - \frac{\epsilon_1(B_0, h(B_0))}{\partial_{\theta}\epsilon_0(\theta, h(B_0))\Big\vert_{B_0}}
=\ \text{Chi}(2 B_0)+(2 B_0-\text{Shi}(2 B_0)) \coth (B_0)-\log (\sinh (2 B_0))-\gamma_E\,,
\ee
where $\text{Chi}(x)$ is the hyperbolic $\cos$ integral function 
\be
 {\rm Chi}(x)  = \gamma_E+\log(x)+ \int_0^x \frac{\cosh t - 1}{t} \rd t \,.
\ee
Given  that $\epsilon_0(B_0)=0$, the next-to-leading term of the the free energy reads
\be
\ba
\CF_1(h(B_0))= -\frac{m}{2 \pi} \int_{-B_0}^{B_0} \rd \theta \cosh(\theta) \ \epsilon_1(\theta,h(B_0))\,.
\label{eq:CF1_GNa}
\ea
\ee
Plugging (\ref{eps1_gn}) in \eqref{eq:CF1_GNa} and computing the integral gives the second equation of (\ref{eq:LNPTC17}).

At higher order in $\Delta$ the computation becomes analytically prohibitive. 
On the other hand, it is straightforward to proceed numerically and, for a given $B_0$, automatize the iteration procedure to compute higher order terms in $\Delta$.
We have been able to compute with high precision, for different values of $h$, $\CF_k$ up to $k=28$.
This allowed us to study the large order behavior of the series. We get\footnote{The presence of a period of oscillation in the $1/N$ coefficients makes less straightforward the determination of the large order behavior. We have made use of a program written by Jie Gu and based on the work \cite{HunterGuerrieri1980}.} 
\be
\CF_k \propto  \rho^{-k} \sin (k  \, \vartheta)\,,
 \ee
 where $\rho$ and $\theta$ are two parameters that we can numerically evaluate.
For example, for $h=3$ we get
\be
\rho= 0.50 \pm 0.02 \qquad \text{and} \qquad \vartheta= 0.35 \pm 0.07 \,.
 \ee
This result confirms that the $1/N$ series of $\CF(h)$ is convergent in the GN model, in agreement with what found
in appendix \ref{app:analytic}. The radius of convergence $\rho$ should equal $1/2$ independently of $h$, while we did not investigate the possible dependence on $h$ of $\vartheta$.
The value $\Delta=1/2$ corresponds to $N=4$, so we see that for any integer value $N>4$, where the fundamental fermions are stable
and the TBA equations \eqref{chi-ie} and \eqref{intetwo} apply, the free energy can be recovered from its $1/N$ expansion. 

It is now natural to ask if the analytic continuation of the series in $\Delta$ beyond $|\Delta|\geq 1/2$ contains any physical information.
This analytic continuation can be obtained by considering Pad\'e approximants $\text{P}\CF_{[m/n]}(\Delta,h)$ of the series of $\CF_{k}$ we computed. It is known that
for convergent series, parametrically diagonal Pad\'e approximants converge (in capacity) to the exact function and the location of their poles and zeros  
define an appropriate locus of branch-cuts connecting  branch-point singularities \cite{STAHL1997139} (see e.g. app. D of \cite{DiPietro:2020jne} for a brief overview and \cite{baker1996pade} for a comprehensive introduction). Moreover, as we will see below, $\CF(\Delta, h)$, at fixed $h$, is analytic at $\Delta= \infty$ and non-vanishing, hence diagonal approximants are the optimal choice to reconstruct the function.

We calculated $\text{P}\CF_{[14/14]}(\Delta,h)$ for a given set of values of $h$, and compared the result with $\CF(\Delta,h)$ computed by directly solving (numerically) \eqref{intetwo} at given $h$ and $\Delta$. 
We find full agreement for all values of $h$ sampled and for  $0 < \Delta < \frac{1}{2}$, and consider it a sanity check of the correctness of the coefficients $\CF_{k}$. 
The location of the poles and zeros of $\text{P}\CF_{[14/14]}(\Delta,h)$ shows that the point $\Delta = 1/2$ is a branch-point of $\CF(\Delta)$. On the other hand, no singularities appear for $\Delta < 0$ (as expected from the form of the GN kernel) and hence we can reliably continue $\CF(\Delta)$ for $\Delta<0$ using its approximant $\text{P}\CF_{[14/14]}(\Delta,h)$.
The two interesting points  to discuss are $\Delta = 1/2$ ($N=4$) and $\Delta =- \infty$ ($N=2$).

For $N=4$ the stable particles in the model are the kinks and their mass equals
\be
m_\rk =\frac{m}2\,.
\label{eq:N41}
\ee
Since the fermions are exactly at threshold and are marginally unstable, we can use \eqref{intetwo} to compute the free energy by choosing either kinks or fermions
as particles populating the vacuum,  but some care is needed.
It is useful to briefly review how the analysis goes  \cite{fnw1}. 

Kinks are in the $(\pm 1/2,\pm 1/2)$ spinorial representation of $O(4)$, so their charges are half those of the fermions.
When $h/2 > m/2$ (or $h>m$) the vacuum is populated by kinks with $O(4)$ components $(1/2,1/2)$ and $(1/2,-1/2)$, which do not interact with each other. 
The $S$-matrix is identical for the two chiralities and the associated kernel is 
\be
K_\rk(\theta) = \frac{1}{\pi^2} \sum_{n=1}^\infty  (-1)^{n+1} \frac{n}{n^2+(\theta/\pi)^2}\,.
\label{eq:N43}
\ee
If we consider kinks, the associated TBA equation is
\be
\epsilon_\rk(\theta) -  \int_{-B}^B\! \! K_\rk(\theta-\theta') \epsilon_\rk(\theta') \rd\theta' = \frac 12 (h-m \cosh\theta) \,,
\label{eq:N46}
\ee
where $\epsilon_\rk$ describes the excitation of the kink holes. Given $\epsilon_\rk(\theta)$, the free energy is computed as
\be
\label{fh-betheKinks}
\CF(h)=  - 2 {m_\rk \over 2 \pi}\int_{-B}^B \rd \theta  \,  \epsilon_\rk(\theta) \cosh \theta=  -  {m \over 2 \pi}\int_{-B}^B \rd \theta  \, \epsilon_\rk(\theta) \cosh \theta\,,
\ee
where the factor 2 counts the two-fold degeneracy of the kink and exactly compensates for the 1/2 factor in the mass.
The kink kernel \eqref{eq:N43} is naively $1/2$ of the fermion kernel in \eqref{eq:AEGN2} for $\Delta = 1/2$.
If we take the limit carefully, however, we also get a $\delta$ function because
\be
\lim_{y \rightarrow 0} \frac{1}{\pi} \frac{y}{y^2 + x^2} = \delta (x) \,.
\label{eq:delta}
\ee
Hence
\be
\lim_{\Delta \rightarrow 1/2} K_\rf(\theta) =  -\delta(\theta) + 2 K_\rk(\theta)\,,
\label{eq:N45}
\ee
where we denote by $K_\rf$ the fermion kernel. So, for $\Delta \rightarrow 1/2$, the TBA equation \eqref{intetwo} for the fermion excitation holes $\epsilon_\rf$  becomes
\be
\epsilon_\rf(\theta) -  \int_{-B}^B\! \!  \Big(- \delta(\theta) + 2 K_\rk(\theta)\Big) \epsilon_\rf(\theta') \rd\theta' 
= 2 \epsilon_\rf(\theta) -2   \int_{-B}^B\! \!   K_\rk(\theta) \epsilon_\rf(\theta') \rd\theta' 
= h-m \cosh\theta \,,
\label{eq:N47}
\ee
which is identical to \eqref{eq:N46} with
\be
\epsilon_\rk(\theta) = \epsilon_\rf(\theta) \,.
\ee
Note that we would {\it not} get the correct result by setting $N=4$ directly in the fermion case, because in this way we would not detect the 
$\delta(\theta)$ term in \eqref{eq:N45}. On the other hand, the analytic continuation of $\text{P}\CF_{[14/14]}(\Delta,h)$, computed using fermion states for $\Delta <1/2$,
gives the correct result at $\Delta = 1/2$. 

The  $N=4$ model is also equivalent to a pair of decoupled sine-Gordon models: 
\be
{\cal L}  = \sum_{j=1,2} \bigg( \frac 12 \frac{8\pi}{b^2} (\partial \phi_j)^2 + \frac{\pi}{4} \Big(\frac{8\pi}{b^2}-1\Big) \cos \sqrt{8\pi} \phi_j \Big) \bigg)\,,
\label{eq:SGtwo}
\ee
where $b$ is the inverse radius of the compact scalars.
For $b^2> 4\pi$ the only asymptotic states in the sine-Gordon models are given by kinks and anti-kinks.  In presence of a chemical potential $h$, the vacuum gets populated by kinks (and no anti-kinks) with a kernel whose Fourier transform is given by \cite{Zamolodchikov:1995xk}
\be 
\widetilde K_{{\rm SG}}(\omega,p) = \frac{\sinh \frac{\pi (p+1) \omega}{2}}{2\cosh \frac{\pi \omega }{2} \sinh  \frac{\pi p \omega }{2} }\,,
\ee
where the parameter $p$ is defined in terms of $b$ as follows:
\be
\frac{8\pi}{b^2} \equiv \frac{p+1}{p}\,. 
\ee
On the other hand, the Fourier transform of the kink kernel \eqref{eq:N43} reads
\be
\widetilde K_\rk(\omega)  \equiv \int_{-\infty}^\infty \! \re^{\ri \omega \theta} \, K_\rk(\theta) \rd\theta = \frac{1}{1+\re^{\pi |\omega|}}\,.
\label{eq:N43a} 
\ee
Interestingly enough, 
\be
\lim_{p\rightarrow \infty} \widetilde K_{{\rm SG}}(\omega,p) = \widetilde K_\rk(\omega) \,,
\label{eq:N44} 
\ee
so there is an equivalence of the free energy $F(h)$ in the two models provided 
\be
b^2= 8 \pi
\ee
in the two sine-Gordon models. Note that when $b^2\rightarrow 8\pi$, the coupling of the sine-Gordon interaction at the same time becomes marginal and vanishes.
From \eqref{eq:SGtwo} it might seem that we get in this limit a pair of decoupled free field scalars, but in fact this is an artefact. The only asymptotic states are kinks and these
are still interacting, as evident from the non-triviality of the kernel \eqref{eq:N44}.\footnote{See e.g. \cite{Amit:1979ab} for a detailed analysis of the sine-Gordon model when $b^2\sim 8\pi^2$.}
In the correspondence the kink mass of the GN model is mapped to the mass of the sine-Gordon kink: $m_\rk = m^{\rm N=4}_{\rm SG}$.

\begin{figure}[t!]
 \centering
\includegraphics[height=0.53\textwidth]{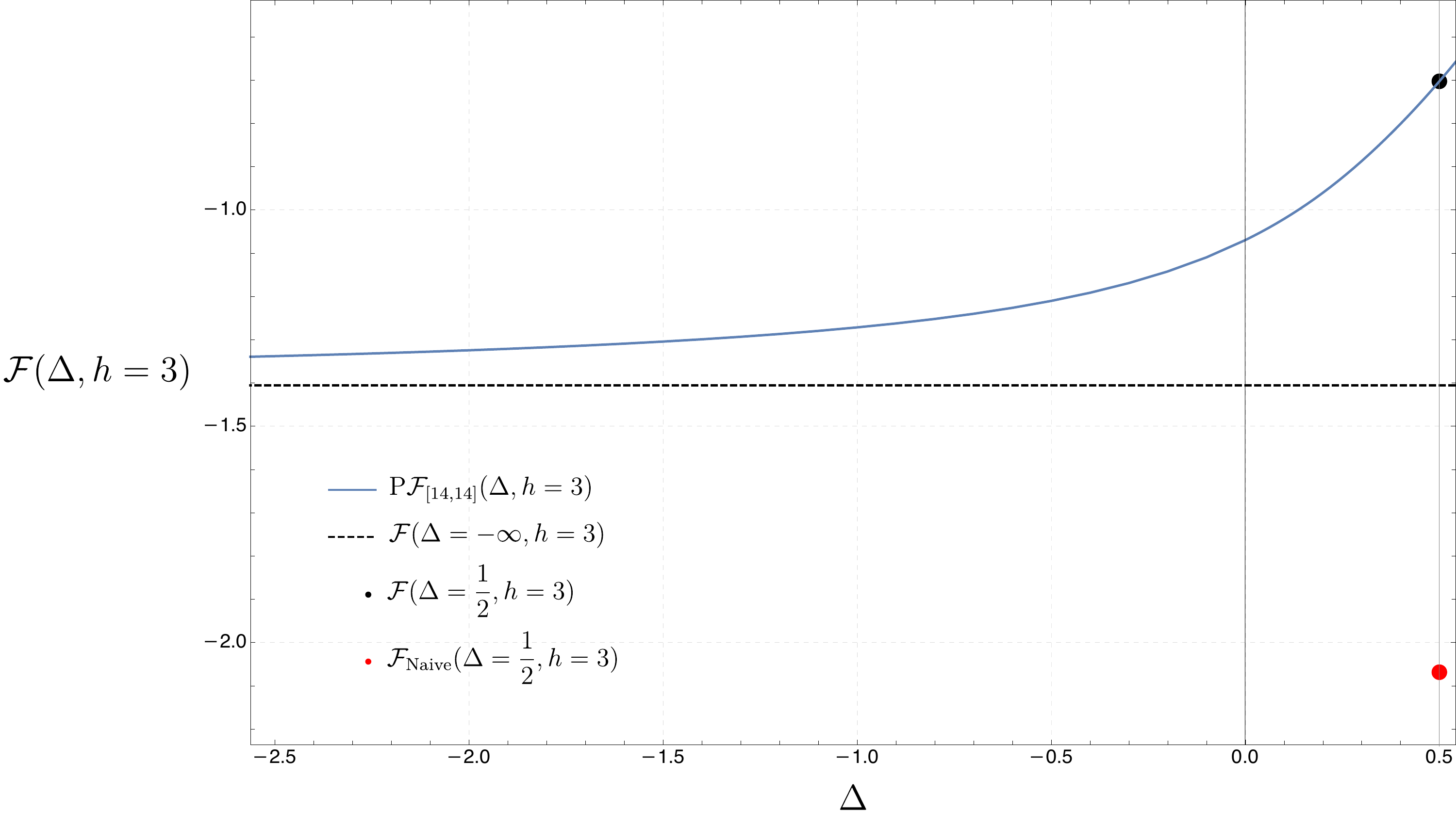}
\caption{Free energy as as function of $\Delta$ at fixed $h=3$. The red dot corresponds to the (wrong) free energy one would get by setting $\Delta = 1/2$ in the kernel appearing in \eqref{intetwo}. The black dot
is the correct value found using \eqref{fh-betheKinks}.}
 \label{fig:Gn_P}
\end{figure}	

For $N=2$, i.e. $|\Delta|=\infty$, the TBA equations \eqref{intetwo} are physically meaningless, since
fundamental particles are no longer asymptotic states. Yet, we can wonder if its analytic continuation is of physical interest and is related in
some way to the actual $N=2$ Gross-Neveu model. The latter is nothing else than the Thirring model,
famously dual to the sine-Gordon theory \cite{Coleman:1974bu}. If we approach the infinite limit from negative values of $\Delta$,
we see from direct inspection of either \eqref{eq:AEGN1} or \eqref{eq:AEGN2} that the GN kernel is analytic at infinity (recall that the digamma function $\psi(z)$
is meromorphic with simple poles at $z=-n$, with integer $n\geq 0$). By direct inspection we can also check that this kernel is a contraction:
\be
 \int_{-B}^B\! \!  K(\theta-\theta') \rd\theta'  < 1 \,.
 \label{eq:N48} 
\ee
All iterated kernels are hence bounded and the solution of the TBA equation is analytic in $\Delta$. 
By taking the analytic continuation of \eqref{eq:AEGN3} for $\Delta <0$ and then the limit $\Delta\rightarrow -\infty$ we immediately see that the Fourier transform of the kernel equals the kernel for the kink in the sine-Gordon theory as $p\rightarrow \infty$ (and the kernel for kinks in the $N=4$ GN model).
Quite interestingly, the in principle meaningless $|\Delta|=\infty$ point of the GN TBA equation \eqref{intetwo} is related to the free energy $\CF(h)$ of a sine-Gordon model at $b^2= 8 \pi$!
In the correspondence the fermion mass appearing in \eqref{intetwo}  is identified with the mass of the sine-Gordon kink: $m_\rf = m_{\rm SG}^{\rm N=2}$.

We can match the values of $h$ in the $N=2$ and $N=4$ theories, since $h$ multiplies a conserved quantity and does not renormalize. In particular, we can compute the ratio $h^{{\rm N=2}} /h^{{\rm N=4}}$. 
Recall that $h$ is the chemical potential for a single $U(1)$ current of the form $\bar \chi^1 \gamma^\mu \chi^2$.
Bosonizing we have  
\be
\bar \chi^1 \gamma^\mu \chi^2  =  \frac{i}{\sqrt{\pi}} \epsilon^{\mu \nu} \partial_\nu \phi \,.
 \label{eq:N49} 
\ee
When $N=2$ the scalar is free and its Lagrangian reads
\be
\frac{1}{2} \frac{4\pi}{b^2} (\partial \phi)^2 +  \frac{h}{\sqrt{\pi}} \partial_x \phi \underset{b^2=8\pi}{\longrightarrow}  \frac{1}{4} (\partial \phi)^2 +  \frac{h}{\sqrt{\pi}} \partial_x \phi \,,
 \label{eq:N50} 
\ee
while for $N=4$ the Lagrangian of the two scalars at $b^2=8\pi$ reads\footnote{Note that the relation between the inverse radius of the sine-Gordon and the GN coupling is different in the two cases, namely
\begin{equation}
b^2_{N=4}= \frac{8\pi}{1+\frac{g^2}{2\pi}}~,~~b^2_{N=2}= \frac{4\pi}{1+\frac{g^2}{2\pi}}~.
\end{equation}
Both for $N=2$ and for $N=4$ the kernel on the GN side (for fermions and for kinks, respectively) coincides with the kernel of the sine-Gordon kink, when their corresponding $b^2$ parameters are set to $8\pi$. Curiously, this corresponds to $g^2 = 0$ for $N=4$ and $g^2=-\pi^2$ for $N=2$.
} 
\be
\sum_{j=1,2} \bigg( \frac{1}{2} (\partial \phi_j)^2  +\frac{h}{\sqrt{2\pi}} \partial_x \phi_j \bigg) \,.
 \label{eq:N51} 
\ee
Rescaling the fields to have a canonically normalized kinetic term in \eqref{eq:N49} we get
\be
h^{{\rm N=2}} =2  h^{{\rm N=4}}\,. 
\label{eq:N52} 
\ee
When $h<m$ the vacuum is empty and $\CF(h=m)=0$ for both $N=2$ and $N=4$. This implies that the kink mass ratio is the same as in \eqref{eq:N52}, $m_{\rm SG}^{N=2} = 2 m_{\rm SG}^{N=4}$ and hence
that the free energies of the two models are related, namely
\be
\CF(\Delta=-\infty,h)  = 2\CF(\Delta=1/2,h) \,.
\label{eq:N53} 
\ee

\begin{table}[t]
  \centering
  \begin{tabular}{c|c c c}
    \hline
               & $\Delta=1/2$                       & $\Delta=-\infty$                      \\\hline
    $\CF_{\text{TBA}}(\Delta,h_0=3)$ & -0.7028\dots & -1.4056352\dots \\\hline
    $\text{P}\CF_{[14/14]}(\Delta,h_0=3)$           & -0.7029(2)                     & -1.4056353(10)                        \\\hline
  \end{tabular}
  \caption{Comparison between the numerical values of the free energy at $h_0=3$ obtained by a direct numerical evaluation of the TBA equation and by using the analytic continuation given by $\text{P}\CF_{[14/14]}(\Delta,h_0)$ for $\Delta=1/2$ and $\Delta = -\infty$.}
  \label{tab:gn_F_Delta}
\end{table}

In figure \ref{fig:Gn_P} we plot $\text{P}\CF_{[14/14]}(\Delta,h_0)$ as a function of $\Delta$ at fixed $h_0$. At $\Delta = 1/2$ the black dot corresponds to the (correct) free energy
numerically computed using \eqref{fh-betheKinks}, while the red dot is the (wrong) value one would get by naively setting $\Delta=1/2$ in the fermion kernel appearing in \eqref{intetwo}.
We see that the analytic continuation given by $\text{P}\CF_{[14/14]}(\Delta,h_0)$ gives the correct value. The dashed black line corresponds to the asymptotic value for $\Delta = -\infty$ which
should equal to $2\CF(\Delta=1/2,h)$, according to \eqref{eq:N53}. In table \ref{tab:gn_F_Delta} we compare the numerical values of $\CF(h_0)$ obtained by a direct numerical evaluation of the TBA equation 
and by using the analytic continuation given by $\text{P}\CF_{[14/14]}(\Delta,h_0)$ for $\Delta=1/2$ and $\Delta = -\infty$. The two results are in total agreement.

\section{Conclusions}

\label{sec:conclusions}

The $1/N$ expansion provides a non-perturbative resummation of the conventional perturbation theory. 
In this paper we have discussed the interplay between resurgence and the $1/N$ expansion in three integrable theories with a continuous global symmetry: the 
non-linear sigma model, the principal chiral field and the Gross-Neveu models. All these theories have marginal interactions, they are UV free and are affected by IR renormalon singularities.
A notable observable is the relative free energy ${\cal F}(h)$ defined in \eqref{eq:freeenergy}. Its special role comes from the fact that it is  possible to compute it exactly using TBA techniques  \cite{pw} and has
a non-trivial structure (unlike, e.g., $S$-matrix elements in integrable theories).
Standard large $N$ QFT and/or TBA techniques also allow to analytically determine the first $1/N$ coefficients ${\cal F}_k(h)$ defined in \eqref{eq:FNLSM}-\eqref{eq:FGN}.

We have computed ${\cal F}_0$ and ${\cal F}_1$ in the NLSM, given respectively in \eqref{leading-fh} and \eqref{eq:obsfinal}, and determined ${\cal F}_0$ analytically in the PCF model (for the choice of charges in \cite{pcf}), given in \eqref{f0b}. Crucial for the latter computation has been the observation that the NLSM and the PCF kernels can be expanded in $1/N$ if the non-analytic term is treated separately.
In this way we have also been able to check ${\cal F}_1$ in the NLSM by using TBA techniques. 
These expressions, as well as the previously known coefficients ${\cal F}_{0,1}$ in the GN model \cite{fnw1,fnw2}, have then been compared 
to the asymptotic series expansion, one gets in terms of the coupling constant defined in \eqref{eq:LNPTC6}. While the perturbative asymptotic expansion agrees with 
the ones previously determined \cite{mr-ren}, we get a plethora of non-perturbative trans-series terms which are associated to the non-Borel summability of the series
due to the presence of IR renormalons. 

The final results turned out to be different in the three models.
In the NLSM $\CF_0$ contains a non-perturbative term which cannot be captured from the (trivial) perturbative expansion. 
On the other hand, the median Borel resummation of the perturbative series reconstructs the full next-to-leading coefficient ${\cal F}_1$, see fig.\ref{fig:NLSM}.
The series for $\CF_1$ is non-Borel resummable because of the presence of an IR renormalon, yet somehow unexpectedly no non-perturbative terms are missed.
In the PCF model the expansion of $\CF_0$ gives rise to the non-trivial trans-series \eqref{ts-F0}, with a perturbative series affected by an infinite number of IR renormalon singularities.
In this case, resurgence techniques work nicely and allow us to reconstruct the full answer from the perturbative series.  
In the GN model the expansion of both $\CF_0$ and $\CF_1$ give rise to trans-series \eqref{eq:LNPTC16} and \eqref{eq:LNPTC26} which can {\it not} be 
reconstructed from the perturbative series only, using resurgence.

We also studied the behavior of the $1/N$ series for $\CF(h)$. The non-analyticity of the kernel at $N=\infty$ for the NLSM and the PCF models suggest that the $1/N$ expansion of $\epsilon(\theta)$ and $\chi(\theta)$ 
should be divergent asymptotic. This points towards a divergent $1/N$ expansion of $\CF(h)$, as well as of its Legendre transform $e(\rho)$. 
In contrast, the $1/N$ expansion of  $\CF(h)$ (and $e(\rho)$) in the GN model is expected to be convergent.
We have numerically computed higher values of ${\cal F}_k$ in each model  (for some values of $h$) in order to verify these expectations. 
Our results for the NLSM and PCF models are inconclusive. The number of coefficients ${\cal F}_k$ we computed does not allow us to establish whether the series are convergent or divergent asymptotic.
The first possibility is not in contradiction with the $1/N$ non-analyticities of the kernel, because $\CF(h)$ is obtained by integrating $\epsilon(\theta)$ over rapidities and
we cannot exclude that these non-analyticities are smoothed out by the integration procedure. It would be nice to settle this issue in future studies.
In the Gross-Neveu model the expected convergence of the $1/N$ series is numerically confirmed. We analytically continued the series beyond its radius of convergence (see fig.\ref{fig:Gn_P}), where the TBA equations \eqref{intetwo} and  \eqref{chi-ie} no longer make sense, and showed how this continuation gives values of $\CF(h)$ in complete agreement with those obtained for the sine-Gordon theories dual to the GN models with $N=2$ and $N=4$.

There are several directions worth exploring in future studies. From the point of view of the general theory of resurgence, our most important finding 
is its breakdown in certain models, when combined with the $1/N$ expansion. By a breakdown of resurgence we mean that 
the structure of non-perturbative corrections at each order in the $1/N$ expansion can not 
be predicted from the study of the perturbative series only. A better understanding of this breakdown is perhaps the most important problem open by our investigation. There are two possibilities here. One possibility is that this is a feature of the $1/N$ expansion 
and does not apply at finite $N$. It might happen that, when fixing the order in the $1/N$ expansion, the resulting perturbative series in the coupling constant is not sufficiently generic and cannot be used to predict non-perturbative corrections. This would be somewhat similar to the ``Cheshire cat resurgence" in supersymmetric theories \cite{cheshire}. The other possibility is that the phenomenon we have found is generic in theories with renormalons. The detailed study performed for the $O(4)$ sigma model in \cite{abbh1,abbh2} validates standard resurgence expectations and seems to favor the first possibility. Clearly, additional studies are necessary in order to clarify this fundamental issue. 
A detailed resurgent analysis of the free energy of the GN model at finite $N$, along the lines 
of \cite{abbh1,abbh2}, would be very useful. It would be also 
important to understand why resurgence seems to be so successful in the PCF model, at least at leading order in $1/N$, but 
fails in the GN model, though both models belong to the same universality class of integrable, gapped, and UV-free theories. In particular, it would be interesting to clarify if this is related to the different analyticities properties in $1/N$ of the kernels in the two theories. More generally, it would be important to study observables other than $\CF(h)$ which can be computed exactly in the $1/N$ expansion and can be 
analytically decoded in terms of trans-series.

Perhaps it is useful to distinguish three different levels of validity of the theory of resurgence, in order to understand what is at stake. The first, more general level of validity is the statement that observables in quantum theory are given by ambiguity-free Borel--\'Ecalle resummations of trans-series. 
This statement is probably true and it is implicit in many of the early studies of renormalons, like e.g. \cite{David:1982qv,David:1983gz}. All of our results in this paper, including the example of the GN model, vindicate this first level of validity. The second level of validity is the stronger statement that the trans-series is {\it fully} determined by its perturbative part, up to the numerical values of the trans-series parameters. It is this 
second level of validity that breaks down in some of the examples that we have studied, when restricting to a fixed order in the $1/N$ expansion. 
Finally, a third and largely independent issue is whether renormalon singularities and the associated trans-series 
have a semi-classical interpretation, in terms of 
expansions around saddle-points of a classical action. It has been argued that, after a twisted compactification, 
the renormalon sectors of the PCF model and the NLSM can be interpreted 
semiclassically \cite{du-on,cherman-dorigoni-dunne-unsal}. Note however that the successful examples 
of resurgence that we have considered (like e.g. the PCF model at large $N$) are independent of such 
a semiclassical interpretation. It is perfectly conceivable (and, in our opinion, quite likely) that, for theories 
with renormalons at infinite volume, one does not have a semiclassical 
interpretation of the trans-series, but some version of resurgence will be still valid. In that scenario, 
a crucial open question will be  
to find a generalization of perturbation theory which makes it possible to calculate 
the trans-series from first principles, 
and without relying on resurgence properties or integrability. The use of the OPE, as in 
QCD sum rules, goes along this 
direction, but it is clear that a more general procedure has to be devised in order to compute 
general observables for which OPE techniques are in principle not applicable, as it is the case of 
the free energy studied in this paper. 
 
The results that we have obtained apply to the TBA renormalization scheme defined in \eqref{eq:LNPTC6}
and might not be valid in others. It would be useful to better understand if and to what extent resurgence methods depend on the renormalization scheme (e.g., do they apply in $\overline{\rm MS}$?), 
given also the impact that the choice of scheme can have when resumming perturbative series even in absence of renormalons, see e.g. \cite{Sberveglieri:2019ccj,Sberveglieri:2020eko}.

Finally, it would be very interesting to extend these considerations to non-integrable theories. One possibility would be to compute ${\cal F}(h)$ in the quartic linear $O(N)$ model when $m^2<0$ at some order in $1/N$ and check if the result can be reconstructed from a perturbative expansion around the naive vacuum, where IR renormalons have been shown to appear \cite{mr-new}.

\section*{Acknowledgments}

We would like to thank Ramon Miravitllas Mas and Tom\'as Reis for useful discussions and 
comments on the manuscript. LD, GS, and MS are partially supported by INFN Iniziativa Specifica ST\&FI. LD also acknowledges support by the program ``Rita Levi Montalcini'' for young researchers. The work of MM has been supported in part by the ERC-SyG project ``Recursive and Exact New
Quantum Theory" (ReNewQuantum), which received funding from the
European Research Council (ERC) under the European Union's Horizon
2020 research and innovation program, grant agreement No. 810573.

\appendix

\section{$0d$ reduction of quartic vector models}

\label{app:OI}

In this appendix we present the details necessary to reproduce the $1/N$ large order behavior \eqref{eq:OI15} of the coefficients in the $0d$ reduction 
of quartic vector models and provide some further technical comments. To make contact with QFT models, it is useful to introduce an Hubbard-Stratonovich like parameter $\sigma$ to rewrite the quartic term $(\boldsymbol{x}\cdot \boldsymbol{x} )^2$ as
\be
\re^{-\frac gN ( \boldsymbol{x}\cdot \boldsymbol{x} )^2} = \frac{1}{\sqrt{2\pi}} \int_{-\infty}^{\infty} \!\! \rd\sigma \, \re^{-\frac{\sigma^2}{2} + \ri \sqrt{\frac{2g}{N}} \sigma\,  \boldsymbol{x}\cdot \boldsymbol{x}  } \,.
\label{eq:HS1}
\ee
Inserting in \eqref{eq:OI1} and integrating over $\boldsymbol{x}$ gives 
\be
I(m,g) = \sqrt{\frac{N}{2\pi}}2^{-\frac N2}  \int_{-\infty}^{\infty} \!\! \rd\sigma \, \re^{-N K(\sigma)}\,,
\label{eq:HS2}
\ee
where
\be
K(z) = \frac{z^2}2 + \frac 12 \log \Big(\frac{m}{2}- \ri \sqrt{2g} z\Big)\,,
\label{eq:HS3}
\ee
and we rescaled $\sigma\rightarrow \sqrt{N} \sigma$. The function $K$ is complex and the original contour 
of integration is not a downward flow, so that we have to 
decompose it in terms of Lefschetz thimbles. We get two critical points 
\be
z_c^{(\pm)} =  \frac{\ri}{4\sqrt{2g}} \big(-m\pm \sqrt{m^2+16g}\big)\,.
\label{eq:HS4}
\ee
The function $K$ has a branch-cut singularity at $z_{bc} = -\ri m/\sqrt{8g}$. The point $z_c^{(-)}$ sits on top of the branch-cut and $K(z)$ has two saddles for each of the infinite Riemann sheets associated to the log function. The deformed contour passing through $z_c^{(+)}\equiv z_c$ is a regular Lefschetz thimble. As shown in \cite{power}, this is a sufficient condition for Borel summability to the exact result  of the  asymptotic saddle point series expansion around $z_c$. The large-order behavior of the coefficients $c_p$ around a saddle $z_c$ is given by the lowest-order coefficients of the series associated to the so-called adjacent saddles $\hat z_c$ \cite{Berry2}. 
In our case, we have one adjacent saddle $\hat z_c = z_c^{(-)}$  which contributes twice according to the two different analytic continuations of the log function in $K(z)$: 
\be
K^{\eta} (\hat z_c) = {\rm Re} \, K(\hat z_c) + \frac{\ri \pi \eta}{2}\,, \quad \eta = \pm  \,.
\ee
The large order behavior is given by eq.(20) of \cite{Berry2} as follows:
 \be
c_{p} = \frac{\hat I_c}{2\pi \ri} \sum_{\eta = \pm} \eta  \frac{\Gamma(p)}{(K^{\eta}(\hat z_c)-K(z_c))^p} \left(1+{\cal O} \Big(\frac 1p\Big)\right)\,,
\label{eq:HS7}
\ee
where
\begin{align}
\hat I_c^{}    =  \sqrt{\frac{K^{\prime\prime}(z_c)}{K^{\prime\prime}(\hat z_c)}} &   = \frac{\sqrt{m^2+16g}-m}{4\sqrt{g}}  \,, \\
K^{\eta}(z_c^{(-)})-K(z_c^{(+)}) & =  \rho \, \re^{\ri \eta \theta} \,, \quad \quad \rho = \sqrt{Z^2 + \frac{\pi^2}{4}} \,, \quad \theta = \arccos\Big(\frac{Z}{\rho}\Big)\,,
\label{eq:HS8}
\end{align}
and
\be
Z = - \frac{m \sqrt{m^2+16g}}{16g} -\frac 12 \log \frac{\sqrt{m^2+16g}+m}{\sqrt{m^2+16g}-m}\,.
\label{eq:OI14}
\ee
For $p\gg1$ we then get \eqref{eq:OI15}, which agrees with the earlier result (2.16) of \cite{Hikami:1978ya} for $m=1$.
 Note that the large order behavior for $m=1$ and $m=-1$ are related, since $\theta\rightarrow \pi - \theta$  and $\hat  I_c\rightarrow 1/\hat  I_c$ when $m\rightarrow -m$.
 We see that the large order coefficients oscillate with a period given by $\theta$. 
 
 When $\theta = \pi/2$, $c_p=0$ for even $p$ and one has to look at the sub-leading ${\cal O}(1/N)$ corrections. 
This situation is realized when $m=0$. In this case, the integral \eqref{eq:OI1} simply equals to 
\be
I(0,g) = \Big(\frac{N}{16g}\Big)^{\frac{N}{4}} \frac{\sqrt{\pi}}{\Gamma(\frac{N+2}{4})}\,,
\label{eq:HS9}
\ee
and \eqref{eq:OI15} simplifies to
\be
c_{p}(m=0) = \frac{\Gamma(p)}{\pi} \Big(\frac{2}{\pi}\Big)^p \sin \frac{\pi p}{2}\left(1+{\cal O} \Big(\frac 1p\Big)\right) \,.
\label{eq:HS10}
\ee
The sub-leading contributions are captured by the full resurgent (asymptotic) formula \cite{Berry2}
 \be
c_{p} \sim \frac{\hat I_c}{2\pi \ri} \sum_{\eta = \pm} \eta \sum_{q=0}^\infty \frac{\Gamma(p-q) \hat c_{q}^{\eta} }{(K^{\eta}(\hat z_c)-K(z_c))^{p-q}} \,,
\label{eq:HS11}
\ee
where $\hat c_{q}^{\eta}$ are the first terms of the expansion around the adjacent saddle $\hat z_c$, normalized so that  $\hat c_0^{\eta} = 1$. Interestingly, 
$c_{q}^{\eta}= c_{q}$, because these saddles are all equivalent for $m=0$. We easily get
\be
c_{p} = \frac{\Gamma(p)}{\pi} \Big(\frac{2}{\pi}\Big)^p  \Bigg(\sin \frac{\pi p}{2}\bigg(1-\Big(\frac{\pi}{2}\Big)^2\frac{1}{72(p-1)(p-2)}\bigg)-   \cos \frac{\pi p}{2} \bigg(\frac{\pi}{2}\frac{1}{6 (p-1)}\bigg)
+  {\cal O} \Big(\frac{1}{p^3}\Big)\Bigg) \,.
 \label{eq:HS10b}
\ee
This series alternates every two terms and has odd terms parametrically larger than the even ones by a factor of $p$, for large $p$.

\section{Existence, uniqueness and  analyticity in $1/N$ of the TBA equations}

\label{app:analytic}

Both TBA equations \eqref{chi-ie} and \eqref{intetwo}  are instances of a class of equations of the form 
\be
f(\theta) -  \int_{-B}^B\! \! K(\theta-\theta') f(\theta') \rd\theta' = L(\theta)\,,
\label{eq:AELSM0}
\ee
where $L(\theta)$ is a given continuous function, $K(\theta)$ is an integral kernel, and $f(\theta)$ is the function to be determined.  Equations of this kind are known as non-homogeneous Fredholm linear integral equations.
In this appendix we would like to show the existence and uniqueness of the solutions of \eqref{chi-ie} and \eqref{intetwo} in the three models considered in the paper, as well as
some analyticity properties in $1/N$ of the kernel $K(\theta)$.

Let us start by briefly reviewing basic mathematical facts. The above equation \eqref{eq:AELSM0} can be compactly written as a fixed point equation $Tf = f$, where
\be
Tf(\theta) = L(\theta)  +  \int_{-B}^B\! \! K(\theta-\theta') f(\theta') \rd\theta' \,.
\label{eq:AELSM0a}
\ee
Importantly, the operator $T$ is a contraction if
\be
\sup_{-B \leq \theta \leq B} \, \left| \int_{-B}^B\! \! K(\theta-\theta')\rd\theta' \right|  \equiv k <  1\,.
\label{eq:AELSM0b}
\ee
Indeed, for arbitrary functions $f_1$ and $f_2$ we have
\be
|T f_1 - T f_2|  =  | K f_1 - K f_2 | \leq k |f_1 - f_2| < |f_1 - f_2|\,.
\label{eq:AELSM0c}
\ee
If $T$ is a contraction, by the contraction theorem (also known as Banach fixed point) the solution $Tf = f$ exists and is unique, and it can be  written schematically as
\be
f = \sum_{p=0}^\infty K^p L\,.
\label{eq:AELSM0d}
\ee
The key quantity to study is the kernel $K$, which will be discussed separately for each model in 
the next subsections. Our goal will be two-fold: we will first show that $K$ is a contraction, hence proving the existence and uniqueness of the solution, in each case. Secondly, we will study the analyticity in $1/N$ of the solution. Thanks to \eqref{eq:AELSM0d}, the analyticity of the solution in a certain region can be established by proving that all the iterated kernels $K^p$ are analytic in that region (in particular it is not enough to establish this property for $K$ alone). We will see that there is analyticity in a disk close to the origin of the $1/N$ plane, but not including it, for the NLSM and PCF, while in the case of GN the solution is analytic in the origin within a certain radius. The analysis will be quite detailed for the NLSM and more sketchy for the PCF and GN models.

\subsection{Non-Linear sigma model}

For the non-linear sigma model  the kernel reads
\be
K(\theta) =  \frac{1}{4\pi^2} \left(\psi\Big(1+\frac{\ri \theta}{2\pi}\Big) -\psi\Big(\frac 12+\frac{\ri \theta}{2\pi}\Big) 
 +\psi\Big(\frac 12 + \Delta+\frac{\ri \theta}{2\pi}\Big) -\psi\Big(\Delta+\frac{\ri \theta}{2\pi}\Big) \right) + c.c. \;,
\label{eq:AELSM1}
\ee
where $\psi$ is the digamma function and $\Delta = 1/(N-2)$.
The Fourier transform $\widetilde K(\omega)$ of $K(\theta)$ admits a simple analytic expression \cite{hn}, which is straightforwardly obtained by expanding the digamma functions.
For $\Delta >0$ it reads
\be
\widetilde K(\omega) \equiv \int_{-\infty}^\infty \! \re^{\ri \omega \theta} \, K(\theta) \rd\theta = \frac{1+ \re^{\pi |\omega| (1-2\Delta)}}{1+\re^{\pi |\omega|}}\,.
\label{eq:AELSM4}
\ee
The kernel \eqref{eq:AELSM1} is point-wise positive definite, i.e. $K(\theta) \geq 0$ for any $\theta\in \mathbb{R}$.\footnote{It is also positive definite in the sense that
\be
\int_{-\infty}^\infty \!\! d\theta d\theta' \; K(\theta-\theta') f(\theta) f(\theta') \geq 0
\ee
for any square integrable function $f(\theta)$.}
This allows us to immediately prove that the kernel is a contraction and there exists a unique solution to \eqref{eq:AELSM0} for any finite positive $N$. Indeed,
\be
\sup_{-B \leq \theta \leq B} \,  \int_{-B}^B\! \! | K(\theta-\theta')| \rd\theta' <    \int_{-\infty}^\infty\! \! K(\theta-\theta') \rd\theta'  = \widetilde K(0) = 1\,.
\label{eq:AELSM4a}
\ee
Following the appendix B of \cite{ll}, 
we can also prove some analyticity properties in $\theta$ of $f(\theta)$, by showing that  the kernel $K$ and all its derivatives are bounded, and so are the iterated kernels $K^n$. 
We omit this analysis and instead focus on analyticity in $\Delta$ for small values of $\Delta$, which is the limit relevant for large $N$. 
Let then $\Delta = z/2$ be complex and focus on a small disc $D$ around the point $0\leq \lambda\ll 1$, defined as $z= \lambda + \alpha + i \beta$ and
$\alpha^2+\beta^2 \leq \delta^2$, where $0\leq \delta \leq \lambda/\sqrt{2}$. The kernel \eqref{eq:AELSM1} can be conveniently written as
\be
\pi^2 K(\theta,z) = \frac{z}{z^2+(\theta/\pi)^2} + F(\theta,z)\,,
\label{eq:AELSM5}
\ee
where
\be
F(\theta,z) \equiv \sum_{n=1}^\infty (-1)^n \bigg(\frac{z+n}{(z+n)^2+(\theta/\pi)^2} - \frac{n}{n^2+(\theta/\pi)^2} \bigg)\,.
\label{eq:AELSM6}
\ee
Note that the first term in \eqref{eq:AELSM5} coincides with the integral kernel appearing in the Bethe ansatz solution of the Lieb--Liniger model \cite{ll}, while the second term
is an analytic function of $z$ for small $z$. For sufficiently small $z$, we have
\be
|F(\theta,z)|   = |z F'(\theta) + {\cal O}(z^2)| \leq (\lambda+\delta) |F'(\theta) | \,,
\label{eq:AELSM7}
\ee
where
\be
F'(\theta) = -\frac{\pi^2}{2\theta^2} +\frac{\pi^2}{8} \Big({\rm sech}^2 (\theta/2)+{\rm csch}^2 (\theta/2) \Big)\,.
\ee
We also have \cite{ll}
\be
\bigg|  \frac{z}{z^2+(\theta/\pi)^2} \bigg| \leq \frac{\lambda+\delta}{(\lambda-\delta)^2+(\theta/\pi)^2}  \leq \frac{\lambda+\delta}{(\lambda-\delta)^2} \,.
\label{eq:AELSM8}
\ee
For sufficiently small $\lambda$, we then get
\be
| K(\theta,z)| \leq \frac{1}{\pi^2} \Big(\frac{\lambda+\delta}{(\lambda-\delta)^2} +  (\lambda+\delta)  |F'(\theta) |  \Big) \,.
\label{eq:AELSM9}
\ee
Let us now define the iterated kernel ($z$ dependence omitted for simplicity)
\be
K^{(p+1)}(\theta-\theta') \equiv \int_{-B}^B K^{(p)}(\theta-\theta'') K(\theta''-\theta')\rd\theta''\,, \quad K^{(1)}(\theta)\equiv K(\theta) \,, \quad p\geq 1\,, 
 \label{eq:AELSM10}
\ee
and suppose that for a certain $p\geq 1$ and for any $-B\leq \theta\leq B$
\be
|K^{(p)}(\theta) | \leq  \frac{C^{p-1}}{\pi^2} \frac{\lambda+\delta}{(\lambda-\delta)^2} \,.
 \label{eq:AELSM11}
\ee
Then
\begin{align}
|K^{(p+1)}(\theta-\theta')| &=   \left|  \int_{-B}^B K^{(p)}(\theta-\theta'') K(\theta''-\theta')\rd\theta'' \right|\leq \frac{C^{p-1}}{\pi^2} \frac{\lambda+\delta}{(\lambda-\delta)^2} \left| \int_{-B}^B  K(\theta''-\theta')\rd\theta'' \right| \nonumber \\ & \leq  \frac{C^{p-1}}{\pi^2} \frac{\lambda+\delta}{(\lambda-\delta)^2} \Big(\frac{1}{\pi^2}  \int_{-B}^B   \frac{\lambda+\delta}{(\lambda-\delta)^2+\delta \theta^2}  \rd  \theta'' + \frac{\lambda+\delta}{\pi^2}  \int_{-B}^B  |F'(\delta \theta)|    \rd \theta''  \Big)\,,
 \label{eq:AELSM12}
\end{align}
where $\delta \theta \equiv (\theta''-\theta')/\pi$. Let us first set $F=0$, in which case we recover the same kernel as in  \cite{ll}. Performing the integral, we have, for $\delta \ll \lambda$,
\be
|K^{(p+1)}(\theta-\theta')|  \leq   \frac{C^{p-1}}{\pi^2} \frac{\lambda+\delta}{(\lambda-\delta)^2} \Big[1-\frac{1}{\pi} \Big( \arctan\Big(\frac{\pi \lambda}{B-\theta}\Big)   + \arctan\Big(\frac{\pi \lambda}{B+\theta}\Big)\Big) \Big] \,.
 \label{eq:AELSM13}
\ee
For any positive $B$, $\theta \leq B$ and  (small) $\lambda$ strictly greater than 0,  the square bracket is bounded by $(1-\epsilon)$, with $0<\epsilon<1$. Then, if we choose
\be
C  = 1- \epsilon\,,
 \label{eq:AELSM14}
\ee
the relation \eqref{eq:AELSM11}, if valid for $p$, is also valid for $p+1$. Since it applies for $p=1$, it follows that it is valid for any $p\geq 1$. It then follows that the resolvent kernel
\be
\kappa(\theta,z) \equiv \sum_{p=0}^\infty K^{(p+1)}(\theta,z)\,,
 \label{eq:AELSM15}
\ee
and the solution of \eqref{eq:AELSM0}
\be
f(\theta) = L(\theta) + \int_{-B}^B\!\! \kappa(\theta-\theta') L(\theta') \rd\theta'
 \label{eq:AELSM16}
\ee
exists and is analytic for $\Delta > 0$. The point $\Delta=0$ is excluded, because for $\lambda=0$ we have $C=1$ and the resolvent series does not converge.
This reproduces the analysis in appendix B of \cite{ll}.

Let us now come back to the situation of interest with $F'\neq 0$. The second integral in \eqref{eq:AELSM12} is easily bounded by a finite constant $M$. For instance, we have
\be
|F^\prime(\theta)| < \frac{\pi^2}{12} \frac{1}{1+ \theta^2/8}
\ee
and 
\be
\frac{1}{\pi^2} \int_{-B}^B  |F'( \theta''-\theta')|  \rd  \theta'' < \frac{1}{\pi^2} \int_{-\infty}^\infty    \frac{\pi^2}{12} \frac{\rd \theta}{1+ \theta^2/8} = \frac{\pi}{3\sqrt{2}} \,.
 \label{eq:AELSM17}
\ee
We then get
\be
|K^{(p+1)}(\theta-\theta')|  \leq   \frac{C^{p-1}}{\pi^2} (1-\epsilon + \lambda M )  \,.
 \label{eq:AELSM18}
\ee
For any finite $M$ there exists a sufficiently small $\lambda$ (and $\delta \ll \lambda$) such that the bracket in \eqref{eq:AELSM18}
equals $(1-\epsilon')$, with $\epsilon'>0$.  By choosing $C=1-\epsilon'$ we see that the resolvent kernel and the solution are analytic for small $\Delta>0$ but not at $\Delta=0$. 
The analyticity region is given by $0< |\Delta| < 1/2$, with $\Delta = -1/2$ a non-analytic point, as can be inferred from \eqref{eq:AELSM6}.

\subsection{Principal Chiral Field}

In the PCF model the kernel equals
\be
K(\theta)  =
 \frac{1}{4\pi^2} \bigg(2
\psi\Big(1 +\frac{\ri \theta}{2\pi}\Big) -  \psi\Big(1-\bar\Delta +\frac{\ri \theta}{2\pi}\Big) - \psi\Big(\bar\Delta+ \frac{\ri \theta}{2\pi}\Big)  \bigg) + c.c. \;,
\label{eq:PCF1}
\ee
where  $\bar\Delta=1/N$. It can also be rewritten as
\be
\pi^2 K(\theta,2\bar\Delta) = \frac{2\bar\Delta}{(2\bar\Delta)^2+(\theta/\pi)^2} + F(\theta,2\bar\Delta)\,,
\label{eq:PCF4}
\ee
where 
\be
F(\theta,z) \equiv \sum_{n=1}^\infty  \bigg(\frac{2n+z}{(z+2n)^2+(\theta/\pi)^2}+\frac{2n-z}{(z-2n)^2+(\theta/\pi)^2} - \frac{4n}{(2n)^2+(\theta/\pi)^2} \bigg)\,.
\label{eq:PCF5}
\ee
For $0< \bar\Delta < 1$ its Fourier transform reads
\be
\widetilde K(\omega)  \equiv \int_{-\infty}^\infty \! \re^{\ri \omega \theta} \, K(\theta) \rd\theta =  \frac{\re^{-2\pi |\omega|(1-\bar \Delta)}+  \re^{-2\pi |\omega|\bar \Delta}
-2\re^{-2\pi |\omega|}}{1-\re^{-2\pi |\omega|}} \,.
\label{eq:PCF2} 
\ee
Like in the NLSM, the kernel is point-wise positive definite and is a contraction:
\be
\sup_{-B \leq \theta \leq B} \,  \int_{-B}^B\! \! | K(\theta-\theta')| \rd\theta' <    \int_{-\infty}^\infty\! \! K(\theta-\theta') \rd\theta'  = \widetilde K(0) = 1\,.
\label{eq:PCF3}
\ee
We can similarly study the analyticity in $1/N=\bar\Delta$ of $K$. We will be very brief since the analysis is similar to the one performed in the NLSM.
Let $\bar \Delta = z/2$ be complex and focus on a small disc $D$ around the origin, defined as in the NLSM case. 
The function $F(\theta,z)$ is analytic around $z=0$ and can be expanded for small $z$ as 
\be
|F(\theta,z)|   = |z^2 F^{\prime\prime}(\theta) +{\cal O}(z^3)| < (\lambda+\delta)^2 |F''(\theta) | \,.
\label{eq:PCF6}
\ee
For sufficiently small $z$, the analyticity properties of the PCF kernel coincide with those of both the NLSM model and the Lieb--Liniger model. 
The iterated kernels $K^{(p+1)}$ are bounded as in \eqref{eq:AELSM12}, with the replacement $(\lambda+\delta) |F'(\delta \theta)|\rightarrow (\lambda+\delta)^2 |F''(\delta \theta)|$
in the last term of the second row of \eqref{eq:AELSM12}. The integral involving $|F''(\delta \theta)|$ is easily bounded by a finite constant, so we conclude
that the kernel in the PCF model is analytic for $0<|\bar \Delta| <1$, but not at $\bar \Delta =0$ and at $\bar \Delta = \pm 1$, as evident from \eqref{eq:PCF5}.

\subsection{Gross-Neveu model}

In the Gross-Neveu model the kernel equals 
\be
K(\theta) 
=  \frac{1}{4\pi^2} \bigg(
 \psi\Big(\frac{\ri \theta}{2\pi}\Big)- \psi\Big(\frac 12 +\frac{\ri \theta}{2\pi}\Big) +  \psi\Big(\frac 12-\Delta +\frac{\ri \theta}{2\pi} \Big)
- \psi\Big(1-\Delta +\frac{\ri \theta}{2\pi}\Big) 
 \bigg) + c.c. \;,
\label{eq:AEGN1}
\ee
where $\Delta = 1/(N-2)$, which can also be written as
\be
\pi^2 K(\theta) =   \sum_{n=1}^\infty (-1)^{n}  \left(\frac{n-2\Delta}{(2\Delta-n)^2 + (\theta/\pi)^2}  - \frac{n}{n^2 + (\theta/\pi)^2}\right)
\label{eq:AEGN2} \,.
\ee
Its Fourier transform for $\Delta <1/2$ reads
\be
\widetilde K(\omega)  \equiv \int_{-\infty}^\infty \! \re^{\ri \omega \theta} \, K(\theta) \rd\theta = \re^{-\pi|\omega|} \frac{\re^{-2\pi |\Delta| |\omega|}-1}{1+\re^{-\pi |\omega|}} \,. 
\label{eq:AEGN3} 
\ee
The kernel \eqref{eq:AEGN1} is not point-wise positive, as in the NLSM and PCF models, but it is still a contraction. 
For small $\Delta$ this can be established by noticing that the kernel vanishes  for $\Delta = 0$. For any $B$ and $-B\leq \theta \leq B$, for small $\Delta$, we have
\be
 \int_{-B}^B\! \! | K(\theta-\theta')| \rd\theta'  \approx \Delta   \int_{-B}^B\! \! | K'(\theta-\theta',\Delta=0)| \rd\theta'    <  2\Delta M \,,
 \label{eq:AEGN4} 
\ee
where $M$ is finite. So, for sufficiently small $\Delta$, $K$ is a contraction. Numerically we see that $K$ is a contraction for arbitrary $\Delta$, not necessarily small.
So the unique solution \eqref{eq:AELSM0d} is guaranteed to exist. The analyticity in $\Delta$ is trivial in the GN model (see also Appendix A of  \cite{fnw1}). Let again be $\Delta = z/2$ and $z=\lambda + \alpha + \ri \beta$, where $\alpha$ and $\beta$ span a disc of radius $\delta$ around the point $\lambda$, with $\lambda>0$. 
Suppose that for a certain $p\geq 1$ 
\be
|K^{(p)}(\theta) | \leq  \sigma^{p}\,, \quad \sigma \equiv (\lambda+\delta) M\,.
 \label{eq:AEGN5}
\ee
Then
\be
|K^{(p+1)}(\theta-\theta')| =  \Big|  \int_{-B}^B K^{(p)}(\theta-\theta'') K(\theta''-\theta')\rd\theta'' \Big|\leq \sigma^{p} \int_{-B}^B  |K(\theta''-\theta')| \rd\theta''  \leq \sigma^{p+1} \,.
\label{eq:AEGN6}
\ee
The relation \eqref{eq:AEGN5}, if valid for $p$, is then also valid for $p+1$. Since it applies for $p=1$, it follows that it is valid for any $p\geq 1$. 
We see that $K$ and the associated solution $f(\theta)$ are analytic for small $\Delta>0$, {\it including} $\Delta=0$.  By looking at \eqref{eq:AEGN2} we can determine the radius of convergence
of the small $\Delta$ expansion. The above expression is analytic up to the point $2\Delta = z < 1$. Replacing $z=1+w$ in \eqref{eq:AEGN2}  gives
\be
\pi^2 K(\theta) = \frac{w}{w^2 + (\theta/\pi)^2} - F(\theta,w) 
\label{eq:AEGN7} \,.
\ee
with the function $F$ as in \eqref{eq:AELSM6}. We see that around $z=1$ the kernel is locally identically to the one of the NLSM close to the origin. 
Hence the point $z=1$ is non-analytic. 
We conclude that the large $N$ expansion should be convergent with a radius of convergence around $\Delta=0$ 
\be 
\rho  = \frac 12 \,.
\ee

\bibliographystyle{JHEP}
\bibliography{Refs}

\end{document}